\documentclass[superscriptaddress,aps,twocolumn,longbibliography]{revtex4-1}
\usepackage{graphicx}
\graphicspath{{./}}
\usepackage{bm,psfrag,amsmath,amssymb,epsfig,grffile,float}
\usepackage{tabularx,graphicx,epsf,color}
\usepackage{latexsym}
\usepackage{amssymb}
\usepackage{amsmath}
\usepackage{amsfonts}
\usepackage{upgreek}
\usepackage{bm}
\usepackage{multirow}
\usepackage{enumitem}
\usepackage{color}
\usepackage{hyperref}
\hypersetup{
colorlinks = true,
linkcolor = [rgb]{0.70,0.13,0.13},
citecolor = [rgb]{0.13,0.55,0.13},
urlcolor = [rgb]{0.25, 0.41, 0.88}}

\usepackage{tikz}
\usetikzlibrary{positioning}

\usepackage{amsmath}
    \usepackage{amssymb}
    \usepackage{amsthm}
    \usepackage{amsfonts}
    \usepackage{dsfont}
    \usepackage[bb=boondox]{mathalfa}

\usepackage{mathtools}

\def\sss#1{{\scriptscriptstyle #1}}

\def\ssf#1{\sss{\textsf{#1}}}
\def\tket#1{{|  #1 \rangle}}
\def\tbra#1{{\langle #1|}}

\def\sket#1{{|  \, #1 \,  \rangle}}
\def\sbra#1{{\langle \,  #1 \, |}}

\def\sketbra#1#2{\sket{#1}\sbra{#2}}
\def\sexpect#1#2#3{{\langle \, #1 \, | \,  #2  \, | \, #3 \, \rangle}}
\def\vrh{\varrho}
\def\vep{\varepsilon}
\def\Ups{\Upsilon}
\def\Rc{{\rm c}}
\def\CW{{\cal W}}

\def\CS{{\cal S}}
\def\CO{{\cal O}}
\def\CL{{\cal L}}
\def\CM{{\cal M}}
\def\CLt{\widetilde\CL}
\def\CLH{\CL\ns_H}
\def\CLD{\CL\ns_D}
\def\SA{{\textsf{A}}}
\def\SB{{\textsf{B}}}
\def\SE{{\textsf{E}}}
\def\SU{{\textsf{U}}}
\def\SS{{\textsf{S}}}
\def\Ng{{N\ns_{\rm g}}}
\def\Nv{{N\ns_{\rm v}}}

\def\Bdel{{\boldsymbol{\delta}}}

\def\Bn{{\boldsymbol{n}}}
\def\Bm{{\boldsymbol{m}}}
\def\Ba{{\boldsymbol{a}}}

\def\Br{{\boldsymbol{r}}}
\def\Bs{{\boldsymbol{s}}}
\def\BS{{\boldsymbol{S}}}
\def\BR{{\boldsymbol{R}}}

\def\Bx{{\boldsymbol{x}}}
\def\By{{\boldsymbol{y}}}
\def\Bp{{\boldsymbol{p}}}
\def\BM{{\boldsymbol{M}}}
\def\BN{{\boldsymbol{N}}}
\def\Bmu{{\boldsymbol{\mu}}}
\def\Bnu{{\boldsymbol{\nu}}}

\def\xhat{{\hat\Bx}}
\def\yhat{{\hat\By}}

\def\Bsigma{{\boldsymbol{\sigma}}}
\def\Bmu{{\boldsymbol{\mu}}}
\def\vrh{\varrho}

\def\frac#1#2{{\textstyle{#1 \over #2}}}
\def\half{\frac{1}{2}}
\def\nd{^{\vphantom{\dagger}}}
\def\ns{^{\vphantom{*}}}
\def\np{^{\vphantom{\prime}}}
\def\yd{^\dagger}
\def\etc{{\it etc.\/}}
\def\etal{{\it et al.\/}}
\def\ie{{\it i.e.\/}}

\def\Ie{{\it I.e.\/}}
\def\viz{{\it viz.\/}}

\def\CC{{\,,\,}}

\def\util{{\tilde u}}

\def\Mzero{\mathbb{0}}
\def\Mone{\mathbb{1}}
\def\Nc{{N\ns_\Rc}}
\def\Xtil{{\widetilde X}}
\def\Ytil{{\widetilde Y}}
\def\Ztil{{\widetilde Z}}
\def\Qtil{{\widetilde Q}}
\def\Wtil{{\widetilde W}}
\def\Phit{{\widetilde\Phi}}

\def\Gat{{\widetilde\Gamma}}
\def\tht{{\tilde\theta}}
\def\mut{{\tilde\mu}}
\def\RLambda{{\rm\Lambda}}

\def\Detil{{\widetilde\Delta}}
\def\detil{{\tilde\delta}}
\def\Lat{{\widetilde\RLambda}}
\def\Tra{\mathop{\textsf{Tr}}}

\def\MZ{{\mathbb Z}}
\def\RI{{\rm I}}
\def\RP{{\rm P}}
\def\RPi{{\rm\Pi}}
\def\tra{^\ssf{T}}
\def\Rep{\textsf{Re}\,}
\def\Imp{\textsf{Im}\,}
\def\half{\frac{1}{2}}

\def\Wx{{W_x}}
\def\Wy{{W_y}}
\def\Wtx{{\Wtil_x}}
\def\Wty{{\Wtil_y}}
\def\Nx{{N_x}}
\def\Ny{{N_y}}
\def\munu{\sket{\Bmu}\sbra{\Bnu}}

\def\vphi{\varphi}
\def\vvphi{{\vec\vphi}}
\def\QAA{Q^{\SA\SA}}
\def\QAB{Q^{\SA\SB}}
\def\QBA{Q^{\SB\SA}}
\def\QBB{Q^{\SB\SB}}

\usepackage[mathlines]{lineno}

\usepackage{tcolorbox} 

\begin{document}

	\title{A study of dissipative models based on Dirac matrices}
	\author{Jyotsna Gidugu}
	\email{jgidugu@ucsd.edu}
	\author{Daniel P. Arovas}
	\email{darovas@ucsd.edu}

	\affiliation{Department of Physics, University of California San Diego, La Jolla, CA 92093, USA}
	
	\date{\today}
 
	\begin{abstract}

We generalize the recent work of Shibata and Katsura \cite{Shibata2019}, who considered
a $S=\half$ chain with alternating $XX$ and $YY$ couplings in the presence of dephasing, 
the dynamics of which are described by the GKLS master equation. Their model is equivalent to a non-Hermitian system described by the Kitaev formulation
\cite{Kitaev2006} in terms of a single Majorana species hopping on a two-leg ladder
in the presence of a nondynamical $\MZ\ns_2$ gauge field. Our generalization involves
Dirac gamma matrix `spin' operators on the square lattice, and maps onto a non-Hermitian
square lattice bilayer which is also Kitaev-solvable. We describe the exponentially many
non-equilibrium steady states in this model. We identify how the spin degrees of 
freedom can be accounted for in the 2d model in terms of the gauge-invariant 
quantities and then proceed to study the Liouvillian spectrum. We use a genetic 
algorithm to estimate the Liouvillian gap and the first decay modes for large 
system sizes. We observe a transition in the first decay modes, similar to that
in ref. \cite{Shibata2019}. The results we obtain are consistent with a perturbative 
analysis for small and large values of the dissipation strength.
\end{abstract}
	
\pacs{Valid PACS appear here}
\maketitle

\section{Introduction}
Open quantum systems afford us the opportunity to study phenomena such as relaxational quantum dynamics
for systems coupled to a bath \cite{Breuer2007}.  Typically this involves `integrating out' or eliminating
in some way the bath degrees of freedom, resulting in a dynamics for the system itself in terms of its reduced density matrix: ${\dot\vrh}=\CL\vrh$, where $\CL$ is the Liouvillian operator.  At long times, the system
relaxes to a non-equilibrium steady state (NESS); the existence of a NESS is guaranteed by the dynamics, 
but under special circumstances owing to, for example, extra conserved quantities, the NESS may not be unique. 

For noninteracting systems, hybridization with the bath degrees of freedom still results in a solvable
(quadratic) model \cite{Prosen2008}. For interacting systems, solvable models are rare, and numerical 
approaches are challenging.  This is especially true for density matrix evolution since one must keep track 
of not just populations $\tket{\alpha}\tbra{\alpha}$ but also the coherences $\tket{\alpha}\tbra{\beta}$ with
$\alpha\ne\beta$, effectively squaring the size of the problem {\it vis-a-vis\/} the system's Hilbert
space dimension.

Recently, Shibata and Katsura \cite{Shibata2019} (SK) described a model of open system dynamics based
on the GKLS master equation \cite{ChruPa2017} which, though interacting, is solvable in the sense of
Kitaev's celebrated honeycomb lattice Hamiltonian model \cite{Kitaev2006}.  That is to say, the evolution
of $\vrh(t)$ under the Liouvillian $\CL$ is effectively described by a non-interacting dynamics in the 
presence of a static $\MZ\ns_2$ gauge field.  While in each gauge sector the evolution is described by
a quadratic, albeit non-Hermitian, Hamiltonian, there are exponentially many gauge sectors to evaluate
(which in general have no discrete space group symmetries), and in this sense the general problem is 
intractable.  For the Hermitian Kitaev model, oftentimes the ground state may be ascertained with help
from a remarkable theorem by Lieb \cite{Lieb1989}, which provides valuable information regarding much of 
the gauge-invariant content of the ground state, \ie\ the $\MZ\ns_2$ plaquette fluxes.  For the non-Hermitian
case, however, we know of no generalization of Lieb's theorem which constrains the gauge-invariant content
of, say, the longest-lived decaying density matrix.  Thus, in general one must resort to numerics if one is
interested in the complex spectrum of $\CL$. 

The Shibata-Katsura construction involves a $S=\half$ chain where each site is coupled to an environmental
bath.  Within the GKLS formalism, this results in an effective two leg ladder system, where one leg
corresponds to the bra states and the other to the ket states of the density matrix, and the rungs of the
ladder contribute non-Hermitian terms which result from the effective elimination of the bath degrees of
freedom.  The ladder is thus three-fold coordinated, and the model is constructed so that it satisfies
the Kitaev solvability criteria \S \ref{SKsec}). Our main goal is to introduce and analyze a generalization
of the SK model to two space dimensions, based on a $4\times 4$ gamma matrix generalization of the Hamiltonian
Kitaev model \cite{YZK09,WAH09}.  As the dissipative SK model is described by non-Hermitian Hamiltonian 
evolution on the ladder, our model is described by such an evolution on a square lattice bilayer.
As we shall see, while our model is a direct analog of SK in some sense,
it also entails some important differences -- in particular, an extensive number of conserved quantities
leading to exponentially many NESSes.

We first discuss various preliminaries, including the GLKS master equation, its vectorization and
description in terms of non-Hermitian Hamiltonian evolution on a product Hilbert space, the Shibata-Katsura
model, gamma matrix generalizations of the Kitaev honeycomb model, and finally our extension of SK to
a dissipative square lattice model involving $4\times 4$ Dirac matrix `spin' operators.

{\sl Note:} While this paper was in the final stages of preparation, two analyses of
a largely equivalent model appeared on the arXiv \cite{Dai23,Scheurer23}.

\section{Preliminaries}
\subsection{The GKLS master equation}
An open quantum system $\SS$ is one which unitarily co-evolves with an 
environment $\SE$ under a Hamiltonian $H=H\ns_\SS+H\ns_\SE+ H\ns_{\rm int}$,
where $H\ns_{\rm int}$ couples $\SS$ and $\SE$.
The expectation of any operator $\CO$ restricted to $\SS$ is given by
$\langle\CO(t)\rangle=\Tra \big(\vrh\ns_S(t)\,\CO\big)$, where $\vrh\ns_\SS(t)$ is
the time-dependent reduced density matrix of $\SS$, \ie\ $\vrh\ns_\SS(t)=\Tra\ns_\SE\,\vrh\ns_\SU(t)$, where $\vrh\ns_\SU(t)$ is the full density matrix describing the `universe' $\SU=\SS\cup\SE$.  Under certain
assumptions, the dynamics of the system's reduced density matrix is described
by the GKLS master equation \cite{ChruPa2017,Breuer2007},
\begin{equation}
{d\vrh\over dt} = -i\big[H,\vrh\big] + \sum_a
\!\Big( L\nd_a\,\vrh\,L\yd_a - \half L\yd_a L\nd_a\,\vrh -
\half \vrh\,L\yd_a L\nd_a\Big)\quad,
\label{GKLS}
\end{equation}
Here and henceforth we drop the subscript $\SS$ on $\vrh\ns_\SS$. The
$\{L\ns_a\}$ are the Lindblad jump operators, which describe the effects of
the system-environment coupling on $\vrh$ after the environment is traced out.
$H$ is the `Lamb shift Hamiltonian', which commutes with $H\ns_\SS$ and includes
renormalizations of the system's unperturbed energy levels resulting from the
environmental couplings.  In the absence of all such couplings, we recover
the usual Liouville evolution ${\dot\vrh}=-i[H\ns_\SS,\vrh]$.

The full GKLS evolution in eqn. \ref{GKLS} is of the form ${\dot\vrh}=\CL\vrh$.
Assuming $\CL$ is time-independent, one may formally write
$\vrh(t)=\exp(\CL t)\vrh(0)$, which defines for each $t$ a map $\Phi\ns_t\colon \vrh(0)\mapsto\vrh(t)$ which possesses the following salient properties:
(i) linearity, (ii) trace-preserving, (iii) Hermiticity preserving, and
(iv) complete positivity \cite{Breuer2007}.  Writing $\vrh(t)=\sum_{j,k}
\vrh\ns_{jk}(t)\,\tket{j}\tbra{k}$ in terms of basis states, we may write
${\dot\vrh}\ns_{jk}=\CL\ns_{jk,lm}\,\vrh\ns_{lm}$, where $\CL\ns_{jk,lm}$
is a supermatrix of dimension $N^2$, where $N$ is the dimension of the basis
and $(jk)/(lm)$ are composite indices.  Generically $\CL$ is not a normal matrix,
\ie\ $[\CL,\CL\yd]\ne0$, and its eigenvalues $\Lambda\ns_a$ may be complex.  
However, since the evolution is trace-preserving, one has that $\delta\ns_{jk}$ is a 
left-eigenvector of $\CL$ with eigenvalue zero.  The corresponding right 
eigenvector is the NESS, $\vrh^\ssf{NESS}_{lm}$.
Under special circumstances there may be more than one NESS \cite{Albert2018}. 
Positivity entails that $\Rep\Lambda\ns_a \le 0$ for each eigenvalue of the
Liouvillian $\CL$.

When each jump operator is Hermitian, then from eqn. \ref{GKLS} we have that the
infinite temperature state $\vrh\propto\Mone$ is a valid NESS.  Furthermore, if
$H$ as well as all the jump operators commute with a set of independent projectors
$\{\RP\ns_s\}$ with $s\in\{1,\ldots,K\}$, then any density matrix of the form
\begin{equation}
\vrh=c\ns_0\,\Mone + \sum_{s=1}^K c\ns_s\,\RP\ns_s
\label{Gibbs}
\end{equation}
is also a valid NESS. This shall be the case for the model we investigate below.
Thus we shall describe a system where there is relaxation to a degenerate block
of NESSes.  While such solutions to GKLS depend on the form of $H$ and the jump operators
$\{L\ns_a\}$, they are independent of the various coupling constants (so long as they
remain finite), and we shall consider them all to be infinite temperature states.

\subsection{Equivalent non-Hermitian Hamiltonian}
Any density matrix $\vrh=\sum_{m,n}\vrh\ns_{mn}\,\sket{m}\,\sbra{n}$ may be represented in 
vector form as
\begin{equation}
\vrh\longrightarrow \sket{\vrh} \equiv \sum_{m,n}\vrh\ns_{mn}\,\sket{m}\otimes\sket{n}\quad.
\end{equation}
Thus, the bra vector $\sbra{n}$ is replaced by the corresponding ket vector $\sket{n}$, \ie\ $\sket{m}\sbra{n}\to\sket{m}\otimes\sket{n}$.
If $B$ is any operator, then under vectorization we have
\begin{equation}
\begin{split}
\sbra{n} \,B &=\sum_k \sexpect{n}{B}{k}\sbra{k} \\
&\longrightarrow \sum_k \sket{k}\sexpect{k}{B\tra}{n}=B\tra\,\sket{n}\quad.
\end{split}
\end{equation}
The GKLS master equation eqn. \ref{GKLS} then takes the vectorized form
 \begin{equation}
i\,{d\over dt}\,\sket{\vrh}= \CW\,\sket{\vrh}\quad,
\label{LSE}
\end{equation}
where \cite{Dzhioev_2012}
\begin{align}
\CW&=H\otimes\Mone - \Mone\otimes H\tra +\label{NHM}\\
&\quad i\sum_r\Big(L\ns_r\otimes L^*_r - 
\half\, L\yd_r\, L\ns_r\otimes\Mone - \Mone\otimes \half\, L\tra_r\,L^*_r\Big)\quad.\nonumber
\label{Wdef}
\end{align}
Note that operators $\CO$ acting on the $\sket{n}$ component of the product $\sket{m}\otimes\sket{n}$ appear as transposes $\CO\tra$, 
since they would normally act to the left on $\sbra{n}$\,.  Eqn. \ref{LSE} takes the form of an effective Schr{\"o}dinger equation, with 
$\sket{\vrh(t)}$ evolving according to the non-Hermitian effective Hamiltonian $\CW$ acting on a doubled Hilbert space.
For any operator $\CO$, we may compute the trace in the vectorized representation according to
\begin{equation}
\Tra(\CO\vrh)=\sexpect{\RI}{\CO\otimes\Mone}{\vrh}\quad,
\end{equation}
where $\sbra{\RI}=\sum_n \sbra{n}\otimes\sbra{n}$.  The eigenvalues of $\CW$, which
we denote by $\{E\ns_a\}$, are related to those of the Liouvillian 
by $E\ns_a=i\Lambda\ns_a$.

\subsection{Shibata-Katsura Model}\label{SKsec}
\begin{figure}[t]
\begin{centering}
\includegraphics[width=0.45\textwidth]{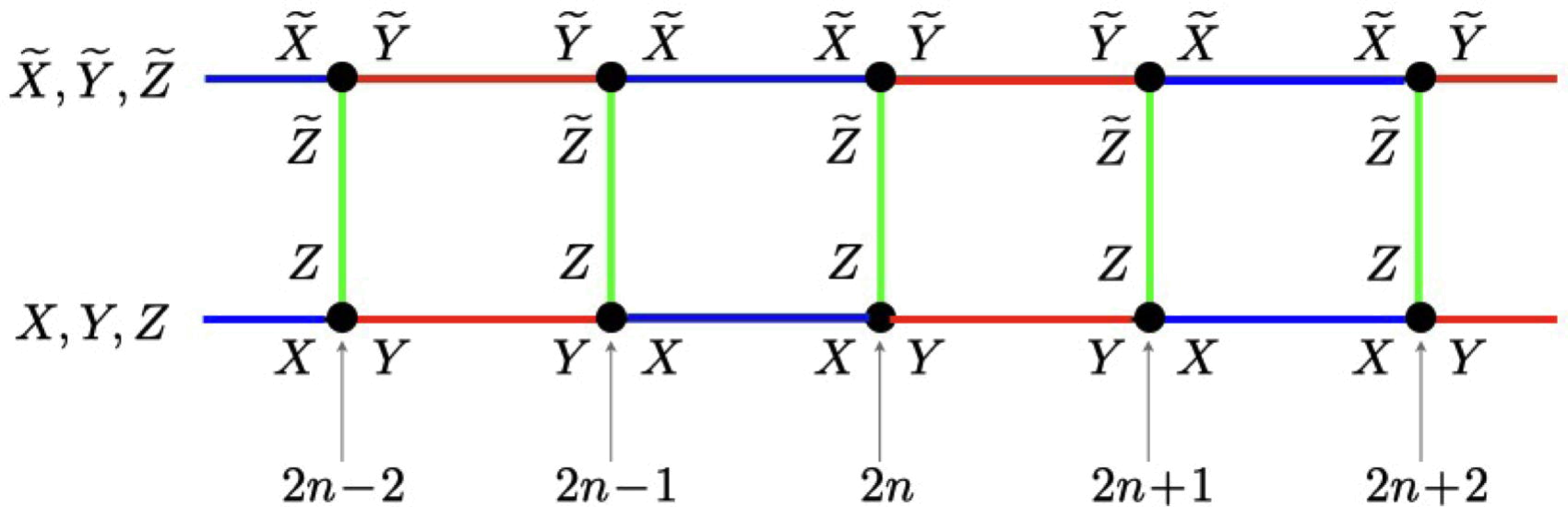}
\end{centering}
\caption{The Shibata-Katsura ladder (see text for description). \label{SKmodel}}
\end{figure}

The Shibata-Katsura (SK) model \cite{Shibata2019} describes a dissipative $S=\half$ chain.  The Hamiltonian is
\begin{equation}
H=\sum_n \big(J\ns_x\,X\ns_{2n-1} X\ns_{2n} + J\ns_y\,Y\ns_{2n}Y\ns_{2n+1}\big)
\end{equation}
and the jump operators are $L\ns_n=\sqrt{\gamma}\, Z\ns_n$\,, with $\gamma > 0$.  Thus, we have
\begin{align}
\CW(\gamma)&=\sum_{n=1}^\Nc\Big(J\ns_x\,X\ns_{2n-1} X\ns_{2n} + J\ns_y\,Y\ns_{2n} Y\ns_{2n+1}- J\ns_x\,\Xtil\ns_{2n-1}\Xtil\ns_{2n} \nonumber\\
&\hskip 0.3in - J\ns_y\,\Ytil\ns_{2n}\Ytil\ns_{2n+1}\Big) + i\gamma\sum_{j=1}^N\Big(Z\ns_j\Ztil\ns_j-1 \Big)\quad,
\end{align}
where the $(X,Y,Z)$ operators act on the first Hilbert space and $(\Xtil,\Ytil,\Ztil)$ act on the copy.
The system is depicted in Fig. \ref{SKmodel} and corresponds to a non-Hermitian two-leg ladder.
$\Nc$ is the number of unit cells, and there are $N=2\Nc$ sites on each leg of the ladder.
Note that $\CW^*(\gamma)=\CW(-\gamma)$, and that if we define $R$ as the reflection operator mapping one leg into the other,
\ie\ $(X\ns_j,Y\ns_j,Z\ns_j)\leftrightarrow(\Xtil\ns_j,\Ytil\ns_j,\Ztil\ns_j)$ for all $j$, then
\begin{equation}
R\,\CW(\gamma)R = -\CW(-\gamma)=-\CW^*(\gamma)\quad.
\end{equation}
This establishes that the eigenvalues of $\CW(\gamma)$ come in pairs $\RLambda^\pm_a=\pm E\ns_a + i\Gamma\ns_a$\,.
Total positivity requires that $\Imp(\Gamma\ns_a)\le 0$.  Any NESS $\vrh\ns_\ssf{NESS}$ satisfies
$\CW(\gamma)\,\sket{\vrh\ns_\ssf{NESS}}=0$.

Introducing on each site four Majorana fermions $\theta^{0,1,2,3}$ and expressing the Pauli matrices therefrom, 
\begin{equation}
X\ns_j = i\theta^0_j\theta^1_j\quad,\quad Y\ns_j = i\theta^0_j\theta^2_j\quad,\quad Z\ns_j = i\theta^0_j\theta^3_j\quad,
\end{equation}
with corresponding expression for $(\Xtil\ns_j,\Ytil\ns_j,\Ztil\ns_j)$, one may express $\CW(\gamma)$ as
\begin{equation}
\begin{split}
\CW(\gamma)&=\sum_{n=1}^\Nc\Big\{iJ\ns_x\big[\mu^x_{2n-1} \theta^0_{2n-1}\theta^0_{2n} 
-\mut^x_j \tht^0_{2n-1}\tht^0_{2n}\big]\\
&\hskip 0.4in + iJ\ns_y\big[\mu^y_{2n} \theta^0_{2n}\theta^0_{2n+1} -
\mut^y_{2n} \tht^0_{2n}\tht^0_{2n+1}\big]\Big\}\\
&\hskip 0.8in - \gamma \sum_{j=1}^N \mu^z_j\theta^0_j \tht^0_j -2i\gamma\Nc\quad,
\label{WSK}
\end{split}
\end{equation}
where
\begin{align}
\mu^x_{2n-1}=-i\theta^1_{2n-1}\theta^1_{2n} \ \ &,\ \ 
\mut^x_{2n-1}=i\tht^1_{2n-1}\tht^1_{2n}  \\
\mu^y_{2n}=-i\theta^2_{2n}\theta^2_{2n+1}   \ \ ,\ \ 
\mut^y_{2n}&=-i\tht^2_{2n}\tht^2_{2n+1} \ \ ,\ \
\mu^z_j=-i\theta^3_j\tht^3_j\nonumber
\end{align}
are $\MZ\ns_2$ gauge fields on the links of the two leg ladder in fig. \ref{SKmodel}.  These gauge fields commute with each other and with
the $\theta^0$ hopping terms, as well as with the constraints
\begin{equation}
\RLambda\ns_j\equiv \theta_j^0\theta_j^1\theta_j^2\theta_j^3=+1 \quad,\quad
\Lat\ns_j\equiv \tht_j^0\tht_j^1\tht_j^2\tht_j^3=+1\quad.
\end{equation}
which must be imposed at each site in order to guarantee $XY=iZ$.  This is the magic of the Kitaev honeycomb lattice model, where the link lattice 
is also tripartite: the Hamiltonian corresponds to a single species ($\theta^0$) of Majorana fermion hopping in the presence of a nondynamical
$\MZ\ns_2$ gauge field.  The gauge-invariant content of the theory is contained in the plaquette fluxes 
$\Phi\ns_{2n-1}=\mu^x_{2n-1}\mu^z_{2n}\mut^x_{2n-1}\mu^z_{2n-1}$ 
and $\Phi\ns_{2n}=\mu^y_{2n}\mu^z_{2n+1}\mut^y_{2n}\mu^z_{2n}$ 
and in the Wilson phases $Q=\prod_{j=1}^N Z\ns_j$ and $\Qtil=\prod_{j=1}^N \Ztil\ns_j$.
With periodic boundary conditions, $Q\Qtil=\prod_{j=1}^N\Phi\ns_j$.

\section{Dirac Matrix SK Model}
\subsection{Gamma matrix Kitaev models}
A Clifford algebra is defined by the anticommutation relations,
\begin{equation}
\big\{ \Gamma^a \, , \, \Gamma^b \big\} = 2\delta^{ab} \qquad \mu,\nu\in \{1,\ldots,n\}\quad.
\end{equation}
When $n=2k$, a representation of the algebra can be constructed by tensor products of $k$ Pauli matrices, {\it viz.\/}
\begin{align}
\Gamma^1 &= X \otimes \Mone\otimes \cdots  \otimes \Mone &
\Gamma^{2k-1} &= Z \otimes Z \otimes \cdots  \otimes X \nonumber\\
\Gamma^2 &= Y \otimes \Mone\otimes \cdots  \otimes \Mone &
\Gamma^{2k} &= Z \otimes Z \otimes \cdots  \otimes Y \\
\Gamma^3 &= Z \otimes X \otimes \cdots \otimes \Mone &
\Gamma^{2k+1} &= Z \otimes Z \otimes \cdots \otimes Z \nonumber
\end{align}
The gamma matrices defined above are all Hermitian.  In even dimensions, we define
\begin{equation}
\Gamma^{2k+1}=(-i)^k\,\Gamma^1\,\Gamma^2\cdots\Gamma^{2k}\quad.
\end{equation}
Introducing $2k+2$ Majorana fermions $\theta^a$ with indices $a\in\{0,\ldots,2k+1\}$ satisfying $\big\{\theta^a,\theta^b\big\}=2\delta^{ab}$, we
define $\Gamma^\mu=i\theta^0\theta^\mu$ with $\mu > 0$.  Analogous to the constraint $\theta^0\theta^1\theta^2\theta^3=1$
when $k=1$, we demand
\begin{equation}
\theta^0\,\theta^1\cdots\theta^{2k+1}=i^{k-1}\quad.
\label{majcon}
\end{equation}
The case $k=1$ yields the $2\times 2$ Pauli matrices, with $\Gamma^1=X$, $\Gamma^2=Y$, and
$\Gamma^3=-i\,\Gamma^1\Gamma^2=Z$.  The case $k=2$ yields the $4\times 4$ Dirac matrices, with
$\Gamma^5=-\Gamma^1\Gamma^2\Gamma^3\Gamma^4$. For general $k$ this yields $2k+1$ matrices of rank $2^k$.
One can then form $\Gamma^{\mu\nu}= i\Gamma^\mu\Gamma^\nu = i\theta^\mu\theta^\nu$ of which there are ${2k+1\choose 2}$ independent 
representatives (take $\mu<\nu$), and next $\Gamma^{\mu\nu\rho}=-i\Gamma^\mu\Gamma^\nu\Gamma^\rho=\theta^0\theta^\mu\theta^\nu\theta^\rho$ and $\Gamma^\mu\Gamma^\nu\Gamma^\rho\Gamma^\sigma=\theta^\mu\theta^\nu\theta^\rho\theta^\sigma$, which yield
${2k+1\choose 3}$ and ${2k+1\choose 4}$ independent terms, respectively, \etc\ Proceeding thusly one obtains at level $k$
a basis of $4^k$ Hermitian matrices of rank $2^k$.

Analogs of Kitaev's honeycomb lattice model using these higher level Clifford algebras have been considered, {\it inter alia\/} in refs. \cite{YZK09,WAH09}, with 
interactions $\Gamma^\mu_i\Gamma^\mu_j$ along the links.  When the underlying 
lattice is such that each site lies at the confluence of $2k+1$ distinctly labeled 
$\mu$--links, the `spin' Hamiltonian is again expressible as a single
species ($\theta^0$) Majorana fermion hopping in the presence of a static $\MZ\ns_2$ gauge field.  Other generalizations, in which multiple
species of Majoranas hop in the same $\MZ\ns_2$ static gauge field and hybridize as well have also been constructed \cite{YZK09,CYF11}.

\subsection{Dirac matrix SK model}
\begin{figure}[t]
\begin{centering}
\includegraphics[width=0.45\textwidth]{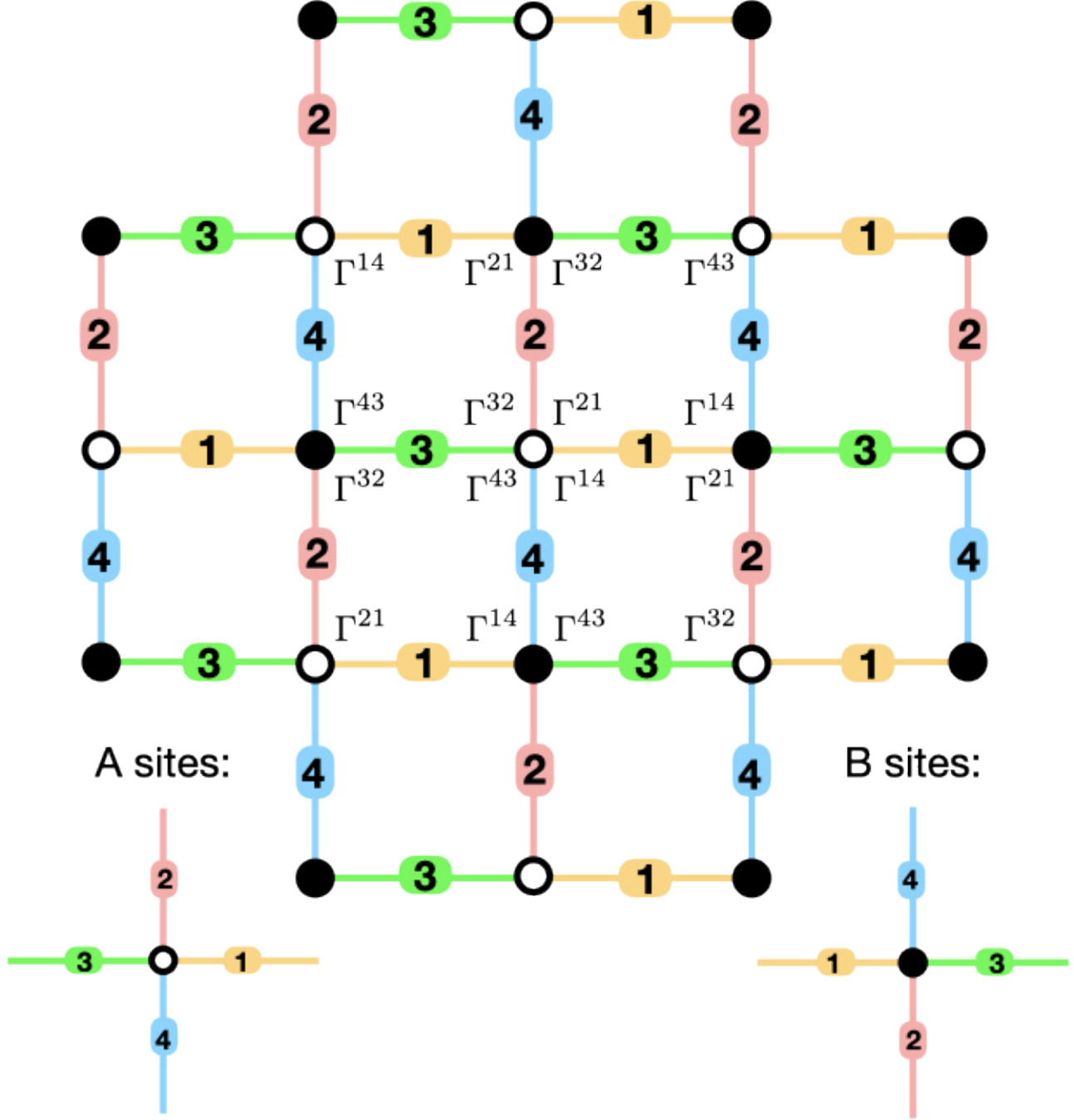}
\end{centering}
\caption{Square lattice Dirac matrix Shibata-Katsura model. See description in \S \ref{GSK}. \label{FdKsquare}}
\end{figure}

We generalize the SK model to a dissipative $\Gamma$--matrix model defined on the square lattice, as depicted
in fig. \ref{FdKsquare}.  We regard the square lattice as bipartite, with elementary direct lattice vectors
$\Ba\ns_{1,2}=\xhat\pm\yhat$.  Our Hamiltonian is
\begin{equation}
\begin{split}
H&=\sum_\BR \Big( J\ns_1\,\Gamma_\BR^1\, \Gamma_{\BR+\xhat}^1 + J\ns_2\,\Gamma_\BR^2\, \Gamma_{\BR+\yhat}^2\\
&\hskip 0.75in +J\ns_3\,\Gamma_\BR^3\, \Gamma_{\BR-\xhat}^3 +J\ns_4\,\Gamma_\BR^4\, \Gamma_{\BR-\yhat}^4\Big)\quad,
\end{split}
\end{equation}
where $\BR=n\ns_1 \Ba\ns_1 + n\ns_2\Ba\ns_2$ with $n\ns_{1,2}\in\MZ$ are the \SA\ sublattice sites, which
are $\Nc$ in number.  We use the symbol $\Br$ to denote a site which may be in either sublattice.
Thus, on each site of the square lattice, a four-dimensional Hilbert space is acted upon by operators
$1\ns_\Br$, $\Gamma^\mu_\Br$, and $\Gamma^{\mu\nu}_\Br$, where $\Gamma^\mu$ are $4\times 4$ Dirac matrices, 
with $\mu\in\{1,\ldots,5\}$.

\begin{figure}[t]
\begin{centering}
\includegraphics[width=0.44\textwidth]{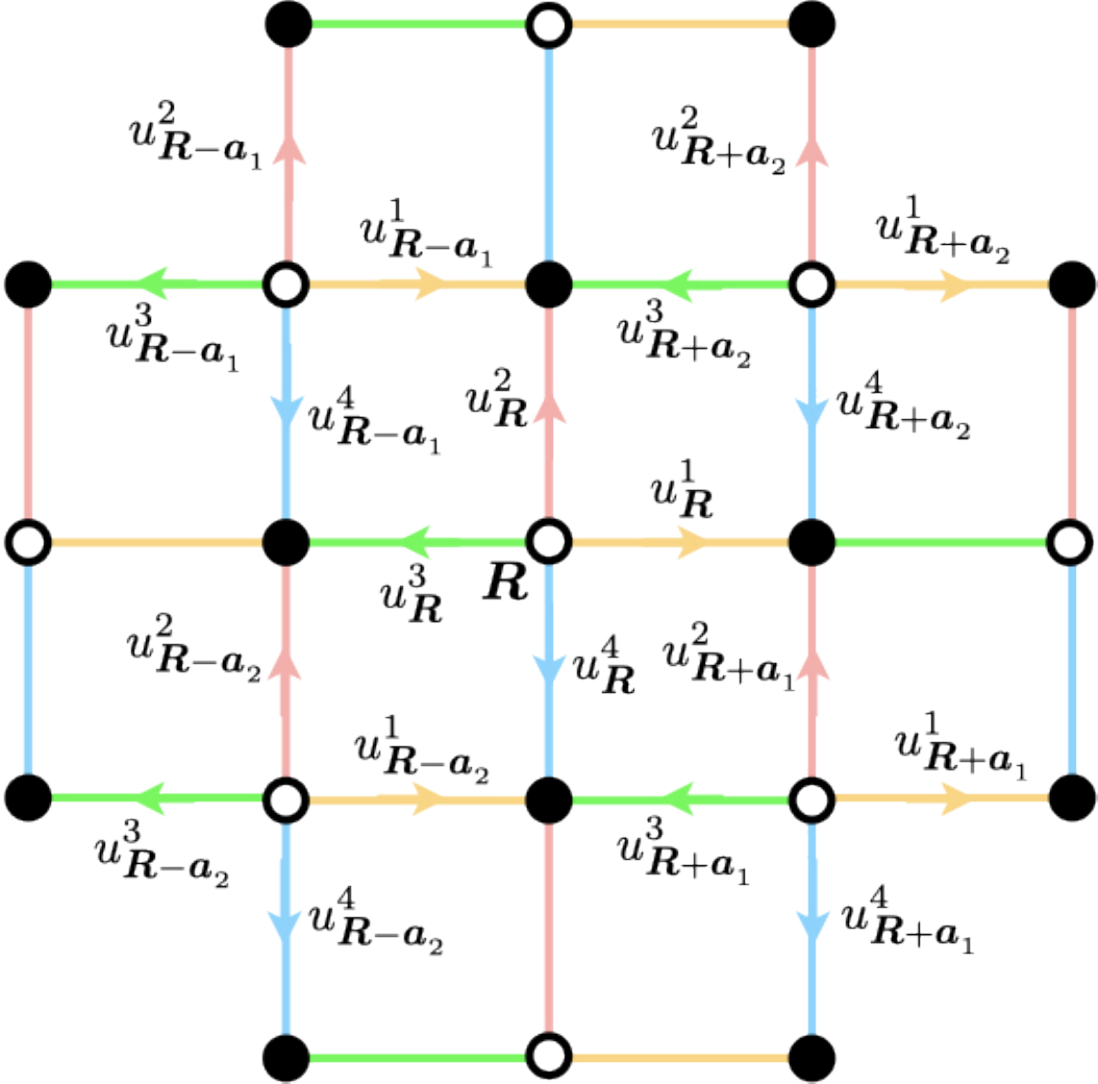}
\end{centering}
\caption{At each $\SA$ sublattice site in the bottom layer, the gauge field $u^\delta_\BR$ points along the nearest neighbor
vector $\Bdel$ toward a neighboring $\SB$ sublattice site.  A corresponding convention pertains for the $\util^\delta_\BR$ 
gauge fields in the top layer. \label{SLGF}}
\end{figure}

Following SK, we take the Lindblad jump operators to be $L\ns_\Br=\sqrt{\gamma}\,\Gamma^5_\Br$ at each site.
The GKLS master equation can then be written as a non-Hermitian Hamiltonian evolution of a model on a
square lattice bilayer, with each layer corresponding to one copy of the Hilbert space.  This Hamiltonian is
\begin{align}
\CW\big(\{J\ns_\delta\},\gamma\big)&=\sum_{\BR\in\SA}\sum_{\delta=1}^4 J\ns_\delta\Big(i u^\delta_\BR\, \theta^0_\BR\,\theta^0_{\BR+\Bdel} - i \util^\delta_\BR\, \tht^0_\BR\,\tht^0_{\BR+\Bdel}\Big)\nonumber\\
&\hskip 0.6in - \gamma\!\!\sum_{\Br\in\SA,\SB}\!\! u^5_\Br\,\theta^0_\Br\,\tht^0_\Br -2i\gamma\Nc \hskip 0.1in,
\label{WSB}
\end{align}
where $\Bdel\in\{\xhat,\yhat,-\xhat,-\yhat \}$ for $\delta\in\{1,2,3,4\}$, respectively, and where (see fig. \ref{SLGF})
\begin{equation}
u^\delta_\BR=-i\theta^\delta_\BR\theta^\delta_{\BR+\Bdel} \quad,\quad
\util^\delta_\BR=-i\tht^\delta_\BR\tht^\delta_{\BR+\Bdel} \quad,\quad
u^5_\Br=-i\theta^5_\Br\tht^5_\Br
\end{equation}
are the nondynamical gauge fields in the bottom, top, and between layer regions. There are $5N$ such gauge fields, 
but as we shall see the number of gauge-invariant quantities is $3N+1$, \ie\ there are $2^{3N+1}$ gauge sectors,
where $N$ is the total number of sites in either layer.

\subsubsection{Conserved quantities}
For the original SK model, the product $Q=Z\ns_1\cdots Z\ns_N$ is conserved as it commutes with $H$ and with each
of the jump operators $\sqrt{\gamma}\,Z\ns_j$.  This means that both $\Mone$ and $Q$ are annihilated by the
Liouvillian $\CL$, and that both
\begin{equation}
\vrh\ns_\pm=2^{-N}\big(\Mone \pm Q\big)
\end{equation}
are thus valid NESSes, for all $\gamma$ \cite{Shibata2019}.

For our model of eqn. \ref{WSB}, there are vastly more conserved quantities.  
With periodic boundary conditions along both axes,
there are $N+1$ gauge-invariant quantities, which are the $N$ plaquette fluxes (see fig. \ref{FdKsquare}),
\begin{equation}
\Phi\ns_\Br\equiv\begin{cases}
-\Gamma^{21}_\Br\,\Gamma^{14}_{\Br+\xhat}\,\Gamma^{43}_{\Br+\xhat+\yhat}\,
\Gamma^{32}_{\Br+\yhat} 
&{\rm if}\ \Br\in\SA\\
-\Gamma^{43}_\Br\,\Gamma^{32}_{\Br+\xhat}\,\Gamma^{21}_{\Br+\xhat+\yhat}\,
\Gamma^{14}_{\Br+\yhat} 
&{\rm if}\ \Br\in\SB\quad,
\end{cases}
\end{equation}
where the $\MZ\ns_2$ flux in plaquette $\Br$ is labeled by the lower left site of the plaquette \footnote{
We find it convenient here to use expressions $\Gamma^{\mu\nu}\equiv -\Gamma^{\nu\mu}$ with $\mu>\nu$}. 
Note that the product $\prod_\Br \Phi\ns_\Br=1$, hence there are $N-1$ independent $\MZ\ns_2$ plaquette fluxes.
In addition, we have the two Wilson phases,
\begin{equation}
\begin{split}
\Wx&=-\Gamma^{13}_{1,1}\,\Gamma^{31}_{2,1} \,\cdots\, 
\Gamma^{13}_{\Nx-1,1} \,\Gamma^{31}_{\Nx,1}\\
\Wy&=-\Gamma^{24}_{1,1}\,\Gamma^{42}_{1,2} \,\cdots\, \Gamma^{24}_{1,\Ny-1}\, \Gamma^{42}_{1,\Ny}\quad,
\end{split}
\end{equation}
where both $\Nx$ and $\Ny$ are taken to be even, and with the total number of sites $N\equiv \Nx \Ny$\,.
(Note that $\Gamma^{31}=-\Gamma^{13}$ and $\Gamma^{42}=-\Gamma^{23}$; we choose to write the Wilson phases as above
because the repetition of consecutive $\Gamma$--matrix indices is a useful mnemonic.)
One can readily check that $\Phi\ns_\Br$ commutes with both $H$ and with all the jump operators.  In addition, the operator $Q=\prod_\Br\Gamma^5_\Br$ also commutes with the Hamiltonian and with all of the jump operators.  However, if we examine the 
product of the $\MZ\ns_2$ fluxes
over the $\SA$ plaquettes alone, \ie\ over those plaquettes with an $\SA$ site in their lower left corner, then from
$\Gamma^{43}\Gamma^{21}=-\Gamma^1\Gamma^2\Gamma^3\Gamma^4 = \Gamma^5$,
we conclude that $\prod_{\BR\in\SA} \phi\ns_\BR=\prod_\Br\Gamma^5_\Br = Q$,
and therefore $Q$ is not an independent conserved quantity.  Finally, as the jump operators are 
all Hermitian, according to eqn. \ref{Gibbs} we have a $2^{N+1}$--dimensional subspace of $T=\infty$ nonequilibrium
steady states, since there are $2^{N+1}$ projectors,
\begin{equation}
\begin{split}
\Pi\ns_{\eta\ns_x,\eta\ns_y,\{\eta\ns_\Br\}} &\equiv \bigg({1+\eta\ns_x\Wx\over 2}\bigg)
\bigg({1+\eta\ns_y \Wy\over 2}\bigg)\\
&\hskip 1.0in\times{\prod_\Br}^\prime \bigg({1+\eta\ns_\Br \phi\ns_\Br\over 2}\bigg)\quad,
\end{split}
\end{equation}
labeled by $\eta\ns_x$, $\eta\ns_y$, and $\{\eta\ns_\Br\}$, each taking the value $\pm 1$, which commute with $H$ and with all the jump 
operators $L\ns_\Br$\,. The prime on the product indicates that the final plaquette with $\Br=(\Nx,\Ny)$ is omitted.
The total number of unnormalized density matrices is $(4^2)^N=16^N$.  \Ie\ any density matrix
of the form
\begin{equation}
\vrh = \sum_{\eta\ns_x}\sum_{\eta\ns_y}\sum_{\{\eta\ns_\Br\}} 
C\ns_{\eta\ns_x,\eta\ns_y,\{\eta\ns_\Br\}}\RPi\ns_{\eta\ns_x,\eta\ns_y,\{\eta\ns_\Br\}} 
\end{equation}
with $\Tra\vrh= \sum_{\eta\ns_x}\sum_{\eta\ns_y}\sum_{\{\eta\ns_\Br\}} 
C\ns_{\eta\ns_x,\eta\ns_y,\{\eta\ns_\Br\}}=1$ and each $C\ns_{\eta\ns_x,\eta\ns_y,\{\eta\ns_\Br\}}\ge 0$
is a valid NESS.

\subsection{Analysis} \label{GSK}
We define a complex fermion living along each link between planes of the bilayer, \viz
\begin{equation}
c\nd_\Br=\half(\theta^0_\Br + i\tht^0_\Br) \qquad,\qquad
c\yd_\Br=\half(\theta^0_\Br - i\tht^0_\Br)\quad,
\end{equation}
and thus
\begin{equation}
\theta^0_\Br=c\yd_\Br+c\nd_\Br\qquad,\qquad
\tht^0_\Br=i(c\yd_\Br-c\nd_\Br)\quad.
\end{equation}
The non-Hermitian Hamiltonian of eqn. \ref{WSB} is then expressed in terms of these complex fermions as
\begin{equation}
\begin{split}
\CW&=\sum_{\BR\in\SA}\sum_{\delta=1}^4 \Big\{ 
iJ\ns_\delta\big(u^\delta_\BR-\util^\delta_\BR\big)\big(c\yd_\BR c\nd_{\BR+\Bdel} + c\yd_{\BR+\Bdel} c\nd_\BR\big)\\
&\hskip 0.35in +iJ\ns_\delta\big(u^\delta_\BR+\util^\delta_\BR\big)\big(c\yd_\BR c\yd_{\BR+\Bdel} - 
c\nd_{\BR+\Bdel} c\nd_\BR\big)\Big\}\\
&\hskip 0.7in +i\gamma\!\!\sum_{\Br\in{\SA,\SB}}\!\! u^5_\Br \big(2 c\yd_\Br c\nd_\Br - 1\big) - 2i\Nc\gamma\quad.
\label{WNH}
\end{split}
\end{equation}

\begin{figure}[t]
\begin{centering}
\includegraphics[width=0.45\textwidth]{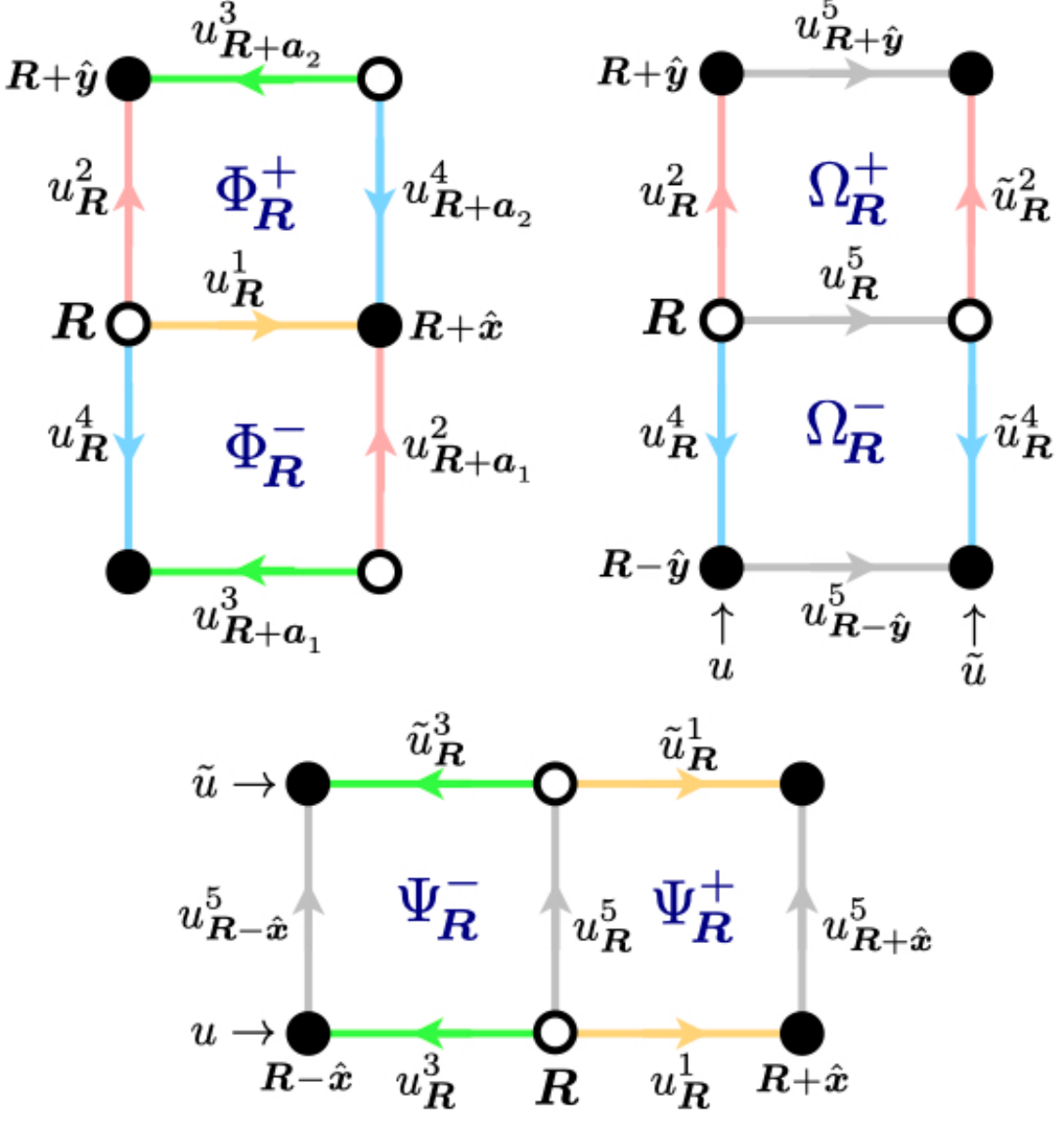}
\end{centering}
\caption{Associated with each \SA\ sublattice site in the bottom layer are eight square plaquette
fluxes: $\Phi^\pm_\BR$, ${\widetilde\Phi}^\pm_\BR$ (not shown), $\Psi^\pm_\BR$, and $\Omega^\pm_\BR$.
\label{Ffluxes}}
\end{figure} 

\subsection{Counting degrees of freedom} \label{counting}
Associated with each \SA\ sublattice site in the bottom layer are
eight square plaquette $\MZ\ns_2$ fluxes (see fig. \ref{Ffluxes}).
These fall into three groups.
First are the fluxes through the $(x,y)$ plaquettes.  For the bottom
layer we have
\begin{align}
\Phi^+_\BR&=u^1_\BR\, u^4_{\BR+\Ba\ns_2} u^3_{\BR+\Ba\ns_2} u^2_\BR
=-\Gamma^{21}_\BR\, \Gamma^{14}_{\BR+\xhat}\, \Gamma^{43}_{\BR+\Ba\ns_2}
\,\Gamma^{32}_{\BR+\yhat}\nonumber\\
\Phi^-_\BR&=u^4_\BR\, u^3_{\BR+\Ba\ns_1} u^2_{\BR+\Ba\ns_1} u^1_\BR
=-\Gamma^{14}_\BR\, \Gamma^{43}_{\BR-\yhat}\, \Gamma^{32}_{\BR+\Ba\ns_1}
\,\Gamma^{21}_{\BR+\xhat}
\label{Phieqn}
\end{align}
with corresponding expressions involving $\Phit^\pm_\BR$, $\Gat\ns_\Br$,
and $\util^\delta_\BR$ in the top layer.  Next, the $(x,z)$
plaquette fluxes $\Psi^\pm_\BR$,
\begin{align}
\Psi^+_\BR&=u^1_\BR\, u^5_{\BR+\xhat} \util^1_\BR\, u^5_\BR
=-\Gamma^{51}_\BR\, \Gamma^{15}_{\BR+\xhat}\, \Gat^{51}_{\BR+\xhat}
\,\Gat^{15}_\BR\label{Psieqn}\\
\Psi^-_\BR&=u^5_\BR\, \util^3_\BR\, u^5_{\BR-\xhat} u^3_\BR
=-\Gamma^{35}_\BR\, \Gamma^{53}_\BR\, \Gat^{35}_{\BR-\xhat}
\,\Gat^{53}_{\BR-\xhat}\quad.\nonumber
\end{align}
Finally, the $(y,z)$ plaquette fluxes $\Omega^\pm_\BR$ are given by
\begin{align}
\Omega^+_\BR&=u^2_\BR\, u^5_{\BR+\yhat} \util^2_\BR\, u^5_\BR
=-\Gamma^{52}_\BR\, \Gamma^{25}_{\BR+\yhat}\, \Gat^{52}_{\BR+\yhat}
\,\Gat^{25}_\BR\label{Gameqn}\\
\Omega^-_\BR&=u^5_\BR\, \util^4_\BR\, u^5_{\BR-\yhat} u^4_\BR
=-\Gamma^{45}_\BR\, \Gat^{54}_\BR\, \Gat^{45}_{\BR-\yhat}
\,\Gamma^{54}_{\BR-\yhat}\quad.\nonumber
\end{align}
There are also the Wilson phases,
\begin{equation}
\begin{split}
\Wx&=u^1_{1,1}\,(-u^3_{3,1})\,u^1_{3,1}\,(-u^3_{5,1})\cdots
u^1_{\Nx-1,1}\,(-u^3_{1,1})\\
&=-\Gamma^{13}_{1,1}\,\Gamma^{31}_{2,1}\cdots\Gamma^{31}_{\Nx,1}\\
\Wy&=u^2_{1,1}\,(-u^4_{1,3})\,u^2_{1,3}\,(-u^4_{1,5})\cdots
u^2_{1,\Ny-1}\,(-u^4_{1,1})\\
&=\Gamma^{24}_{1,1}\,\Gamma^{42}_{1,2}\cdots\Gamma^{42}_{1,\Ny}\quad,
\label{Wloops}
\end{split}
\end{equation}
again with corresponding expressions for $\Wtil\ns_x$ and $\Wtil\ns_y$\,.
At this point it appears that we have $4N+4$ gauge-invariant
$\MZ\ns_2$ degrees of freedom. However, the total flux through each of the
$N$ cubes must be trivial, providing $N$ constraints. There is an additional
constraint $\prod_\BR \Phi^+_\BR \Phi^-_\BR=1$ due to periodic boundary conditions;
the corresponding expression in the top layer does not yield new information given
the condition on each of the cubes. Finally, there are two constraints relating the
products of the Wilson phases in each of the layers to the $\Omega$ and $\Psi$
plaquette fluxes (see eqn. \ref{WLconst} below.) Thus, there are $N+3$ independent constraints, and therefore $3N+1$ independent gauge-invariant configurations of the
fluxes and Wilson phases. We must also acknowledge the constraints imposed by the 
projectors which enforce $\Lambda\ns_\Br=\Lat_\Br=1$, with $\Lambda_\Br=\theta^0_\Br\,\theta^1_\Br\,\theta^2_\Br\,
\theta^3_\Br\,\theta^4_\Br\,\theta^5_\Br=-i$. Taking the product
over all sites, we obtain \cite{YZK09}
\begin{equation}\label{parity_constraint}
\prod_\Br i\theta^0_\Br\tht^0_\Br\times
\prod_{\BR,\delta} u^\delta_\BR \util^\delta_\BR
\times \prod_\Br u^5_\Br = 1\quad.
\end{equation}
This expression includes a product over all the itinerant fermion
parities $2c\yd_\Br c\nd_\Br-1$ as well as over each of the $5N$
$\MZ\ns_2$ gauge fields which reside on the links of the bilayer
structure.  It thereby constrains the parity of the $c$--fermions, which are constructed from $\theta^0$ and $\tht^0$ on
each of the interplane links.  Thus rather than $N$ freedoms for
the dynamical fermion states, there are $N-1$, and the total number
of states in our doubled Hilbert space is $2^{3N+1}\times 2^{N-1}=16^N$, 
which is the correct number of density matrices for an $N$--site system
described by $4\times 4$ gamma matrices \footnote{There is of course
a single remaining constraint associated with the condition
$\Tra\vrh=1$.}.

\begin{figure}[t]
\begin{centering}
\includegraphics[width=0.47\textwidth]{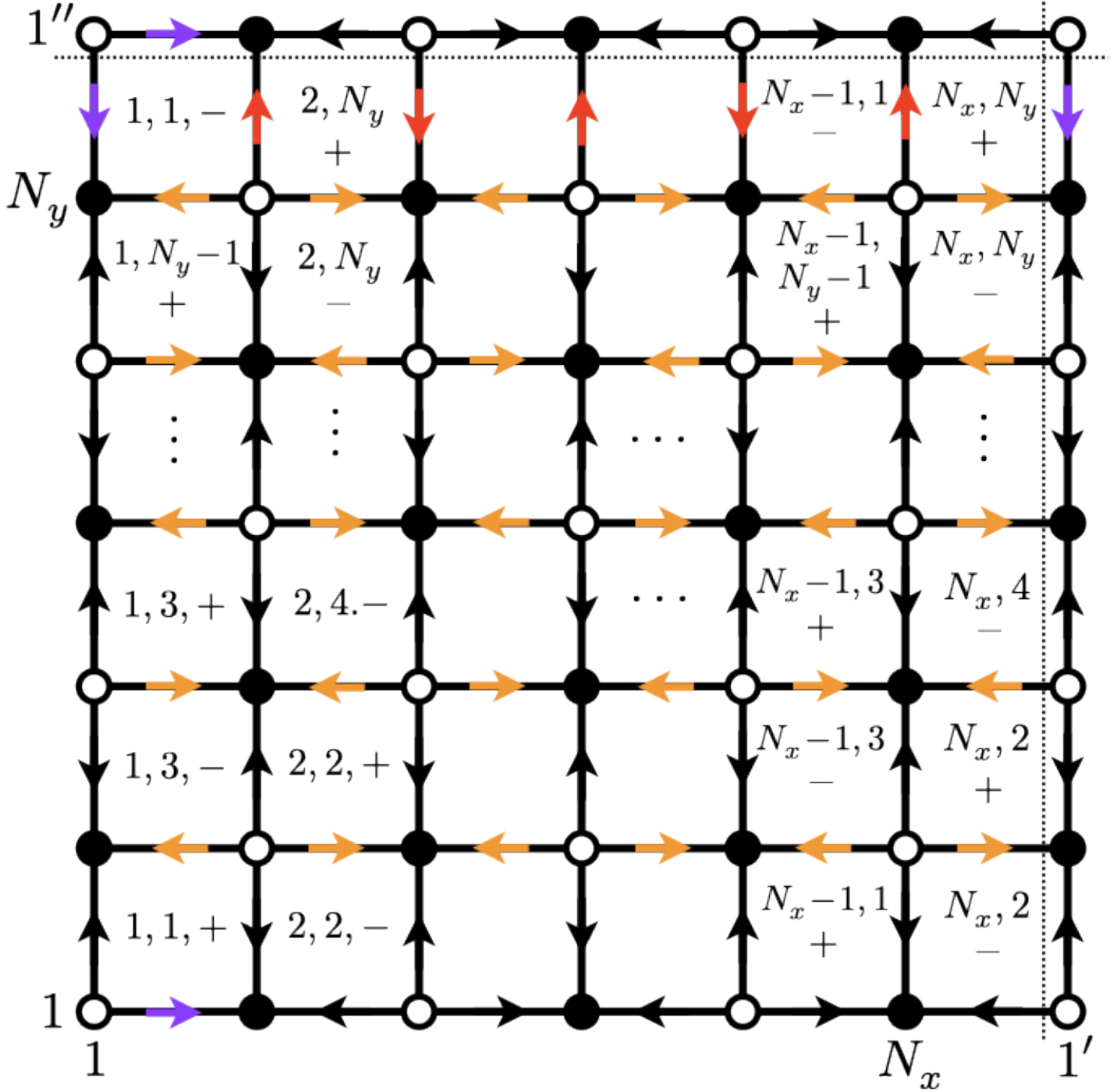}
\end{centering}
\caption{Labels of the unit cells $\Phi^\pm_\BR$.  The black arrows indicate $u^\delta_\BR=+1$ in the direction of the arrow.  There are $N+1$ colored arrows, 
which are determines by the $N-1$ independent plaquette fluxes and the two Wilson
loops. A corresponding assignment pertains to the upper layer with fluxes
$\Phit^\pm_\BR$ and gauge fields $\util^\delta_\BR$.  Dotted lines indicate
periodicity boundaries.
\label{Fxyflux}}
\end{figure} 

\subsection{Choosing a gauge}\label{gauge_fixing}
Given the $3N+1$ independent plaquette fluxes and Wilson phases, how can we pick a gauge?  Let us first consider
the planar fluxes $\Phi^\pm_\BR$ in the bottom layer and the sketch in fig. \ref{Fxyflux}.  The coordinates of the
\SA\ sublattice site in the lower left corner are $\Br=(x,y)=(1,1)$.  The Wilson phase fluxes are defined to be
$u^1_{1,1}\equiv W_x$ and $u^4_{1,1}\equiv -W_y$\,.  We then define the remaining unassigned gauge fields as follows:
\begin{align}
u^3_{2,2}&=\Phi^+_{1,1}\,u^1_{1,1} & u^1_{2,2}&=\Phi^-_{2,2} \nonumber \\
u^1_{1,3}&=\Phi^-_{1,3}\,u^3_{2,2} & u^3_{3,3}&=\Phi^+_{2,2}\,u^1_{2,2} 
\quad\ldots\nonumber \\
u^1_{1,\Ny-1}&=\Phi^-_{1,\Ny-1}\,u^3_{2,\Ny-2} & 
u^3_{3,\Ny-1}&=\Phi^+_{2,\Ny-2}\,u^1_{2,\Ny-2}\nonumber\\
u^3_{2,\Ny}&=\Phi^+_{1,\Ny-1}\,u^1_{1,\Ny-1} & 
u^1_{2,N\ns_1}&=\Phi^-_{2,\Ny}\,u^3_{3,\Ny-1}\nonumber\\
u^2_{2,\Ny}&=\Phi^-_{1,1}\,u^4_{1,1}\, u^3_{2,\Ny} & 
u^4_{3,1}&=\Phi^-_{2,\Ny}\,u^1_{2,\Ny}\, u^2_{2,\Ny}\nonumber
\end{align}
and
\begin{align}
u^3_{4,2}&=\Phi^+_{3,1} & u^1_{\Nx,2}&=\Phi^-_{\Nx,2} \\
u^1_{3,3}&=\Phi^-_{3,3}\,u^3_{4,2}  &
u^3_{1,3}&=\Phi^+_{\Nx,2}\,u^1_{\Nx,2}\nonumber \\
u^3_{4,4}&=\Phi^+_{3,3}\,u^1_{3,3} & 
u^1_{\Nx,4}&=\Phi^-_{\Nx,4}\,u^3_{1,3}\quad\ldots\nonumber\\
u^1_{3,\Ny-1}&=\Phi^-_{3,\Ny-1}\,u^3_{4,\Ny-2} & 
u^3_{1,\Ny-1}&=\Phi^+_{\Nx,\Ny-2}\,u^1_{\Nx,\Ny-2}\nonumber\\
u^3_{4,\Ny}&=\Phi^+_{3,\Ny-1}\,u^1_{3,\Ny-1} & 
u^1_{\Nx,\Ny}&=\Phi^-_{\Nx,\Ny}\,u^3_{1,\Ny-1}\nonumber\\
u^4_{3,1}&=\Phi^+_{2,\Ny}\,u^3_{2,\Ny}\,u^1_{2,\Ny} & 
u^4_{1,1}&=-W_y\quad.\nonumber
\end{align}
Thus, we can iteratively obtain all the unassigned $\MZ\ns_2$ gauge fields $u^\delta_\BR$ from the plaquette phases and the 
Wilson phases.  Again, corresponding expressions hold in the upper layer for the quantitied 
$\big\{\util^\delta_\BR,\Phi^\pm_\BR,\Wtx,\Wty\big\}$.

Next, we consider the $u^5_\BR$ gauge fields and the plaquette fluxes $\big\{\Psi^\pm_\BR,\Omega^\pm_\BR\big\}$. From eqn. \ref{Phieqn} we may iteratively
determine the values of $u^5_{m,k}$ for odd values of $k$ given the
value $u^5_{1,k}\equiv 1$:
\begin{align}
u^5_{2,k} &= u^1_{1,k}\,\util^1_{1,k}\,\Psi^+_{1,k}\cdot u^5_{1,k}\label{oddk}\\
u^5_{3,k} &= u^3_{3,k}\,\util^3_{3,k}\,\Psi^-_{3,k}\cdot u^5_{2,k} \nonumber \\
u^5_{4,k} &= u^1_{3,k}\,\util^1_{3,k}\,\Psi^+_{3,k}\cdot u^5_{3,k}
\quad\ldots \nonumber \\
u^5_{\Nx,k} &= u^1_{\Nx-1,k}\,\util^1_{\Nx-1,k}\,\Psi^+_{\Nx-1,k}
\cdot u^5_{\Nx-1,k} \nonumber \\
u^5_{1,k}&=u^1_{1,k}\,\util^1_{1,k}\,\Psi^-_{1,k}\cdot u^5_{\Nx,k}\quad.\nonumber
\end{align}
For even values of $k$, we have
\begin{align}
u^5_{2,k} &= u^3_{2,k}\,\util^3_{2,k}\,\Psi^-_{2,k}\cdot u^5_{1,k}\label{evenk}\\
u^5_{3,k} &= u^1_{2,k}\,\util^1_{2,k}\,\Psi^+_{2,k}\cdot u^5_{2,k} \nonumber \\
u^5_{4,k} &= u^3_{4,k}\,\util^3_{4,k}\,\Psi^-_{4,k}\cdot u^5_{3,k} 
\quad\ldots \nonumber \\
u^5_{\Nx,k} &= u^3_{\Nx,k}\,\util^3_{\Nx,k}\,\Psi^-_{\Nx,k}
\cdot u^5_{\Nx-1,k} \nonumber \\
u^5_{1,k} &= u^1_{\Nx,k}\,\util^1_{\Nx,k}\,\Psi^+_{\Nx,k}\cdot 
u^5_{\Nx,k}\quad. \nonumber
\end{align}
To obtain $u^5_{1,k+1}$ from $u^5_{1,k}$, we use the relations
\begin{equation}
\begin{split}
u^5_{1,2n}&=u^2_{1,2n-1}\,\util^2_{1,2n-1}\,\Omega^+_{1,2n-1}\cdot
u^5_{1,2n-1}\\
u^5_{1,2n+1}&=u^4_{1,2n}\,\util^4_{1,2n}\,\Omega^-_{1,2n+1}\cdot
u^5_{1,2n}\quad.
\label{kupdate}
\end{split}
\end{equation}
Eqns. \ref{oddk}, \ref{evenk}, and \ref{kupdate} entail the relations
\begin{align}
\prod_{j=1\atop (k\ {\rm odd})}^\Nx \Psi^-_{j,k}\,\Psi^+_{j,k}&=
\prod_{m=1}^{\Nx/2}u^1_{2m-1,k}\,u^3_{2m-1,k}\,\util^1_{2m-1,k}\,
\util^3_{2m-1,k}\nonumber\\
\prod_{j=1\atop (k\ {\rm even})}^\Nx \Psi^-_{j,k}\,\Psi^+_{j,k}&=
\prod_{m=1}^{\Nx/2} u^1_{2m,k}\,u^3_{2m,k}\,\util^1_{2m,k}\,
\util^3_{2m,k}\quad,
\end{align}
for $k$ odd and even, respectively, as well as
\begin{align}
\prod_{k=1\atop (j\ {\rm odd})}^\Ny \Omega^-_{j,k}\,\Omega^+_{j,k}&=
\prod_{n=1}^{\Ny/2}u^2_{j,2n-1}\,u^4_{j,2n-1}\,\util^2_{j,2n-1}\,
\util^4_{j,2n-1}\nonumber\\
\prod_{k=1\atop (j\ {\rm even})}^\Ny \Omega^-_{j,k}\,\Omega^+_{j,k}&=
\prod_{n=1}^{\Ny/2} u^2_{j,2n}\,u^4_{j,2n}\,\util^2_{j,2n}\,
\util^4_{j,2n}\quad,
\end{align}
for $j$ odd and even, respectively. Restricting to the cases $j=1$ and $k=1$, 
we can relate these products to the Wilson phases in eqn. \ref{Wloops}, \viz
\begin{equation}
\begin{split}
\prod_{k=1}^\Ny \Omega^-_{1,k}\,\Omega^+_{1,k}&=\Wx\Wtx\\
\prod_{j=1}^\Nx \Psi^-_{j,1}\,\Psi^+_{j,1}&=\Wy\Wty \quad.
\label{WLconst}
\end{split}
\end{equation}

We showed previously in \S \ref{counting} that, considering all the 
$\MZ\ns_2$ gauge degrees of freedom, we have
a total of $3N+1$ independent plaquette fluxes and Wilson phases.
In each layer, there are $N+1$ free gauge fields $u^\delta_\BR$, as
depicted in fig. \ref{Fxyflux}.  Between the layers, there are $N-1$
free gauge fields $u^5_\Br$, with $u^5_{1,1}\equiv 1$.  Thus, our gauge
assignment accounts for all the independent gauge-invariant quantities.

\subsection{Counting the NESSes}
Referring to eqn. \ref{WNH}, in order to obtain an eigenvalue of zero, we
must have each $u^5_\Br=+1$ and $c\yd_\Br c\nd_\Br=1$.  (The case $u^5_\Br=-1$ for
all $\Br$ is impossible since we have, without loss of generality (\ie\ up to a
gauge transformation), set $u^5_{1,1}\equiv 1$.) We then must eliminate the BCS
pairing terms, which would allow for the simultaneous annihilation of two
neighboring $c$--fermions.  This is accomplished by setting 
$u^\delta_\BR + \util^\delta_\BR=0$
for all $\BR$ and $\Bdel$.  While this may seem inconsistent with the assignment
of the fixed gauge fields (black arrows) in the two layers as depicted in fig. 
\ref{Fxyflux}, in fact we are free to redefine $\util^\delta_\BR\to -\util^\delta_\BR$
for the purposes of counting the NESSes. Thus, there are a total of $N+1$ independent
values of the planar $(\delta\in\{1,2,3,4\})$ gauge fields associated with the NESS
block, and therefore $2^{N+1}$ degenerate NESSes.

It can be seen that for these NESSes $\Phi^+_\BR=\Phit^+_\BR$ and
$\Phi^-_\BR=\Phit^-_\BR$, as well as $\Psi^+_\BR=\Psi^-_\BR=-1$ and
$\Omega^+_\BR=\Omega^-_\BR=-1$, for all $\BR$.  Since 
$\prod_\BR\Phi^+_\BR \Phi^-_\BR=1$, this accounts for $N-1$ freedoms associated with
the plaquette fluxes.  The Wilson phases $\Wx$ and $\Wy$ are also free (but $\Wtx$
and $\Wty$ are then fixed by eqn. \ref{WLconst}), and so again we see that there
are $2^{N+1}$ NESSes.

We numerically verified this counting for the case $\Nx=\Ny=2$ ($N=4$) by choosing the
$J\ns_\delta$ couplings to be all different.  However, when $J\ns_1=J\ns_2=J\ns_3=J\ns_4$
there is an enlarged translational symmetry, and we find a degeneracy of $90$ rather
than $2^{N+1}=32$.  We also find that these additional degenerate states do not satisfy
the flux conditions described in the previous paragraph.

\section{Computational Results}
To calculate the spectrum within a given gauge sector, we use Prosen's method for
complex antisymmetric matrices \cite{Prosen2008}.  We note that the constraint
in equation \ref{parity_constraint} is implemented as a constraint on the parity 
of the complex fermions that arise from Prosen's generalized Bogoliubov
transformation.

\begin{figure}[!t]
\begin{centering}
\includegraphics[width=0.45\textwidth]{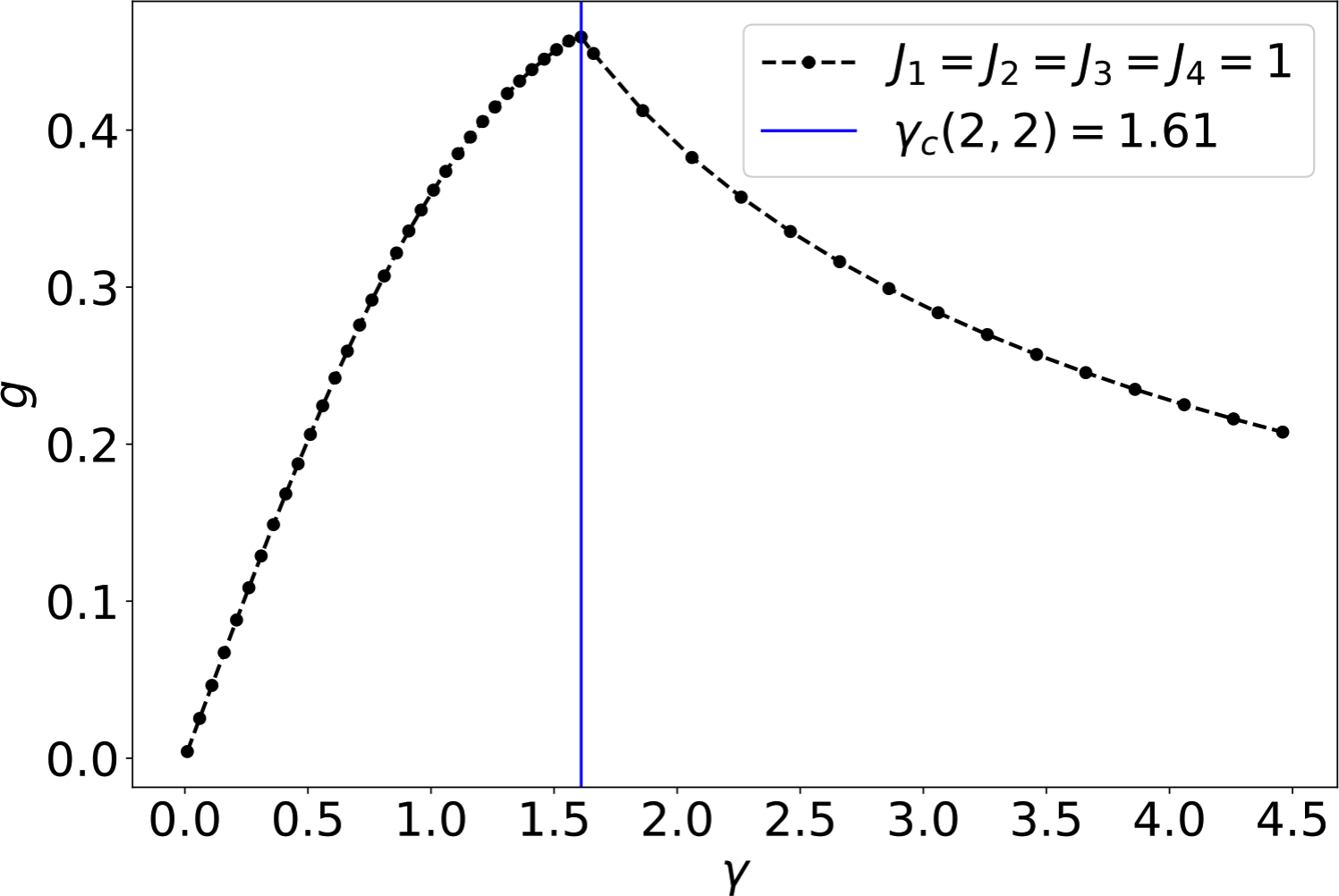}
\end{centering}
\caption{Liouvillian gap in the 2d generalized SK model with periodic boundary conditions for $\Nx=\Ny=2$ and all $J\ns_\delta=1$: The Liouvillian gap, $g$, as a function of $\gamma$. There is a transition in the first decay modes at the cusp seen at $\gamma=\gamma_\Rc(2,2)$, depicted by the blue vertical line.
\label{2d_gap_decay_modes_2by2_1}}
\end{figure}

\begin{figure}[!b]
\begin{centering}
\includegraphics[width=0.4\textwidth]{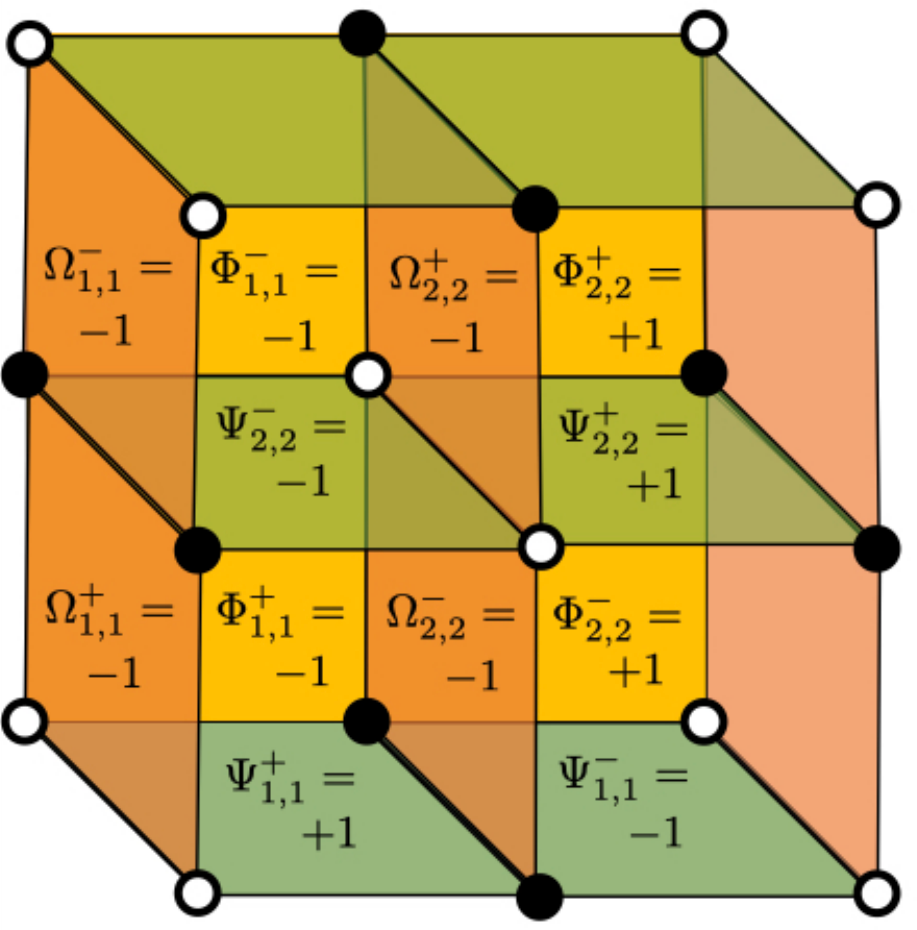}
\end{centering}
\caption{A first decay mode of the 2d generalized SK model with periodic boundary conditions for $\Nx=\Ny=2$ and $J_1=J_2=J_3=J_4=1$ corresponding to the `phase' where $\gamma\leq\gamma_\Rc(2,2)$. For the mode shown in this figure, we have $\Wx=1$, $\Wy=-1$, $\Wtx=-1$, $\Wty=-1$. There are 15 other configurations of the flux plaquettes and Wilson phases corresponding to the same eigenvalue as the first decay mode shown here.
\label{FDMsg}}
\end{figure}

\begin{figure}[!t]
\begin{centering}
\includegraphics[width=0.4\textwidth]{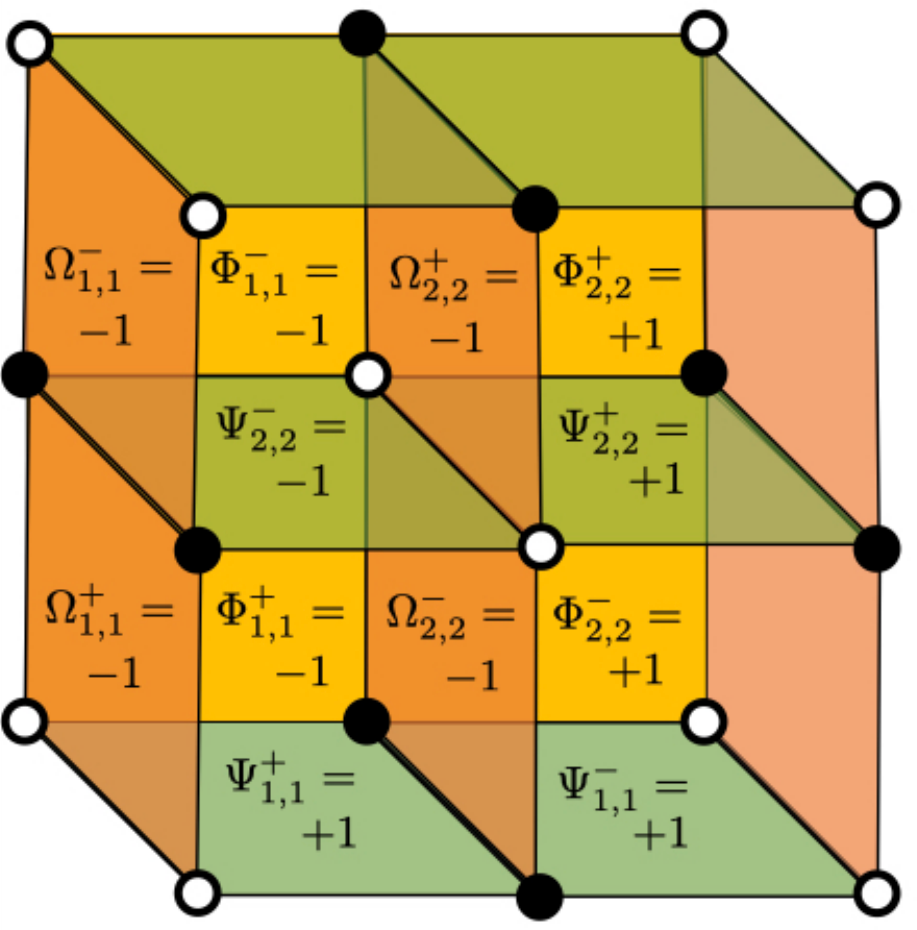}
\end{centering}
\caption{A first decay mode of the 2d generalized SK model with periodic boundary conditions for $\Nx=\Ny=2$ and $J_1=J_2=J_3=J_4=1$ corresponding to the `phase' where $\gamma\geq\gamma_{c}\left( 2 , 2\right)$. For the mode shown in this figure, we have $\Wx=1$, $\Wy=-1$, $\Wtx=1$, $\Wty=-1$. There are 79 other configurations of the flux plaquettes and Wilson phases corresponding to the same eigenvalue as the first decay mode shown here. 
\label{FDMlg}}
\end{figure}

Implementing the field assignments mentioned in section \ref{gauge_fixing}, we 
calculate the gap $g$ to the smallest relaxation rate, \ie\ the negative of the
real part of the eigenvalues of $\CL$, by searching over all the sectors of the
$\CL$ corresponding to the different plaquette flux and Wilson phase configurations,
for the case $\Nx=\Ny=2$. For this system there are $N=4$ sites and thus $2^{16}$
(unnormalized) density matrices. We observe a transition in the first decay modes. 
The plot showing $g$ as a function of $\gamma$ for $J_1=J_2=J_3=J_4=1$ is shown in fig. \ref{2d_gap_decay_modes_2by2_1}. The gauge-invariant quantities corresponding to the first decay modes are shown in figures \ref{FDMsg} and \ref{FDMlg}.

The results obtained from $\Nx=\Ny=2$ system could be subject to finite-size effects. Hence we proceed to estimate the gap for higher system sizes, using other methods to optimize for the gap since it is computationally intensive to examine all $2^{3N+1}$
of the gauge sectors. We first tried looking at all configurations with a fixed number
$\Nv$ of $\MZ\ns_2$ defects -- plaquettes and Wilson phases whose values are reversed
relative to a given NESS configuration.  (A reversed-flux plaquette is a
$\MZ\ns_2$ vortex.)  There being $\Ng=3N+1$ gauge degrees of freedom, the number
of such configurations $\Ng \choose \Nv$ rapidly becomes computationally unwieldy
with growing $\Ng$ and $\Nv$.  We searched exhaustively for the smallest nonzero
relaxation rates for up to $\Nv=4$ total $\MZ\ns_2$ defects for $\Nx=\Ny=4$ and
only up to $\Nv=2$ for $\Nx=\Ny=6$ relative to a particular NESS
(one with all gauge-invariant $\MZ\ns_2$ data set to $-1$).
We also performed a Monte Carlo searches using both simulated annealing
and a genetic algorithm (GA) capable of finding states with arbitrary numbers 
of defects.  These both yielded similar results, and below we show data only
for the GA computations when comparing with the $\Nv$--limited searches.

\begin{figure}[!t]
\begin{centering}
\includegraphics[width=8.5cm,height=5.5cm]{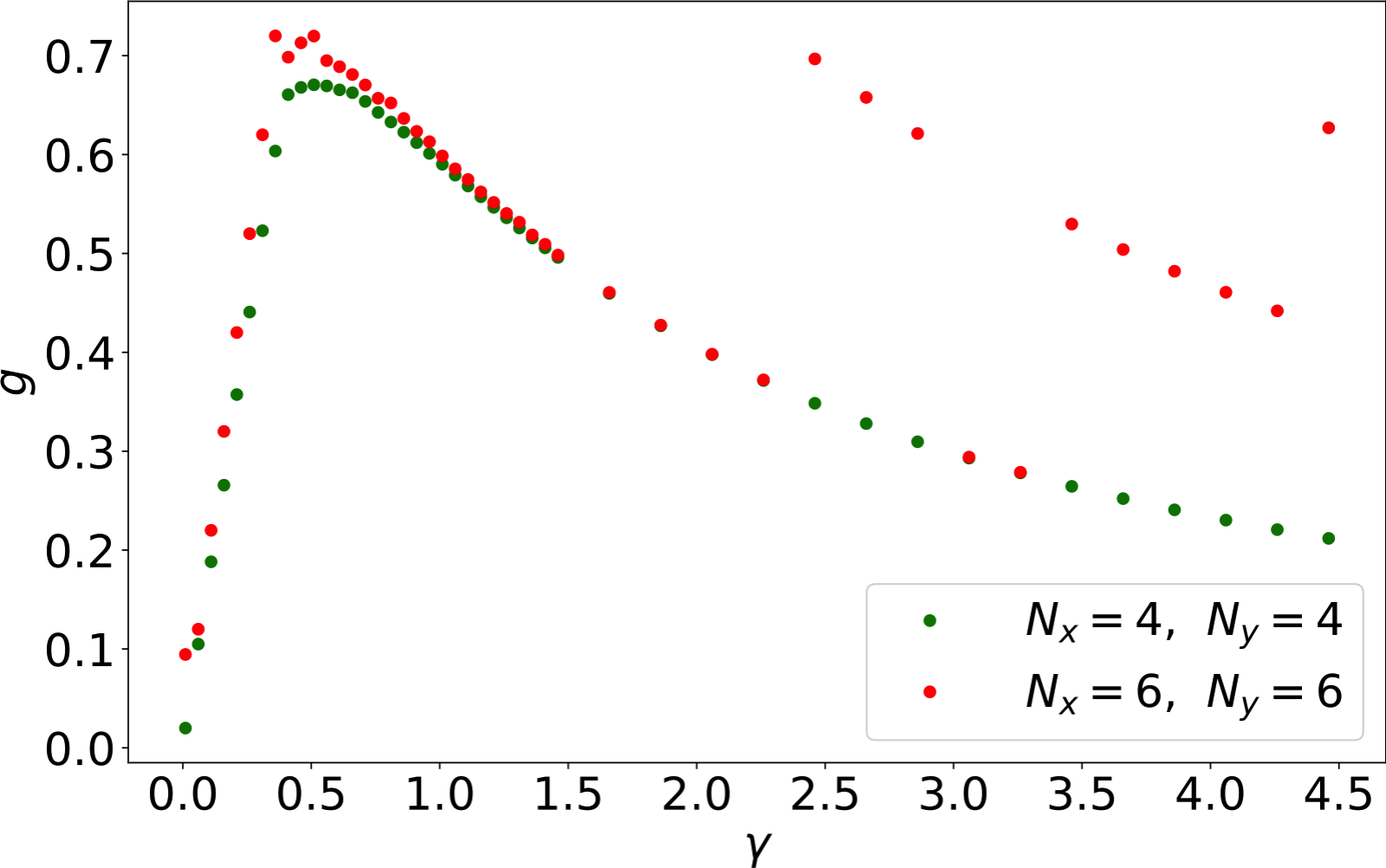}
\end{centering}
\caption{Minimal relaxation rate $g$ {\it versus\/} $\gamma$ obtained over 
different runs of the genetic algorithm (all $J\ns_\delta=1$) for $4\times 4$
and $6\times 6$ system sizes.  The population size was 100 and the number of
runs was 10 for $4\times 4$ and 5 for $6\times 6$. 
\label{genetic_algo_min_data}}
\end{figure}

Some details regarding the GA are provided in \S \ref{appendix_2d_model_genetic}
below. We found the results to be satisfactory both in terms of convergence of
the longest nonzero relaxation rate $g$ as well as the computational run time for
systems up to size $6\times 6$ ($2^{144}$ density matrices).  We cannot estimate
the full spectrum of first decay modes through this method, \ie\ enumerating
all their degeneracies as in the $2\times 2$ case, however.
The results obtained by taking the minimum value from different runs (see \S \ref{appendix_2d_model_genetic}) of the genetic algorithm are shown in fig.  \ref{genetic_algo_min_data}.  The result for $\Nx=\Ny=6$ is subject to more 
error since we used a fewer number of runs than in the $4\times 4$ case. This 
estimate can be improved by using the decay modes obtained from the genetic 
algorithm.

\begin{figure}[!t]
\begin{centering}
\includegraphics[width=8.5cm,height=5.5cm]{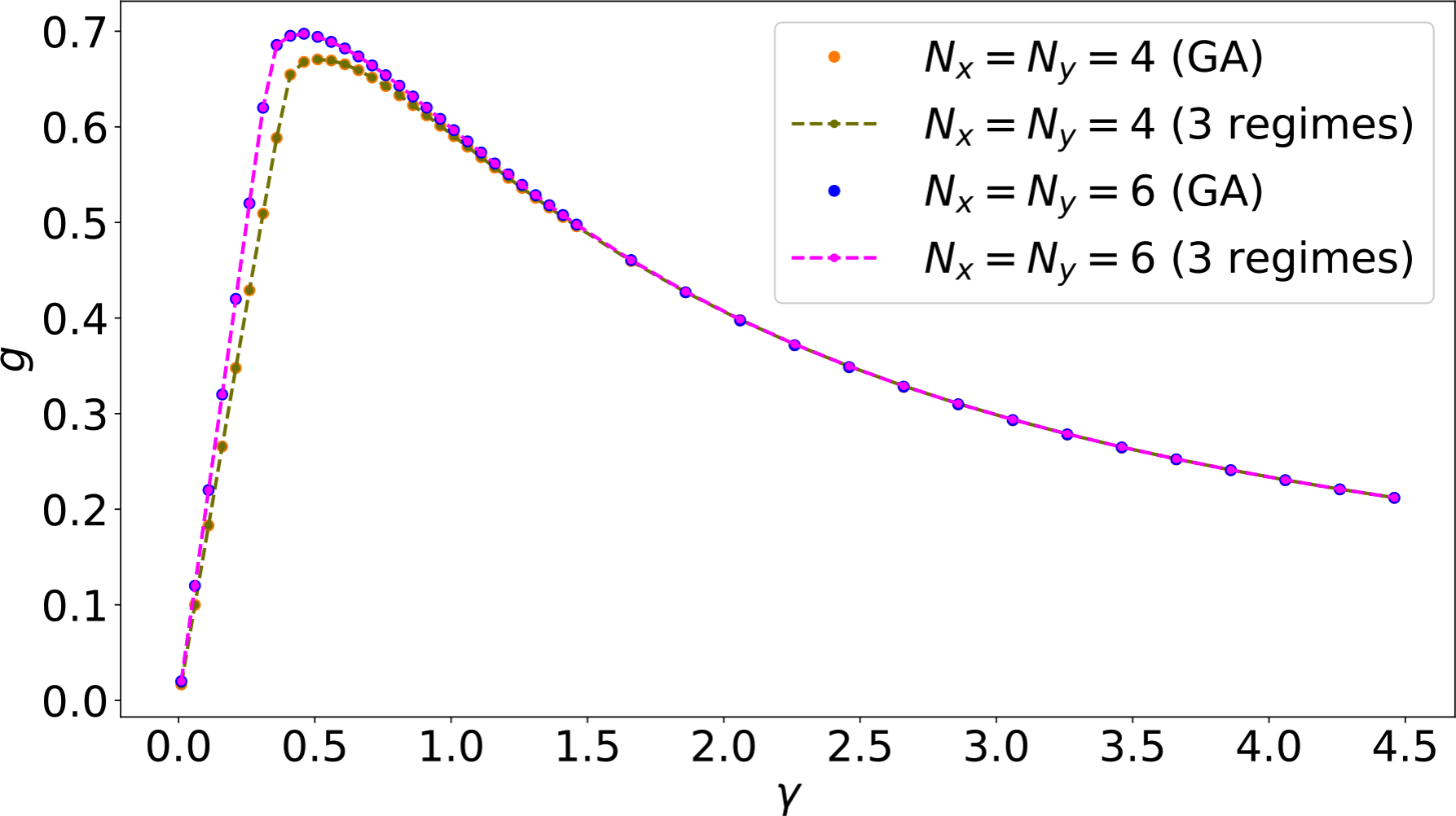}
\end{centering}
\caption{GA and the regimes are described in the text. All $J\ns_\delta=1$. 
The regimes for size $4\times 4$ are given by $\gamma<0.41$, $\gamma=0.41$
(a single point) and $\gamma>0.41$. The regimes for size $6\times 6$ are 
separated at $\gamma^*_1=0.36$ and $\gamma^*_2=0.41$.
\label{2d_gap_first_decay_modes}}
\end{figure} 

To obtain better estimates, we collect the best individuals from different 
GA runs (field configurations with minimum gap, \ie\ the
configurations corresponding to $g_{\text{min}}$ in fig. \ref{2d_genetic_algo_runs_sample} in \ref{appendix_2d_model_genetic}), and for different values of $\gamma$. We then use
this set of field configurations as our pool to be tried for each 
value of $\gamma$ in order to obtain an estimate of the minimum gap, $g$, by optimizing
the relaxation rate gap with respect to allowed configurations.  This yields the
curve labeled GA in fig. \ref{2d_gap_first_decay_modes}. For the system sizes
we have examined, the $g(\gamma)$ curves all exhibit a linear behavior at
small $\gamma$, crossing over to a $1/\gamma$ behavior at large $\gamma$, as
found by SK for their model \cite{Shibata2019}.  While the gauge configurations 
obtained in this manner can vary from one $\gamma$ value to the next, we found that
the curves are largely unchanged by partitioning the $\gamma$ line into three
regimes, each of which is governed by a particular configuration of the $\Ng$
gauge-invariant quantities, as reflected in the figure.  Note also the relatively
small difference between the $4\times 4$ and $6\times 6$ results.

\begin{figure}[!t]
\begin{centering}
\includegraphics[width=8.5cm,height=5.5cm]{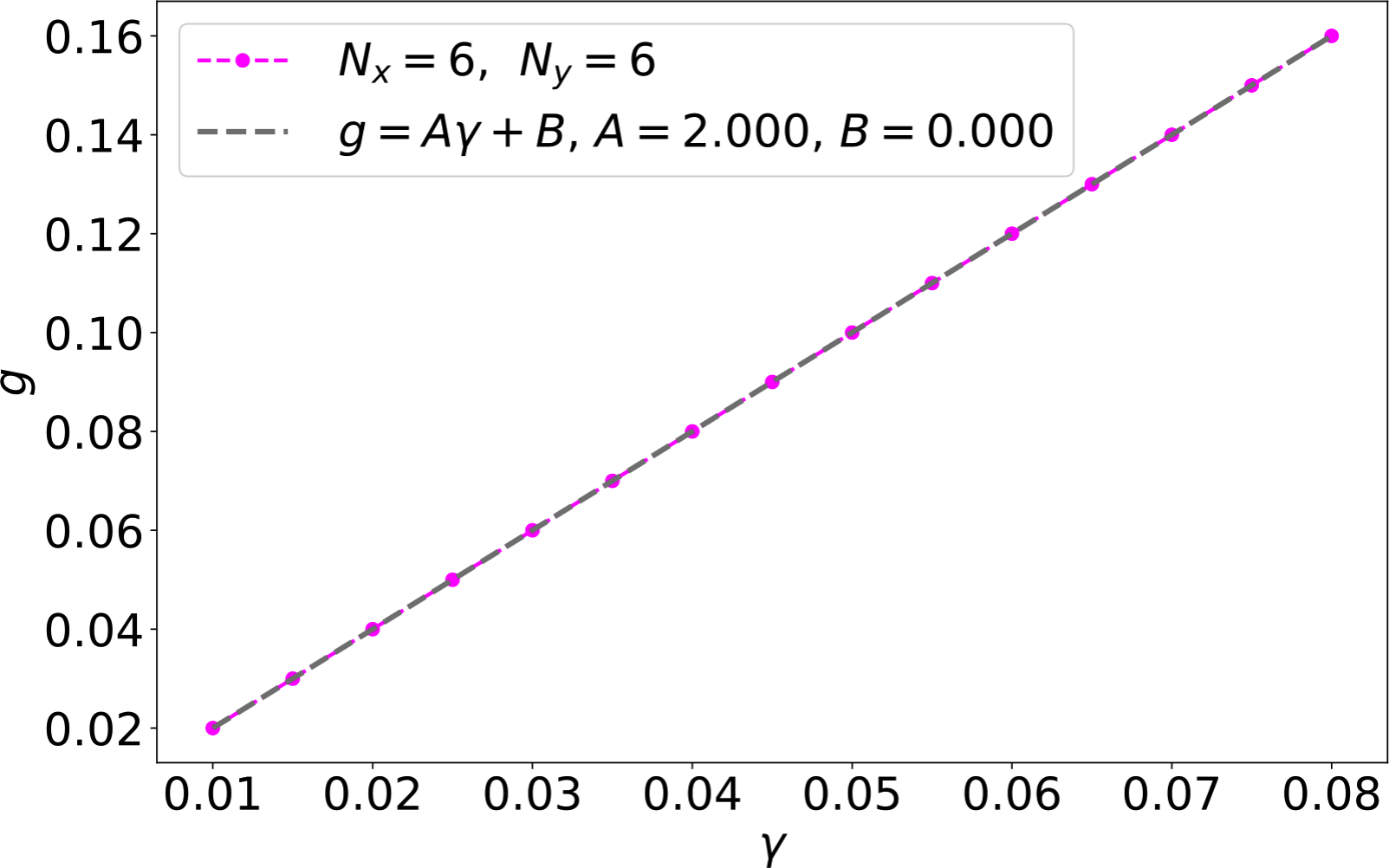}
\end{centering}
\caption{Behavior at small $\gamma$ for the largest system size we used in our calculations ($6\times 6$), with all $J\ns_\delta=1$. We obtained this curve by 
using the first decay modes we used to explain the small $\gamma$ regime in fig. \ref{2d_gap_first_decay_modes}.
\label{small_gamma_2d}}
\end{figure} 

\begin{figure}[!t]
\begin{centering}
\includegraphics[width=8.5cm,height=5.5cm]{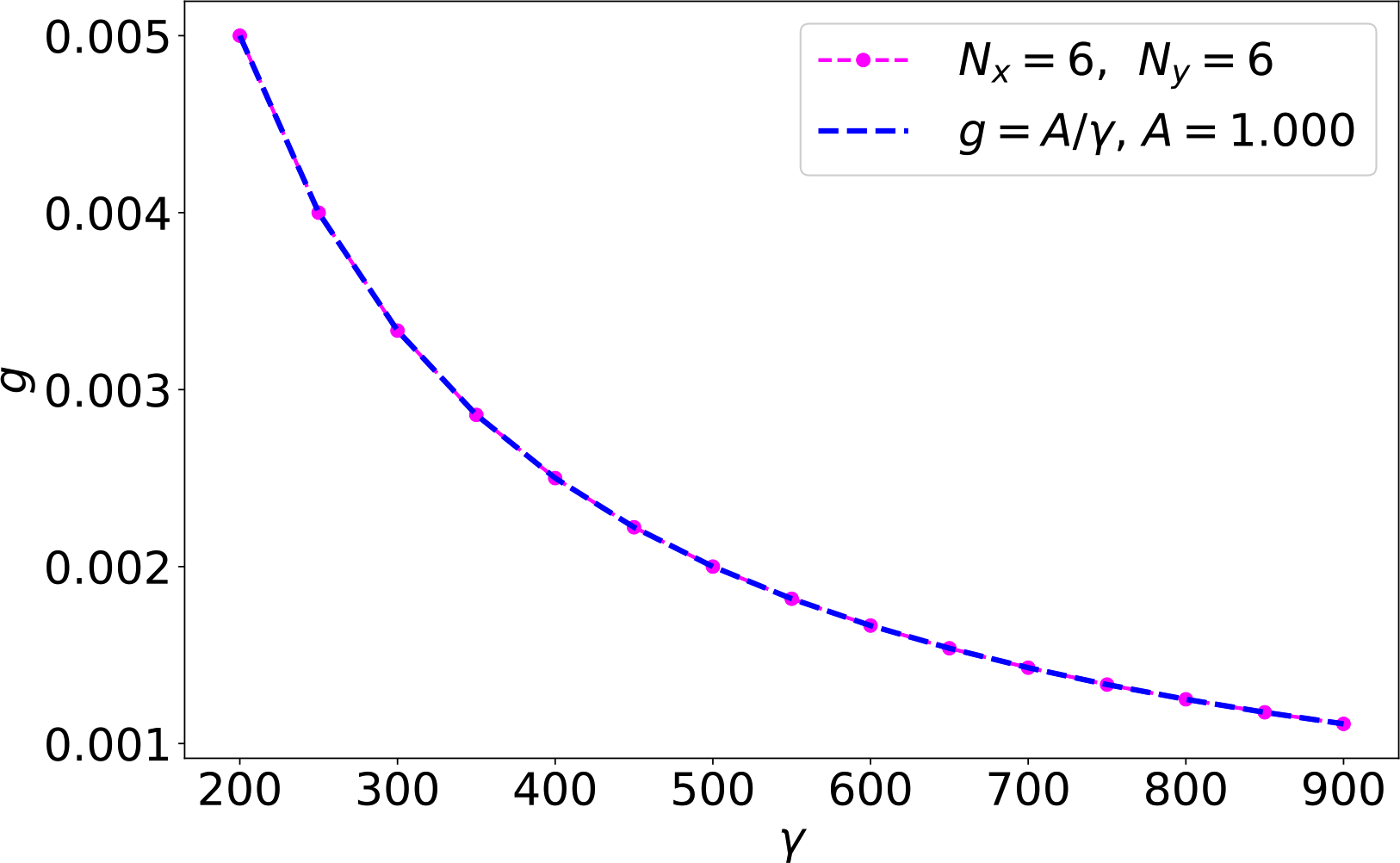}
\end{centering}
\caption{Behavior at large $\gamma$ for the largest system size we used in our calculations ($6\times 6$), with all $J\ns_\delta=1$. We obtained this curve by using the first decay modes we used to explain the large $\gamma$ regime in fig. \ref{2d_gap_first_decay_modes}.
\label{large_gamma_2d}}
\end{figure}

\begin{figure}[!t]
\begin{centering}
\includegraphics[width=8.5cm,height=5.5cm]{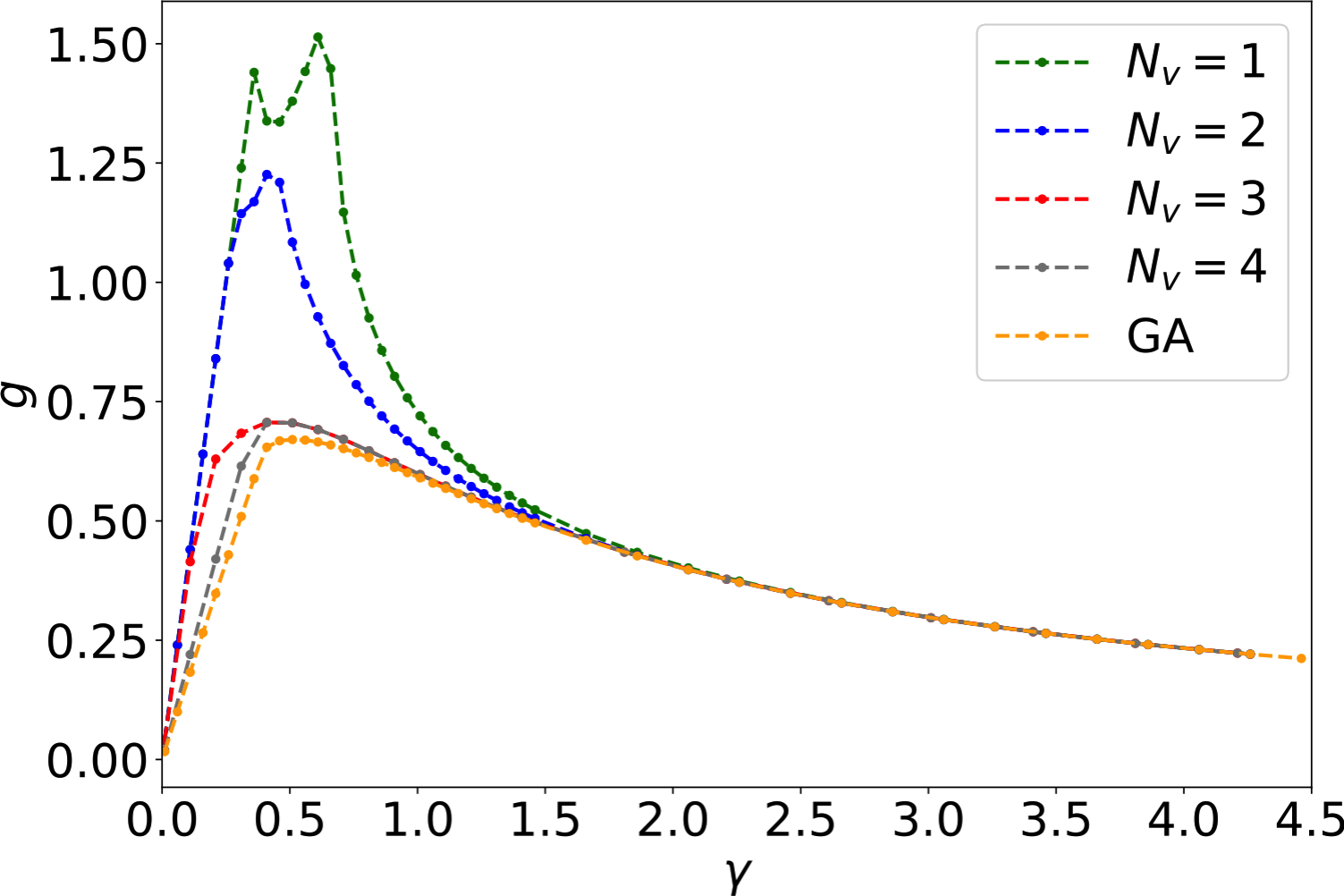}
\end{centering}
\caption{The GA curve and the curves obtained by considering configurations with a given number of vortices $N_v$, as described in the text.  Here all $J_\delta=1$ and the system size is $4 \times 4$.
\label{4_by_4_each_N_v}}
\end{figure}

\begin{figure}[!t]
\begin{centering}
\includegraphics[width=8.5cm,height=5.5cm]{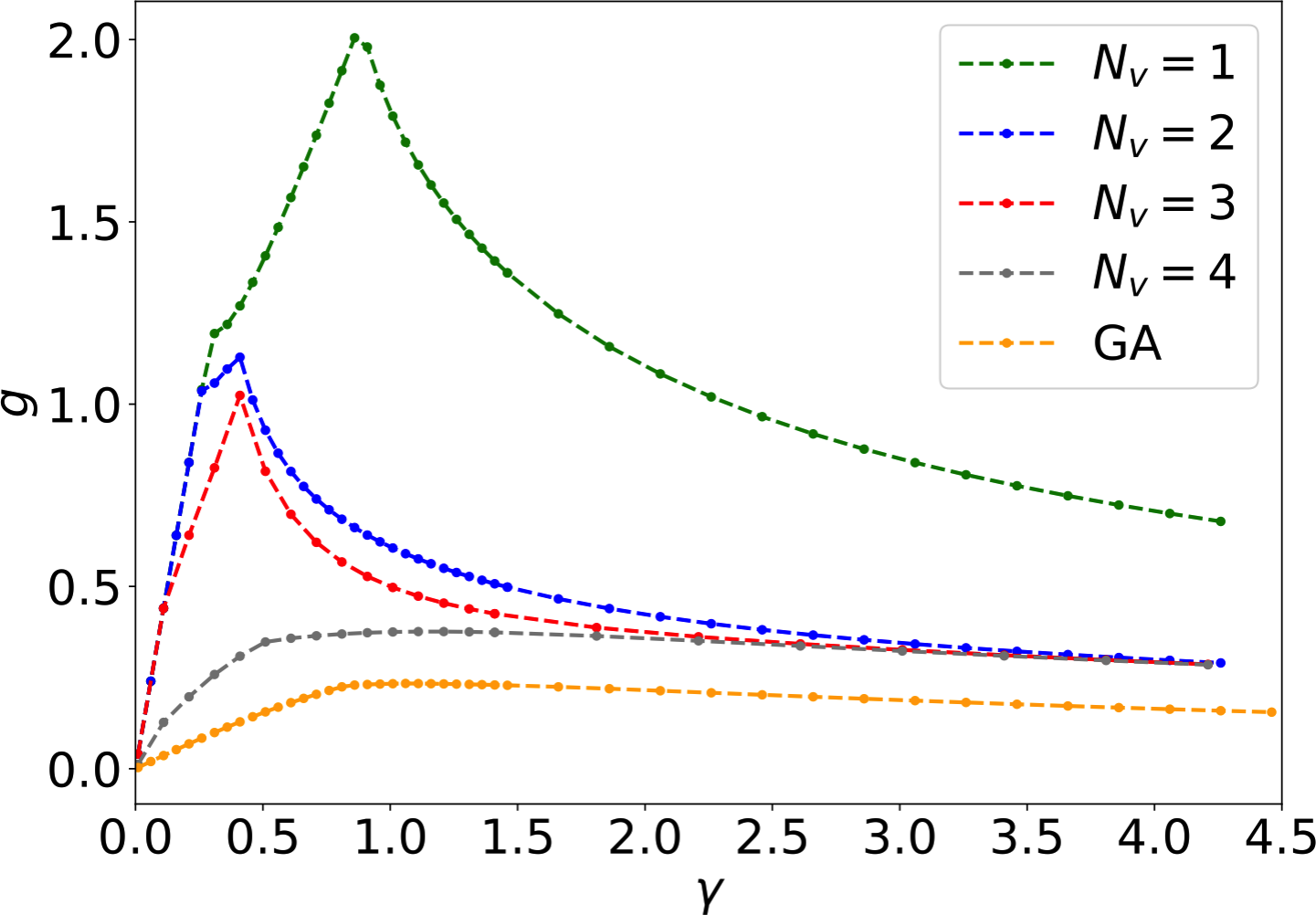}
\end{centering}
\caption{The GA curve and the curves obtained by considering configurations with a given number of vortices $N_v$, as described in the text.  Here $\left(J_1, J_2, J_3, J_4\right)=\left(3,4,1,2\right)$ and the system size is $4 \times 4$.
\label{4_by_4_each_N_v_diff_J_s}}
\end{figure}

\begin{figure}[!t]
\begin{centering}
\includegraphics[width=8.5cm,height=5.5cm]{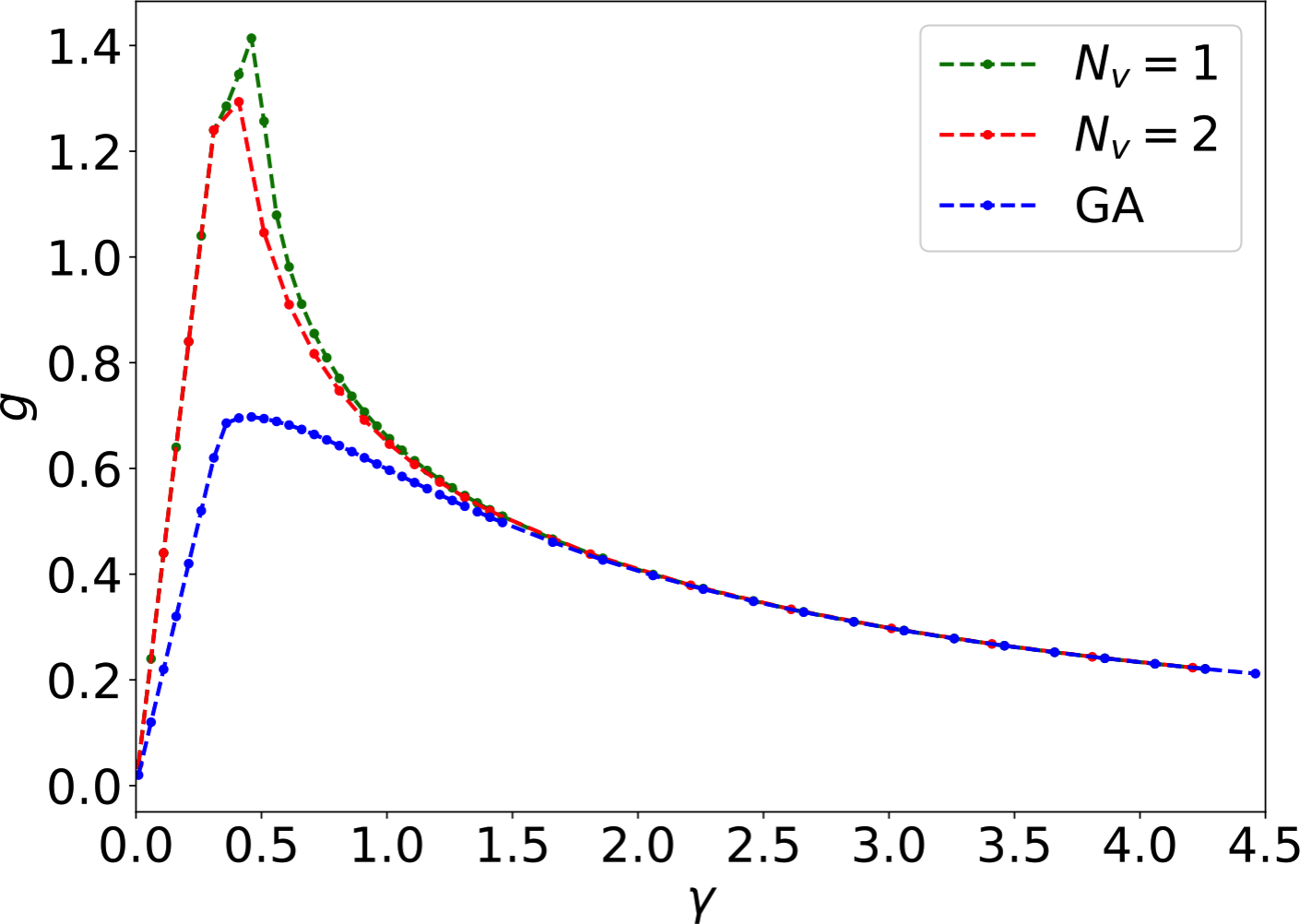}
\end{centering}
\caption{The GA curve and the curves obtained by considering configurations with a given number of vortices $N_v$, as described in the text.  Here all $J_\delta=1$ and the system size is $6 \times 6$.
\label{6_by_6_each_N_v}}
\end{figure}

SK found a sharp transition in the first decay modes between two regimes of
dissipation strength, regardless of system size.  (SK examined their model with open
boundaries, but we have confirmed this result when periodic boundary conditions
are applied to their model as well.) We cannot conclude whether or not this
is the case for our model, but the intermediate regime we find could result from
a failure of the GA to reach the block of true first decay modes.  A clear
intermediate regime is apparent in the $\Nv$--limited data of fig. \ref{4_by_4_each_N_v},
with all $J\ns_\delta=1$, for $\Nv=1,2$ on size $4\times 4$.  (See also fig. \ref{4_by_4_each_N_v_diff_J_s} for the case when all the $J\ns_\delta$ are different.) 
For $\Nv=3,4$ the effect is far less pronounced.  Note that the $\Nv=4$ results are 
in good agreement with the GA results. For the $6\times 6$ case, the $\Nv=1,2$ results 
in fig. \ref{6_by_6_each_N_v} are quite far from the GA curve.

From the genetic algorithm, the optimal flux configurations for the lowest nonzero 
decay seem to contain many defects.  We list some of these configurations in
the appendix \S \ref{configs} below.  For example, for the $4\times 4$ system with
all $J\ns_\delta=1$, the optimal excited state we obtained had 14 defects relative
to the fiducial NESS with all $\MZ\ns_2$ data set to $-1$.  However, since we start 
the GA from a population of random $\MZ\ns_2$ data, it may well be that a configuration
with 14 defects with respect to a particular NESS might be described by fewer defects
with respect to a different state in the $2^{3N+1}$--fold block of NESSes.  We are
now attempting to better clarify the minimal defect content of the degenerate
excited states.

\begin{figure}[!t]
\begin{centering}
\includegraphics[width=8.5cm,height=5.5cm]{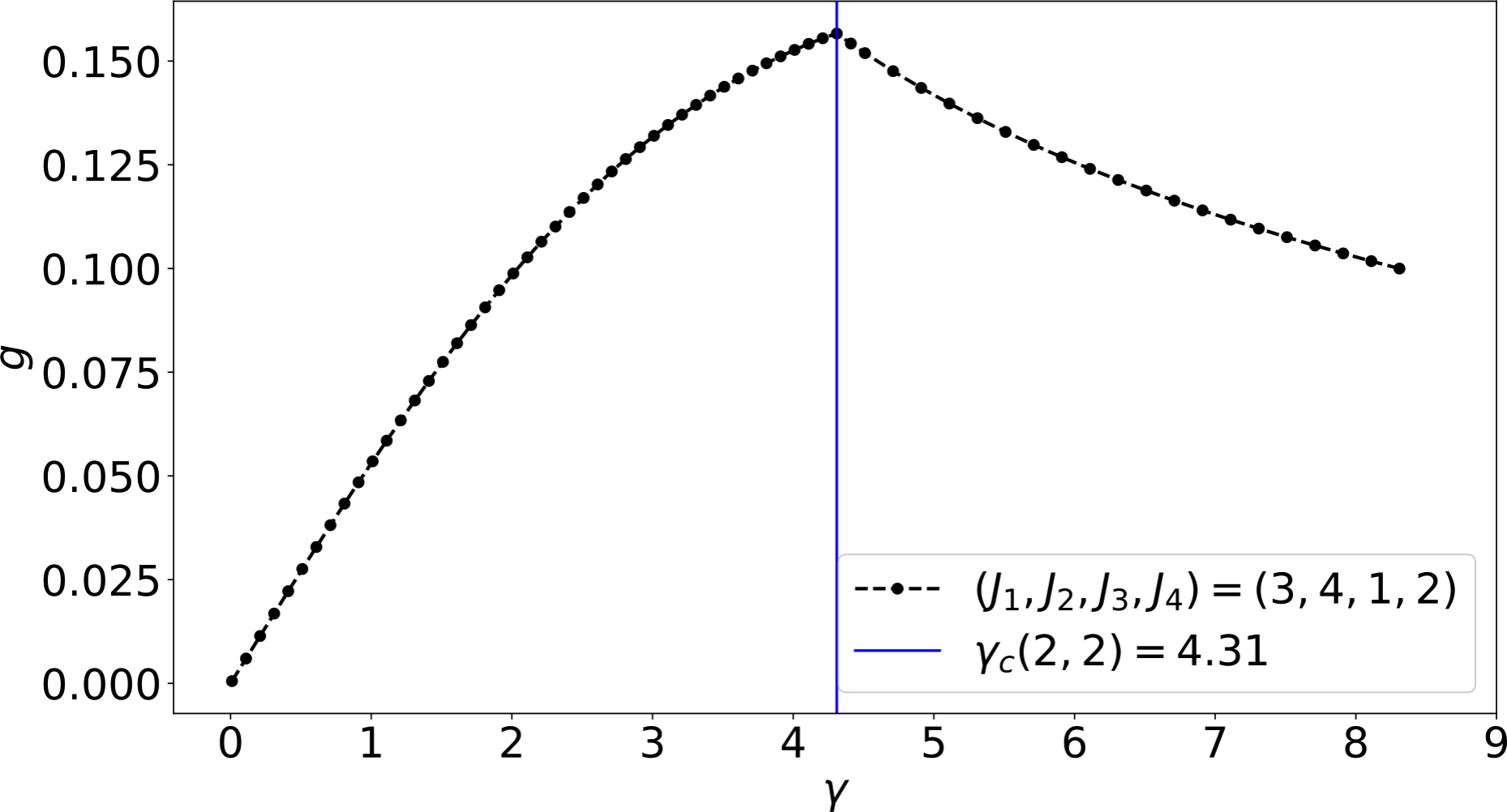}
\end{centering}
\caption{Liouvillian gap for our model with periodic boundary conditions for $\Nx=\Ny=2$ and $(J_1,J_2,J_3,J_4)=(3,4,1,2)$: The Liouvillian gap, $g$, as a function of $\gamma$. There is a transition in the first decay modes at the cusp seen at $\gamma=\gamma_\Rc(2,2)$, depicted by the blue vertical line. We see a transition in the first decay modes. We have 2 first decay modes for $\gamma<\gamma_\Rc(2,2)$ and eight first decay modes for $\gamma\geq\gamma_\Rc(2,2)$.
\label{2d_gap_decay_modes_2by2_1_diff_J_s}}
\end{figure} 

\begin{figure}[!t]
\begin{centering}
\includegraphics[width=8.5cm,height=5.5cm]{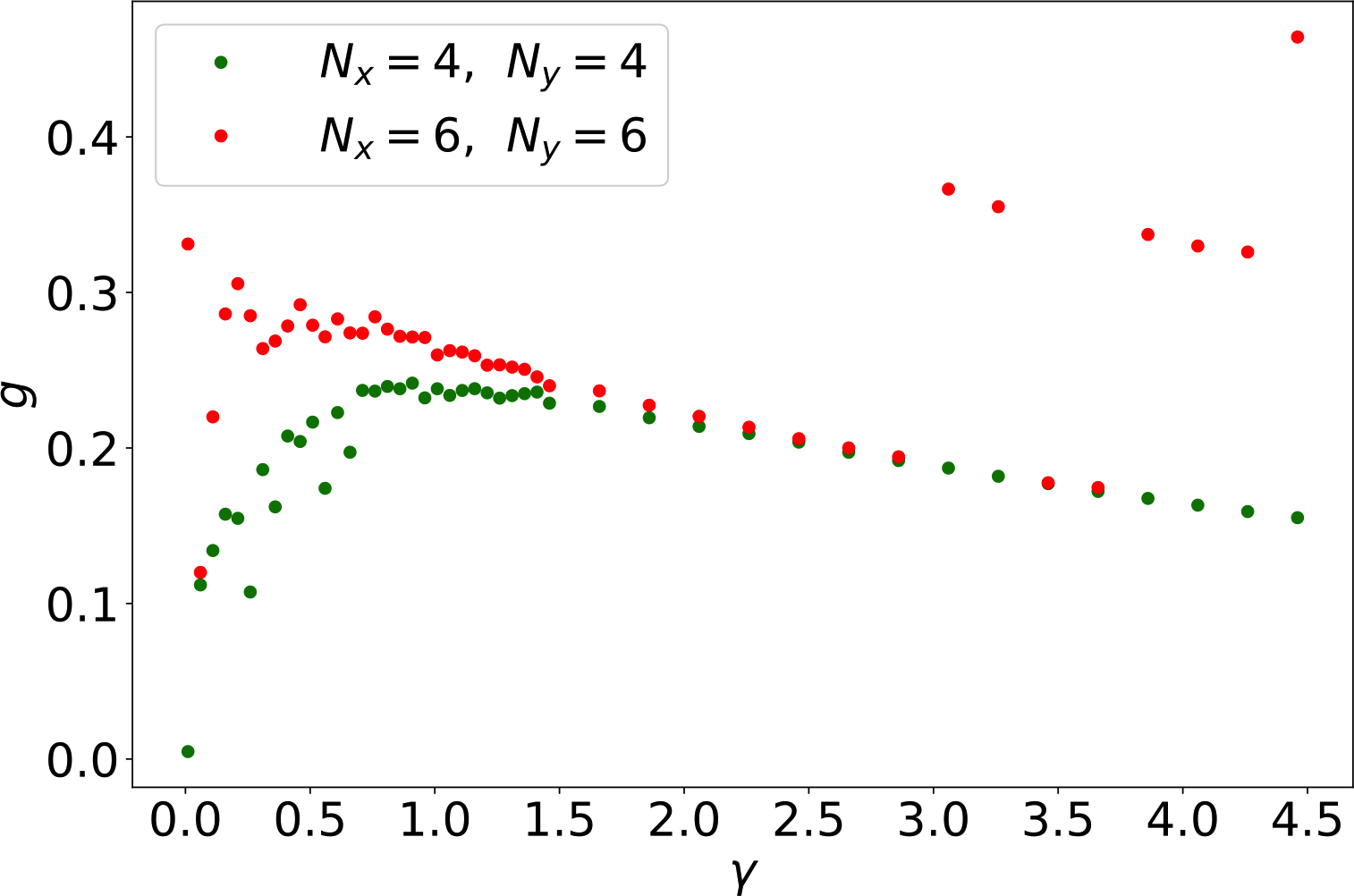}
\end{centering}
\caption{
Minimal relaxation rate $g$ {\it versus\/} $\gamma$ obtained over 
different runs of the genetic algorithm with $(J_1,J_2,J_3,J_4)=(3,4,1,2)$
for system sizes $4\times 4$ and $6\times 6$.  The population size was 100 and 
the number of runs was 10.
\label{genetic_algo_min_data_diff_J_s}}
\end{figure} 

\begin{figure}[!t]
\begin{centering}
\includegraphics[width=8.5cm,height=5.5cm]{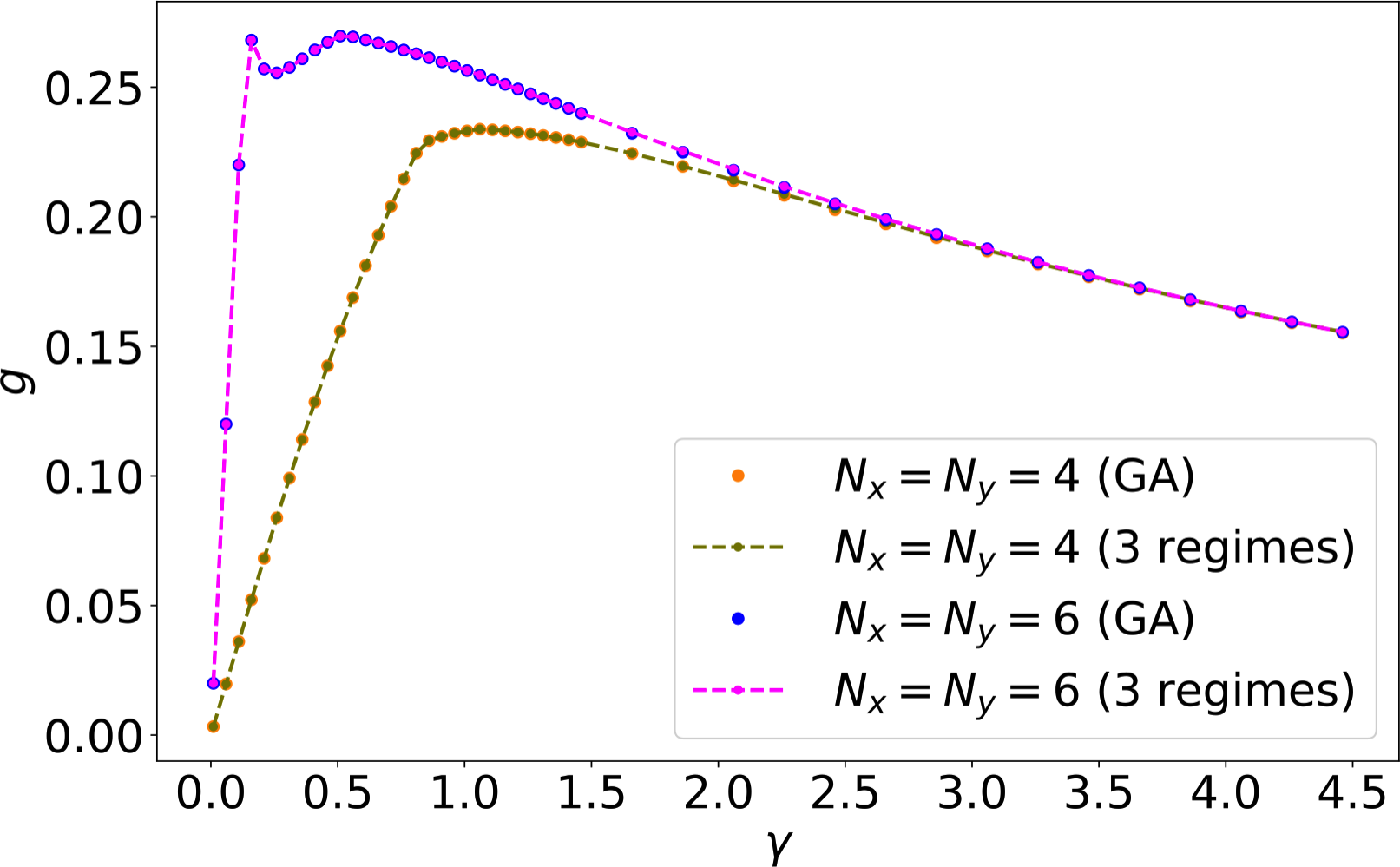}
\end{centering}
\caption{GA and the phases are described in the text. Here, 
$(J_1,J_2,J_3,J_4)=(3,4,1,2)$. The 3 regimes for $4 \times 4$ are separated by $\gamma_1^*=0.86$ and $\gamma_2^*=1.06$ and the three regimes for $6 \times 6$ are separated by $\gamma_1^*=0.16$ and $\gamma_2^*=0.51$.
\label{2d_gap_first_decay_modes_diff_J_s}}
\end{figure} 

We try fitting the $g(\gamma)$ curves to identify their behavior for small and
large values of the dissipation strength $\gamma$ (see figs. \ref{small_gamma_2d} 
and \ref{large_gamma_2d}). The shape of the $g(\gamma)$ curves is similar to 
that found by Shibata and Katsura \cite{Shibata2019}, rising linearly from zero at 
small $\gamma$ and decaying as $1/\gamma$ for large $\gamma$.  In the appendix 
\S \ref{PTH}, we provide analytical support for these behaviors.

We also investigate the behavior of the gap for the case $(J_1,J_2,J_3,J_4)=(3,4,1,2)$,
which breaks certain discrete translation and rotation symmetries present in the model
when all $J\ns_\delta$ are equal.  The results are shown in figs. \ref{2d_gap_decay_modes_2by2_1_diff_J_s}, \ref{genetic_algo_min_data_diff_J_s} and \ref{2d_gap_first_decay_modes_diff_J_s}. GA and the phases are obtained as described above. While the shape of the curves is similar, we find two significant differences.
First, from fig. \ref{2d_gap_first_decay_modes_diff_J_s} it appears that the GA
has not found the lowest decay mode in the small $\gamma$ regime, where the $4\times 4$ 
and $6\times 6$ GA results differ substantially.  Second, as shown in fig.
\ref{4_by_4_each_N_v_diff_J_s}, the $\Nv\le 4$ sector does not yield a good 
approximation to the GA results, as it did for the case with all $J\ns_\delta=1$.

\section{Conclusions}
In this paper we have described a two-dimensional square lattice model of interacting
gamma matrix `spins' coupled to a dissipative environment.  The density matrix
evolution is described by the GKLS master equation ${\dot\vrh}=\CL\vrh$, where $\CL$
is the Liouvillian, which in general has complex eigenvalues $\Lambda\ns_a$.  
This description is equivalent to Schr{\"o }dinger evolution under a non-Hermitian
Hamiltonian $\CW$ on a square lattice bilayer, whose eigenvalues are $E\ns_a=i\Lambda\ns_a$\,. Our model is inspired by, and a generalization of, the dissipative one-dimensional Pauli matrix spin model of Shibata and Katsura
\cite{Shibata2019}. It is in the `solvable' class of models exemplified by Kitaev's
celebrated honeycomb lattice model \cite{Kitaev2006}, equivalent to a single species 
of Majorana fermion hopping in a nondynamical $\MZ\ns_2$ background gauge field. 
It is solvable in the sense that for any given configuration of the gauge-invariant
plaquette fluxes and Wilson phases, the non-Hermitian Hamiltonian $\CW$ is quadratic
and solvable by Prosen's method \cite{Prosen2008}.  However, there are exponentially
many such configurations, and when the gauge field structure is not translationally
invariant, the Hamiltonian must be diagonalized numerically.  Furthermore, there is
no analog of Lieb's theorem \cite{Lieb1989} to assist us in identifying the longest
lived decaying eigenmodes.

In the infinite time limit, the system approaches one of an exponentially large number 
of nonequilibrium steady states, with a spectrum $\{-\Imp E\ns_a\}$ of relaxation
rates. The minimum relaxation rate $g(\gamma)$ is typically achieved for different
$\MZ\ns_2$ flux configurations in the small and large $\gamma$ limits, a feature also
observed by Shibata and Katsura.

\begin{figure}[!t]
\begin{centering}
\includegraphics[width=8.5cm,height=5.5cm]{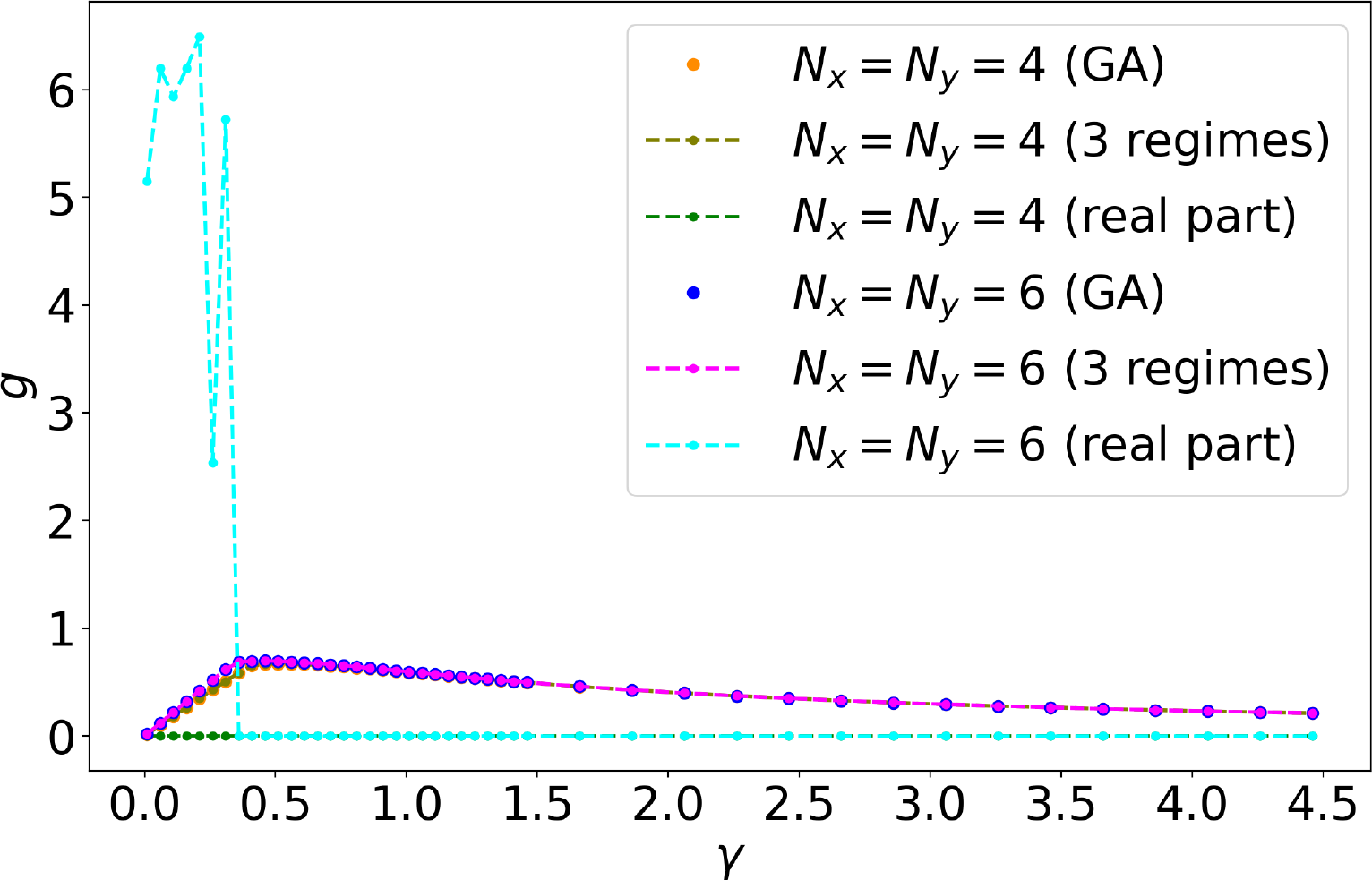}
\end{centering}
\caption{Imaginary ($g$) and real parts of the lowest decay modes as a function of $\gamma$ for $4\times 4$ and $6\times 6$ lattices, with $(J_1,J_2,J_3,J_4)=(1,1,1,1)$.  
$\Rep E_a=0$ in all cases except for $6\times 6$ with $\gamma<\gamma_\Rc$.
\label{ImpL}}
\end{figure}

We have not indicated in our plots the spectrum $\{\Rep E\ns_a\}$ of the
real parts of the eigenvalues of $\CW$.  This is because in almost all cases
studied we have found the imaginary parts of the first decay mode eigenvalues to
be zero.  The only exception we observed was in the $6\times 6$ case with all
$J$ couplings equal and for $\gamma < \gamma_\Rc$\,, as shown in 
fig. \ref{ImpL} \footnote{Modes with equal and opposite nonzero values of $E\ns_a$
are degenerate. We cannot rule out the possibility that there are degenerate lowest
decay modes for the $6\times 6$ lattice with uniform $J\ns_\delta$ with 
$\Rep E\ns_a=0$.} (When all $J$'s are different, we find $\Rep E\ns_a = 0$ for the lowest decay modes, for all $\gamma$ and all sizes.)

\begin{figure}[!t]
\begin{centering}
\includegraphics[width=8.5cm,height=5.5cm]{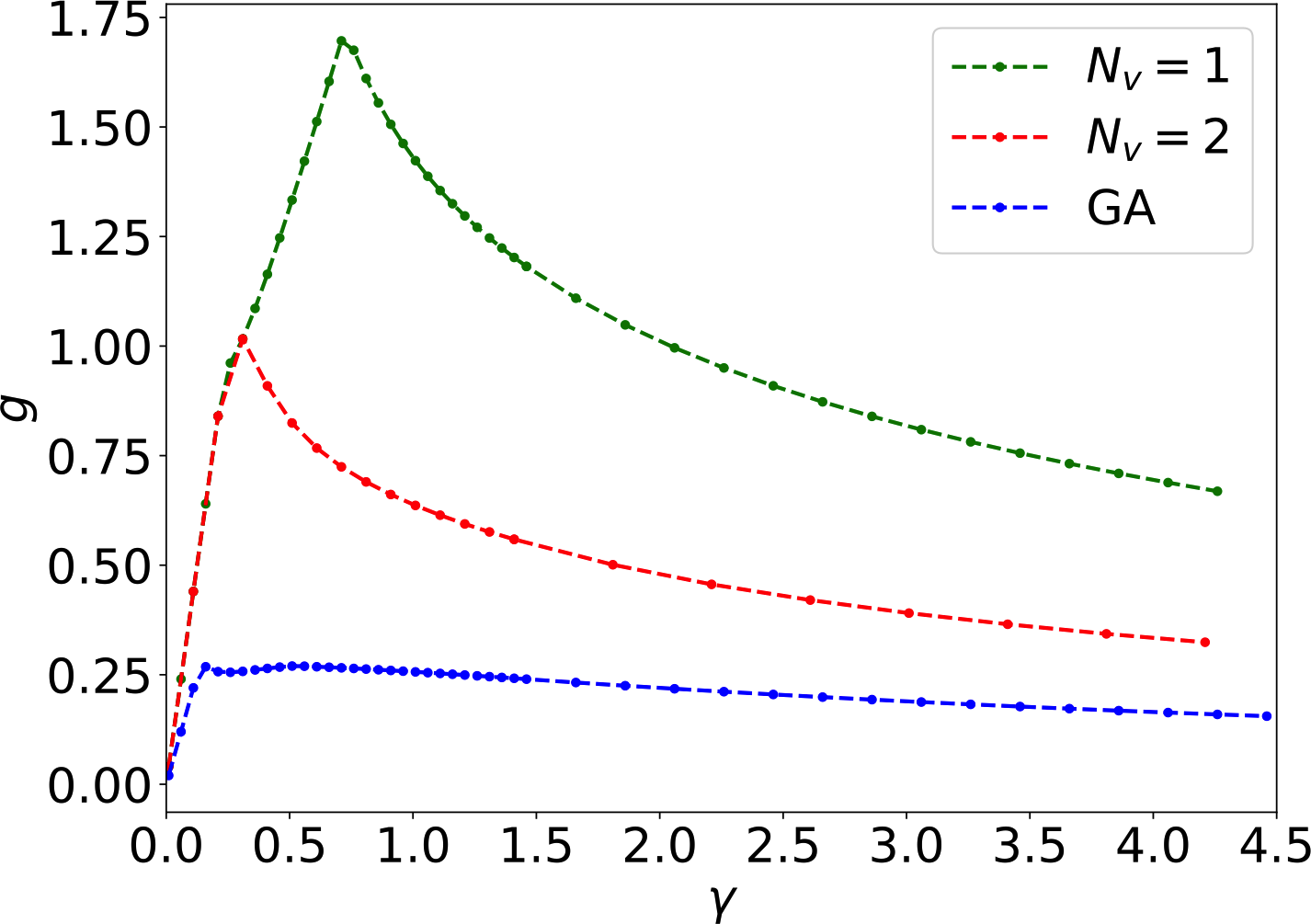}
\end{centering}
\caption{The GA curve and the curves obtained by considering configurations with a given number of vortices $N_v$, as described in the text.  Here all $\left(J_1, J_2, J_3, J_4\right)=\left(3,4,1,2\right)$ and the system size is $6 \times 6$.
\label{6_by_6_each_N_v_diff_J_s}}
\end{figure} 

Our model can further be generalized to other lattices.  The Kitaev solvability of the
SK model is associated with the fact that their model is equivalent to non-Hermitian
Hamiltonian evolution on a two leg ladder, where each site lies at the confluence of
three distinct classes of links. For the dimension $k$ Clifford algebra, we have
$2k+1$ gamma matrices of dimension $2^k$, and a Kitaev Hamiltonian (Hermitian or not)
can be constructed on any lattice where each site lies at the confluence of $(2k+1)$
distinct classes of links \cite{WAH09}. Thus, for $k=2$, our square lattice bilayer
is five-fold coordinated. A corresponding model could thus be constructed on
the kagome lattice, leading to a non-Hermitian Dirac matrix Hamiltonian $\CW$ 
on the kagome bilayer. (Further generalizations of this construction can result
in multiple species of hopping and hybridizing Majoranas in the presence of a
background nondynamical gauge field, as in refs. \cite{YZK09,CYF11}.) Thus,
proceeding to $k=3$ with its seven $8\times 8$ gamma matrices, a corresponding model
can be constructed on a cubic lattice (bipartite NaCl structure) with 
$\Gamma^\delta_\BR\,\Gamma^\delta_{\BR+\Bdel}$ interactions on each class $\delta$ link with $\delta\in\{1,\ldots,6\}$ and Lindblad jump operators
$\sqrt{\gamma}\,\Gamma^7_\Br$ at each site.  Again, there will be an
exponentially large block of NESS density matrices owing to the conserved
plaquette fluxes.

{\sl Note:} Some of this work was presented in a poster at the KITP Workshop, 
{\sl Topology, Symmetry and Interactions in Crystals: Emerging Concepts and 
Unifying Themes} (KITP UC Santa Barbara, April 3--6, 2023) \cite{JGKITP23}.
While this article was in the final stages of preparation, an analysis of
two largely equivalent models appeared on the arXiv \cite{Dai23,Scheurer23}.

\section{Acknowledgements}
We gratefully acknowledge conversations with Tarun Grover and John McGreevy.
We thank Debanjan Chowdhury for alerting us to ref. \cite{Scheurer23}.
This research was funded in part by General Campus Research Award RG104654 from the 
UC San Diego Academic Senate.

\vfill

\bibliographystyle{unsrt}
\bibliography{references.bib}

\begin{thebibliography}{10}

\bibitem{Shibata2019}
Naoyuki Shibata and Hosho Katsura.
\newblock {Dissipative spin chain as a non-Hermitian Kitaev Ladder}.
\newblock {\em Physical review B}, 99(17):174303, 2019.

\bibitem{Kitaev2006}
Naoyuki Shibata and Hosho Katsura.
\newblock {Anyons in an exactly solved model and beyond}.
\newblock {\em Annals of Physics}, 321(1):2, 2006.

\bibitem{Breuer2007}
Heinz-Peter Breuer and Francesco Petruccione.
\newblock {\em {The Theory of Open Quantum Systems}}.
\newblock Oxford University Press, 01 2007.

\bibitem{Prosen2008}
{Third quantization: a general method to solve master equations for quadratic
  open Fermi systems}.
\newblock {\em New Journal of Physics}, 10(4):043026, apr 2008.

\bibitem{ChruPa2017}
Dariusz Chru{\'{s}}ci{\'{n}}ski and Saverio Pascazio.
\newblock {A Brief History of the GKLS Equation}.
\newblock {\em Open Systems and Information Dynamics}, 24(03):1740001, sep
  2017.

\bibitem{Lieb1989}
Elliott~H. Lieb.
\newblock {Two theorems on the Hubbard model}.
\newblock {\em Phys. Rev. Lett.}, 62:1201--1204, Mar 1989.

\bibitem{YZK09}
Hong Yao, Shou-Cheng Zhang, and Steven~A. Kivelson.
\newblock {Algebraic Spin Liquid in an Exactly Solvable Spin Model}.
\newblock {\em Phys. Rev. Lett.}, 102:217202, May 2009.

\bibitem{WAH09}
Congjun Wu, Daniel Arovas, and Hsiang-Hsuan Hung.
\newblock {$\Gamma$-matrix generalization of the Kitaev model}.
\newblock {\em Physical review B}, 79(13):134427, 2009.

\bibitem{Dai23}
Xu-Dong Dai, Fei Song, and Zhong Wang.
\newblock {Solvable BCS-Hubbard Lindbladians in arbitrary dimensions}.
\newblock {arXiv 2306.13148}.

\bibitem{Scheurer23}
Henry Shackleton and Mathias~S. Scheurer.
\newblock {An exactly solvable dissipative spin liquid}.
\newblock {arXiv 2307.05743}.

\bibitem{Albert2018}
Victor~V. Albert.
\newblock {Lindbladians with multiple steady states: theory and applications}.
\newblock {arXiv 1802.00010}.

\bibitem{Dzhioev_2012}
Alan~A Dzhioev and D~S Kosov.
\newblock Nonequilibrium perturbation theory in liouville--fock space for
  inelastic electron transport.
\newblock {\em Journal of Physics: Condensed Matter}, 24(22):225304, may 2012.

\bibitem{CYF11}
Victor Chua, Hong Yao, and Gregory~A. Fiete.
\newblock {Exact chiral spin liquid with stable spin Fermi surface on the
  kagome lattice}.
\newblock {\em Phys. Rev. B}, 83:180412, May 2011.

\bibitem{Note1}
We find it convenient here to use expressions $\Gamma ^{\mu \nu }\equiv -\Gamma
  ^{\nu \mu }$ with $\mu >\nu $.

\bibitem{Note2}
There is of course a single remaining constraint associated with the condition
  $\mathop {\protect \textsf {Tr}}\varrho =1$.

\bibitem{Note3}
Modes with equal and opposite nonzero values of $E^{\protect \vphantom {*}}_a$
  are degenerate. We cannot rule out the possibility that there are degenerate
  lowest decay modes for the $6\times 6$ lattice with uniform $J^{\protect
  \vphantom {*}}_\delta $ with $\protect \textsf {Re}\protect \,E^{\protect
  \vphantom {*}}_a=0$.

\bibitem{JGKITP23}
Jyotsna Gidugu and Daniel~P. Arovas.
\newblock {A study on dissipative models based on $\Gamma$-matrices}.
\newblock
  {https://online.kitp.ucsb.edu/online/qcrystal-c23/gidugu/qcrystal-c23\_poster\_gidugu.pdf},
  2023.

\bibitem{Can19}
Tankut Can, Vadim Oganesyan, Dror Orgad, and Sarang Gopalakrishnan.
\newblock Spectral gaps and midgap states in random quantum master equations.
\newblock {\em Phys. Rev. Lett.}, 123:234103, Dec 2019.

\bibitem{HoJo85}
Roger~A. Horn and Charles~R. Johnson.
\newblock {\em {Matrix Analysis}}.
\newblock Cambridge University Press, 1985.

\end{thebibliography}
\onecolumngrid

\appendix

\section{Perturbation Theory for Large and Small values of $\gamma$}\label{PTH}

Can \etal\ \cite{Can19} considered a model of a random Hermitian Hamiltonian
$H$ in the presence of a single Hermitian jump operator $\sqrt{\gamma}\,L$, 
with both $H$ and $L$ random and chosen from the Gaussian unitary ensemble.
They showed how the spectrum of relaxation times could be computed perturbatively
in the small and large $\gamma$ limits, by perturbing in the dissipator
$\CLD=\half \gamma\big[\,L\CC[\,\bullet\CC L\,]\,\big]$ or in the 
non-dissipative Liouvillian $\CLH=-i\,[\,H\CC\bullet\,]$, respectively.
Thus, at small $\gamma$, the smallest nonzero relaxation rate $\tau^{-1}$
is proportional to $\gamma$, while at large $\gamma$ it is proportional to
$\gamma^{-1}$. In this appendix we perform a related analysis for our model.

\subsection{$\gamma\ll 1$}
For $\gamma\ll 1$ we write the GKLS equation as a classical master equation
for the projectors $\RP\ns_\alpha=\sketbra{\alpha}{\alpha}$, where $\alpha$ is
an eigenstate of the Hamiltonian $H$ \cite{Can19}.  This classical master equation
is written
\begin{equation}
{d\,\RP\ns_\alpha\over dt} = \gamma\CM\ns_{\alpha\beta}\,\RP\ns_\beta\quad,
\end{equation}
with
\begin{equation}
\CM\ns_{\alpha\beta}=\sum_\Br\Big(\big|\sexpect{\alpha}{\Gamma^5_\Br}{\beta}\big|^2 -
\delta\ns_{\alpha\beta}\Big)\quad.
\end{equation}
Clearly this analysis will lead in this limit to real eigenvalues proportional to $\gamma$.  What is the coefficient for
the smallest such eigenvalue?

To analyze this, we start with our Hamiltonian,
\begin{equation}
H=i\sum_{\BR\in\SA}\sum_{\delta=1}^4 J\ns_\delta\,u^\delta_\BR\, \theta^0_\BR\,\theta^0_{\BR+\Bdel}\equiv
{i\over 2}\sum_{\Br,\Br'} J\ns_{\Br,\Br'}\,\theta^0_\Br\,\theta^0_{\Br'}\quad,
\end{equation}
where $J\ns_{\Br,\Br'}$ is a real antisymmetric matrix with $J\ns_{\BR,\BR+\Bdel}=J^\delta\,u^\delta_\BR$.  The $\theta^5_\Br$
Majoranas are cyclic in $H$ and lead to an exponential degeneracy $\sim 2^\Nc$ for each energy level.  Including
the $N-1$ plaquette fluxes and the two Wilson phases, and accounting for the constraint 
\begin{equation}
\prod_\Br i\theta^0_\Br\,\theta^1_\Br\,\theta^2_\Br\,\theta^3_\Br\,\theta^4_\Br\,\theta^5_\Br=1\quad,
\end{equation}
which constrains the combined parity of the $a=0$ and $a=5$ species fermions given the gauge fields $u\ns_{\Br\Br'}$,
we have a total of $4^N$ projectors: $2^{N+1}$ from plaquette fluxes and Wilson phases (see Fig. \ref{Fxyflux}), and $2^{N-1}$ for 
the parity-constrained $a=0,5$ fermions.  This is a subset of the $16^N$ (unnormalized) density matrices.  

Every real antisymmetric matrix can be diagonalized by an orthogonal transformation.  For each gauge field configuration, the
corresponding real antisymmetric matrix $J\ns_{\Br\Br'}$ is brought to block diagonal form by a real orthogonal matrix $Q\ns_{\Br,\Bs}$, 
such that
\begin{equation}
Q\tra J\,Q=\textsf{diag}\>\Bigg\{\begin{pmatrix} 0 & \vep\ns_1 \\ -\vep\ns_1 & 0 \end{pmatrix} , \ldots ,
\begin{pmatrix} 0 & \vep\ns_\Nc \\ -\vep\ns_\Nc & 0 \end{pmatrix}\Bigg\}\quad,
\label{QtJQ}
\end{equation}
where $\{\vep\ns_\BS\}$ are the singular values of $J$ \cite{HoJo85}.
We define a new set of Majoranas $\xi\ns_\Bs=\theta^0_\Br\, Q\ns_{\Br,\Bs}\,$, and we may associate each $\Bs$ with a site on the 
$N\ns_x\times N\ns_y$ square lattice, exactly as is the case for the $\Br$ sites.  Thus, the $\Bs$ values are divided into 
\SA\ and \SB\ sublattices, and writing $\BS\in\SA$ we have $\BS+\xhat\in\SB$.  Each of the $2\times 2$ blocks in eqn. \ref{QtJQ} 
is then associated with $\Bs=\BS$ and $\Bs'=\BS+\xhat$, for some $\BS$.  We now define
\begin{equation}
c\nd_\BS=\half\big(\xi\nd_\BS - i\xi\ns_{\BS+\xhat}\big) \quad,\quad
c\yd_\BS=\half\big(\xi\nd_\BS + i\xi\nd_{\BS+\xhat}\big) \quad,\quad 
d\nd_\BR=\half\big(\theta^5_\BR - i\theta^5_{\BR+\xhat}\big) \quad,\quad 
d\yd_\BR=\half\big(\theta^5_\BR + i\theta^5_{\BR+\xhat}\big) \quad,
\end{equation}
which entail
\begin{equation}
\xi\ns_\BS=c\yd_\BS+c\nd_\BS \quad,\quad \xi\ns_{\BS+\xhat}=i(c\yd_\BS-c\nd_\BS) \quad,\quad
\theta^5_\BR=d\yd_\BR+d\nd_\BR \quad,\quad\theta^5_{\BR+\xhat}=i(d\yd_\BR-d\nd_\BR)\quad.
\end{equation}
The Hamiltonian is then $H=\sum_\BS \vep\ns_\BS\,\big(2\,c\yd_\BS c\nd_\BS-1\big)$.  We may now write
\begin{equation}
\begin{split}
\Gamma^5_\BR&=\sum_\BS\bigg\{ i\, \QAA_{\BS,\BR}\,(c\yd_\BS+c\nd_\BS)(d\yd_\BR+d\nd_\BR)
+\QBA_{\BS,\BR}\,(c\nd_\BS-c\yd_\BS)(d\yd_\BR+d\nd_\BR)\bigg\}\\
\Gamma^5_{\BR+\xhat}&=\sum_\BS\bigg\{ \QAB_{\BS,\BR}\,(c\yd_\BS+c\nd_\BS)(d\nd_\BR-d\yd_\BR)
-i\,\QBB_{\BS,\BR}\,(c\nd_\BS-c\yd_\BS)(d\nd_\BR-d\yd_\BR)\bigg\}\quad,
\end{split}
\end{equation}
where
\begin{equation}
\QAA_{\BS,\BR}=Q\nd_{\BS,\BR} \quad,\quad \QAB_{\BS,\BR}=Q\nd_{\BS,\BR+\xhat} \quad,\quad
\QBA_{\BS,\BR}=Q\nd_{\BS+\xhat,\BR} \quad,\quad \QBB_{\BS,\BR}=Q\nd_{\BS+\xhat,\BR+\xhat}\quad.
\end{equation}
Now consider the matrix elements $\sexpect{\Bm,\Bn}{\Gamma^5_\BR}{\Bm',\Bn'}$ and
$\sexpect{\Bm,\Bn}{\Gamma^5_{\BR+\xhat}}{\Bm',\Bn'}$, where
\begin{equation}
\sket{\Bm,\Bn}\equiv\prod_\BS (c\yd_\BS)^{m\ns_\BS}\,\prod_\BR (d\yd_\BR)^{n\ns_\BR}\>\sket{0}\quad,
\end{equation}
where $\sket{0}$ is the vacuum for $c$ and $d$ fermions.  Then
\begin{equation}
\sum_\Br\big|\sexpect{\Bm,\Bn}{\Gamma^5_\Br}{\Bm',\Bn'}\big|^2 =
\sum_{\BR\ns_1}\sum_{\BS\ns_1} \Big[\big(\QAA_{\BS\ns_1,\BR\ns_1}\big)^2 + \big(\QAB_{\BS\ns_1,\BR\ns_1}\big)^2 +
\big(\QBA_{\BS\ns_1,\BR\ns_1}\big)^2 + \big(\QBB_{\BS\ns_1,\BR\ns_1}\big)^2\Big]\,\detil\ns_{\Bm,\Bm',\BS\ns_1}\,
\detil\ns_{\Bn,\Bn',\BR\ns_1}\quad,
\label{Lmat}
\end{equation}
where we have defined the symbol
\begin{equation}
\detil\ns_{\Bm,\Bm',\BS\ns_1}=\delta_{m'_{\BS_1},1-m\np_{\BS_1}}\times\prod_{\BS\ne\BS\ns_1}\!\!
\delta_{m'_\BS,m\np_\BS}\quad.
\end{equation}
In other words, $\detil\ns_{\Bm,\Bm',\BS\ns_1}=1$ when $m\np_\BS=m\np_\BS$ for all $\BS$ other than $\BS\ns_1$,
where the two occupations are complementary.

When the gauge fields have the periodicity of the lattice, \ie\ when $u^\delta_\BR$=$u^\delta_{\BR'}$ for all $\BR$ and $\BR'$, 
translational invariance allows us to simplify eqn. \ref{Lmat}, in which case
\begin{equation}
\CM\ns_{\Bm\Bn,\Bm'\Bn'}={4\over N}\,\delta\ns_{d(\Bm,\Bm'),1}\,\delta\ns_{d(\Bn,\Bn'),1}-
N\delta\nd_{\Bm,\Bm'}\,\delta\ns_{\Bn,\Bn'}\quad,
\end{equation}
where $d(\Bm,\Bm')$ is the number of locations where the occupation $\Bm\np_\Bs$ differs from $\Bm'_\Bs$\,.
The eigenvalues of $\CM$ are then given by
\begin{equation}
\Lambda(\Bsigma,\Bmu)={4\over N}(\sigma\ns_1+\ldots+\sigma_\Nc)(\mu\ns_1+\ldots+\mu\ns_\Nc) - N\quad,
\end{equation}
where each $\sigma\ns_\BS$ and $\mu\ns_\BR$ are either $0$ or $1$.  When all $\sigma\ns_\BS$ and $\mu\ns_\BR$ are $+1$ or all
are $-1$, the eigenvalue is zero, corresponding to a NESS.  When one of the $\sigma$ or $\mu$ values has a reversed sign, we 
obtain $\Lambda=-2$, corresponding to a Liouvillian eigenvalue of $-2\gamma$.  Numerically, we find that the slope of the smallest
nonzero decay rate $-\Rep\Lambda(\gamma)$ is $1$ rather than $2$.  It may be that for sectors of $\CM$ corresponding to non-translationally invariant flux configurations, where the $Q$--matrices don't reflect such a symmetry, that the coefficient for the
lowest decay rate is smaller, but we do not understand how to arrive at a slope 
of $1$. Another possibility we have not explored is the dynamics of a restricted
class of coherences $\sketbra{\alpha}{\beta}$, where 
$\sket{\alpha}=\sket{\Bm,\Bn\ns_1}$ and $\sket{\beta}=\sket{\Bm,\Bn\ns_2}$. 
There are $2^{5N/2}$ such coherences, all of which commute with $H$, 
arranged into $2^N$ blocks of size $2^{3N/2}$.  The generalization of the matrix
$\CM$ in each block is then
\begin{equation}
\CM\ns_{\Bm,\Bn\np_1,\Bn\np_2,\Bm',\Bn'_1,\Bn'_2}=
\sum_\Br \sum_\BM \sum_{\BN\ns_1}\sum_{\BN\ns_2}\Big(
\sexpect{\BM,\BN\ns_1}{\Gamma^5_\Br}{\Bn,\Bn\ns_1}
\sexpect{\Bm,\Bn\ns_2}{\Gamma^5_\Br}{\BM,\BN\ns_2}
-\delta\ns_{\BM,\Bm}\,\delta\ns_{\BM\ns_1,\Bn\ns_1}\,
\delta\ns_{\BM\ns_2,\Bn\ns_2}\Big)\quad.
\end{equation}
While the the matrix element products
\begin{equation}
\sexpect{\BN\ns_1}{d\yd_\BR\pm d\nd_\BR}{\Bn\ns_1}
\sexpect{\Bn\ns_2}{d\yd_\BR\pm d\nd_\BR}{\BN\ns_2}
\end{equation}
are nonzero only if $\BN\ns_1$ and $\Bn\ns_1$ have complementary occupancies
in the same location $\BR$ as do $\BN\ns_2$ and $\Bn\ns_2$, the presence of
occupation number dependent sign factors complicates the analysis.

\subsection{$\gamma\gg 1$}
We start with the Dirac matrices,
\begin{align}
\Gamma^1=X\otimes 1 &=\begin{pmatrix} 0 & 0 & 1 & 0 \\ 0 & 0 & 0 & 1 \\
1 & 0 & 0 & 0 \\ 0 & 1 & 0 & 0 \end{pmatrix} &
\Gamma^2=Y\otimes 1 &=\begin{pmatrix} 0 & 0 & -i & 0 \\ 0 & 0 & 0 & -i \\
i & 0 & 0 & 0 \\ 0 & i & 0 & 0 \end{pmatrix} &
\Gamma^3=Z\otimes X &=\begin{pmatrix} 0 & 1 & 0 & 0 \\ 1 & 0 & 0 & 0 \\
0 & 0 & 0 & -1 \\ 0 & 0 & -1 & 0 \end{pmatrix} \nonumber \\
\Gamma^4=Z\otimes Y &=\begin{pmatrix} 0 & -i & 0 & 0 \\ i & 0 & 0 & 0 \\
0 & 0 & 0 & -i \\ 0 & 0 & i & 0 \end{pmatrix} &
\Gamma^5=Z\otimes Z &=\begin{pmatrix} 1 & 0 & 0 & 0 \\ 0 & -1 & 0 & 0 \\
0 & 0 & -1 & 0 \\ 0 & 0 & 0 & 1 \end{pmatrix} \quad. & &
\end{align}
We label the eigenstates of $\Gamma^5$ by an index $\mu\in\{1,2,3,4\}$,
with eigenvectors $\psi^{(\mu)}_i=\delta\ns_{i,\mu}$ and with eigenvalues
$\zeta\ns_\mu = \{1,-1,-1,1\}$, respectively.  Since
$\CLD\,\vrh=\gamma\sum_\Br\big(\Gamma^5_\Br\,\vrh\,\Gamma^5_\Br-\vrh\big)$,
any operator $\munu$ is annihilated by $\CLD$ provided
it is an eigenoperator under the action of $\Gamma^5_\Br$ from either the
left or the right, at every site $\Br$.  There are $8^N$ such operators,
since we can freely choose each of the $\mu\ns_\Br$ so long as
$\nu\ns_\Br=\mu\ns_\Br$ or $\nu\ns_\Br=5-\mu\ns_\Br$.  In general, the
eigenoperators of $\CLD$ are arranged into sectors $\Ups\ns_k$, where
$\munu\in\Ups\ns_k$ provided $\zeta_{\mu\ns_\Br}\ne\zeta_{\nu\ns_\Br}$ 
at $k$ locations $\Br$. The eigenvalue under $\CLD$ for any operator 
in sector $\Ups\ns_k$ is then $-2k\gamma$. 

We label the $8^N$ operators in the $\Ups\ns_0$ sector as
$A\ns_\Bp=\munu$, with $p\ns_\Br=\mu\ns_\Br$ if $\nu\ns_\Br=\mu\ns_\Br$ and
$p\ns_\Br=\mu\ns_\Br+4$ if $\nu\ns_\Br=5-\mu\ns_\Br$ at each site $\Br$. 
We also define the $N\cdot 8^N$ operators
\begin{equation}
B^l_\Bp=i\,\munu\, H\ns_l - i\, H\ns_l\, \munu\quad,
\end{equation}
where $l$ denotes one of the $2N$ links $(\BR,\BR+\Bdel)$ and where
$H\ns_l=J\ns_\delta\,\Gamma^\delta_\BR\,\Gamma^\delta_{\BR+\Bdel}$.
For simplicity we shall assume $J\ns_\delta=1$ for each of the four types of links,
although our method described below can easily be applied to the more general case.
Since each of the matrices $\Gamma^{1,2,3,4}$ anticommutes with $\Gamma^5$, 
its application reverses the $\Gamma^5$ eigenvalue, and thus $B^l_\Bp\in S\ns_2$
for all $\Bp$ and links $l$.  We then have
\begin{equation}
\CLH A\ns_\Bp=\sum_l B^l_\Bp\quad.
\end{equation}
We wish to analyze the Liouvillian $\CL=\CLH+\CLD$ when restricted to
the subspace $S\ns_0\cup S'_2$, where $S'_2\subset S\ns_2$ includes
operators $\munu$ where there are differences in the $\Gamma^5$ eigenvalues
of the bra and ket states at two sites from the same link.  In other words,
we restrict the action of $\CL$ to the subspace of operators spanned by
the $A\ns_\Bp$ and the $B^l_\Bp$.  The total dimension of this operator space
is then $(2N+1)\cdot 8^N$, since there are $2N$ values of $l$.

We now need to evaluate $\CLH B^l_\Bp$.  We have
\begin{equation}
\CLH B^l_\Bp = 2\,A^l_\Bp - 2\,A\ns_\Bp + \ldots\quad,
\end{equation}
where $A^l_\Bp\equiv H\ns_l\,\munu\,H\ns_l$, which also this lives in
sector $S\ns_0$ provided $\munu\in S\ns_0$.  The remaining terms include
operators in other sectors $S\ns_{k>2}$ and operators in $S\ns_2$ where
the differences in the $\Gamma^5_\Br$ eigenvalues of the bra and ket states
are at two sites not connected by a link, thus requiring two applications
of $\CLH$ to reach from $S\ns_0$\,.  In this basis, the matrix form of
the projected Liouvillian $\CLt$ is
\begin{equation}
\CLt=\begin{pmatrix} \Mzero & -2R\ns_1 & -2R\ns_2 & \cdots & -2R\ns_M \\
\Mone & -4\gamma\Mone & \Mzero &  & \Mzero \\
\Mone & \Mzero & -4\gamma\Mone & & \Mzero \\
\vdots & & & \ddots & \vdots \\
\Mone & \Mzero & \cdots & & -4\gamma\Mone\end{pmatrix} 
\qquad\Rightarrow\qquad
\omega -\CLt=\begin{pmatrix} \omega\,\Mone & 2R\ns_1 & 2R\ns_2 & \cdots & 2R\ns_M \\
-\Mone & (\omega+4\gamma)\Mone & \Mzero &  & \Mzero \\
-\Mone & \Mzero & (\omega+4\gamma)\Mone & & \Mzero \\
\vdots & & & \ddots & \vdots \\
-\Mone & \Mzero & \cdots & & (\omega+4\gamma)\Mone\end{pmatrix}\quad,
\label{omLt}
\end{equation}
where $M=2N$ is the number of links in the square lattice (with periodic 
boundary conditions.) We can now perform row and column reduction 
on the matrix $\omega-\CLt$ to obtain
\begin{equation}
\omega-\CLt \quad \longmapsto \quad
\begin{pmatrix} \omega & 2\CS\ns_M & 2\CS\ns_{M-1} & \cdots && 2\CS\ns_1 \\
-\Mone & (\omega+4\gamma)\Mone  & \Mzero && & \Mzero \\
\Mzero & \Mzero & (\omega+4\gamma)\Mone && & \Mzero \\
\Mzero & \Mzero & \Mzero  && & \Mzero \\
\vdots&&&\ddots&&\vdots\\
&&&&(\omega+4\gamma)\Mone&\Mzero\\
\Mzero & \cdots & && \Mzero & (\omega+4\gamma)\Mone
\end{pmatrix}\quad,
\label{omLtred}
\end{equation}
where $\CS\ns_k=\sum_{l=1}^k R\ns_l$\,, and the characteristic polynomial is
\begin{equation}
P(\omega)=(\omega+4\gamma)^{(M-1)D}\,
\textsf{det}\big(\omega^2+4\gamma\omega+2\CS\big)\quad,
\end{equation}
where $D=8^N$ and where we define $\CS\equiv\CS\ns_M$.  Thus there are $(M-1)D$ degenerate eigenvalues with 
$\omega=4\gamma$ and $2D$ eigenvalues
\begin{equation}
\omega\ns_{j,\pm}=-2\gamma\pm\sqrt{4\gamma^2-2s\ns_j}\quad,
\end{equation}
where $\{s\ns_j\}$ are the $D$ eigenvalues of the matrix $\CS$.  In the limit
$\gamma\gg s\ns_j$ we then have $\omega\ns_{-,j}=-s\ns_j/2\gamma+\CO(\gamma^{-2})$.
{\sl Note:} Regarding the row and column reduction of $\omega-\CLt$, starting with the expression for $\omega-\CLt$
in eqn. \ref{omLt}, subtract the penultimate $(M^{\rm th})$ block row from the final $\big((M+1)^{\rm th}\big)$ one.  
Then add the last block column from the penultimate block column.  These two operations have the effect of eliminating
the leftmost $-\Mone$ block in the $(M+1)^{\rm th}$ block row and replacing $R\ns_{M-1}$ with $R\ns_{M-1}+R\ns_M$ at
the top of the $M^{\rm th}$ block column.  Iterate this process until obtaining the matrix in eqn. \ref{omLtred}.

\begin{figure}[t]
\begin{centering}
\includegraphics[width=0.75\textwidth]{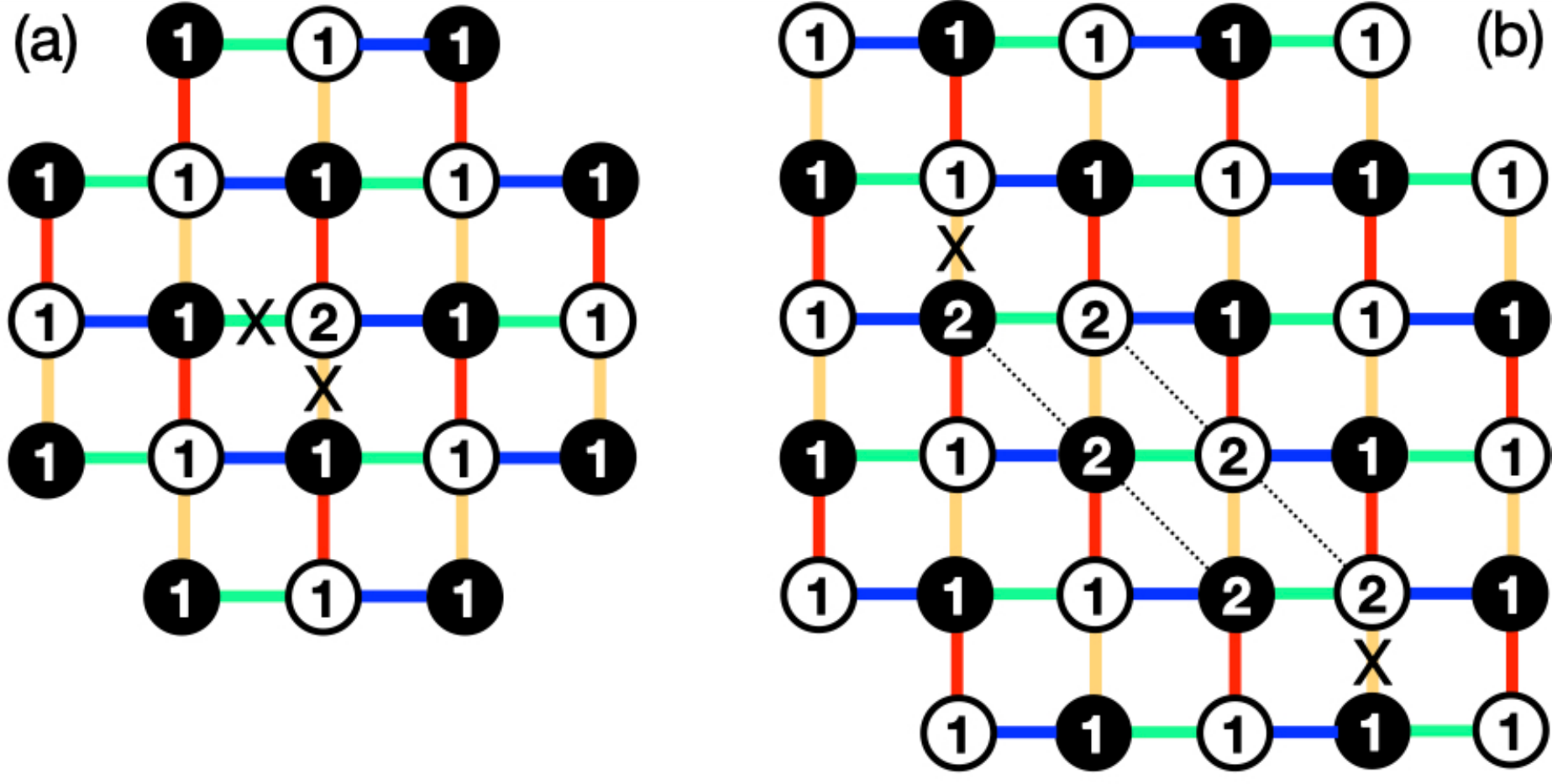}
\end{centering}
\caption{Finite lifetime modes of the Liouvillian $\CL$, with sited labeled by
their $q\ns_\Br$ labels. (a) A configuration with $s=4$, and (b) a string
defect with $s=4$.  In both cases, $\omega=-2/\gamma$ as $\gamma\to\infty$
is the relaxation rate.
\label{Fdefect}}
\end{figure} 

To find the spectrum of $\CS$, we note that
\begin{equation}
\CS(\Bp'\,|\,\Bp)=M \delta\ns_{\Bmu\Bmu'}\delta\ns_{\Bnu\Bnu'}
-\sum_{l=1}^M \sexpect{\Bmu'}{H\ns_l}{\Bmu}\sexpect{\Bnu}{H\ns_l}{\Bnu'}\quad,
\end{equation}
where each $p\ns_\Br=p(\mu\ns_\Br,\nu\ns_\Br)\in\{1,\ldots,8\}$ is a composite
index, as described above.  We find
\begin{equation}
\Gamma^\delta A\ns_p\,\Gamma^\delta = \sum_{p'}\Delta^\delta_{p,p'}\,A\ns_{p'} 
\end{equation}
with
\begin{equation}
\Delta^{1,2}=\Gamma^1\oplus \pm\Gamma^1\qquad,\qquad
\Delta^{3,4}=\Gamma^{54}\oplus \mp\Gamma^{54}\quad.
\end{equation}
For example,
\begin{equation}
\begin{split}
\Gamma^1 A\ns_1\,\Gamma^1 &=\Gamma^1\,\sketbra{1}{1}\,\Gamma^1
=+\sketbra{3}{3}=+A\ns_3\\
\Gamma^2 A\ns_5\,\Gamma^2 &=\Gamma^2\,\sketbra{1}{4}\,\Gamma^2
=-\sketbra{3}{2}=+A\ns_7\\
\Gamma^3 A\ns_6\,\Gamma^3 &=\Gamma^3\,\sketbra{2}{3}\,\Gamma^3 
=-\sketbra{1}{4}=-A\ns_5\\
\Gamma^4 A\ns_3\,\Gamma^4 &=\Gamma^4\,\sketbra{3}{3}\,\Gamma^4 
=+\sketbra{4}{4}=-A\ns_4\quad.
\end{split}
\end{equation}
Note that $\Gamma^{54}=\Mone\otimes X$.  Since $\Gamma^1$ and $\Gamma^{54}$ commute, we 
may find a common basis:
\begin{equation}
\vvphi\ns_1=\begin{pmatrix} 1 \\ 1 \\ 1 \\ 1 \end{pmatrix} \quad,\quad
\vvphi\ns_2=\begin{pmatrix} 1 \\ -1 \\ 1 \\ -1 \end{pmatrix} \quad,\quad
\vvphi\ns_3=\begin{pmatrix} 1 \\ 1 \\ -1 \\ -1 \end{pmatrix} \quad,\quad
\vvphi\ns_4=\begin{pmatrix} 1 \\ -1 \\ -1 \\ 1 \end{pmatrix} \quad.
\end{equation}
In this basis, we have
\begin{equation}
\begin{split}
\Detil^1&=\textsf{diag}(+,+,-,-,+,+,-,-)\\
\Detil^2&=\textsf{diag}(+,+,-,-,-,-,+,+)\\
\Detil^3&=\textsf{diag}(+,-,+,-,-,+,-,+)\\
\Detil^4&=\textsf{diag}(+,-,+,-,+,-,+,-)\quad.
\label{Detil}
\end{split}
\end{equation}
The eigenvalues of $\CS$ are thus given by 
\begin{equation}
s(q\ns_1,\ldots,q\ns_M)=\sum_\BR\sum_{\delta=1}^4 \Big( 1 - \Detil^\delta(q\ns_\BR)
\,\Detil^\delta(q\ns_{\BR+\Bdel})\Big)\quad,
\end{equation}
where each $q\ns_\Br\in\{1,\ldots,8\}$ and the values $\Detil^\delta(q)$ 
are given in eqn. \ref{Detil}.  The $q$ values label the linear combinations of
single-site density matrices associated with the eight common eigenvectors
of the matrices $\Delta^\delta$,
\ie\ $\begin{pmatrix} \vphi\ns_\eta \\ \pm\vphi\ns_\eta\end{pmatrix}$,
with index $\eta\in\{1,2,3,4\}.$

Clearly any configuration with all $q\ns_\Br$ equal will be a NESS, with $\CS$
eigenvalue $s=0$.  In fig. \ref{Fdefect} we sketch the configurations for two
excited (\ie\ decaying) modes, with nonzero eigenvalues of $\CS$.  Panel (a)
depicts a configuration with a single site defect, an isolated $q=2$ state in
a sea of $q=1$.  According to the rules derived here, configuration (a) has
`bad' bonds (labeled with an \textsf{X}) for which $\Detil^\delta(q)\,\Detil^\delta(q')=-1$, resulting in an eigenvalue $s=4$.  
For configuration (b), which features two parallel
diagonal line defects, there are also two bad bonds, but with a string
connecting them, again giving $s=4$.  Both these configurations are highly degenerate.  In the case of open boundary conditions, the string can run
to a boundary, and one has a defect state with $s=2$.  Numerically, though,
we find $s=2$ to be the lowest eigenvalue of $\CS$ even with periodic 
boundary conditions.

In summary, we have analytical arguments for $\omega\propto\gamma$ as
$\gamma\to 0$ and $\omega\propto\gamma^{-1}$ as $\gamma\to\infty$ based on
perturbation theory, but the analytical value of the coefficient is twice
that obtained from the numerical analysis.

\section{Computational Procedure}\label{appendix_2d_model}
\label{appendix_2d_model_genetic}
The genetic algorithm that we use to estimate the was as follows. For a given gauge 
field configuration $F=\big\{\{u^\delta_\BR\},\{u^5_{\Br}\}\big\}$ we find the smallest
relaxation rate gap $g\ns_F$ by solving for the spectrum of $\CW$ using Prosen's method
\cite{Prosen2008}.  In the genetic algorithm, the fitness function is given by the gap
$g\ns_F$ and we try to further minimize this by varying over $F$. We start with a
population of randomly chosen individuals (field configurations). We find individuals
with low values of gap perform crossovers and mutate by flipping the sign at some 
places. We then calculate the gaps for the individuals in the new population and 
repeat the same process. 

\begin{figure}[H]
	\centering
	\begin{tikzpicture}
	\node (img1){\includegraphics[width=8cm,height=6cm]{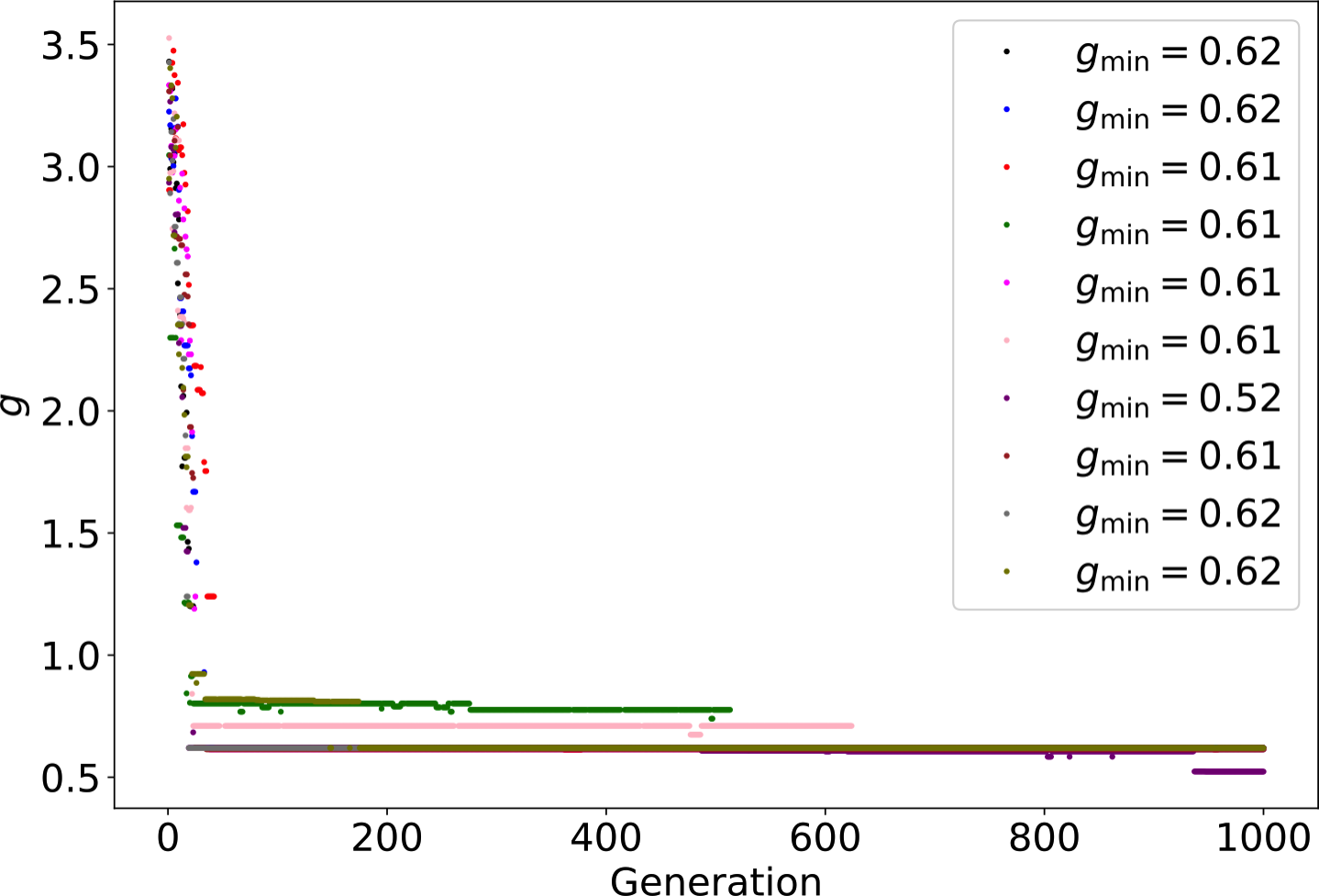}};
	\node (img2) [right= of img1,node distance=0cm,xshift=0cm]{\includegraphics[width=8cm,height=6cm]{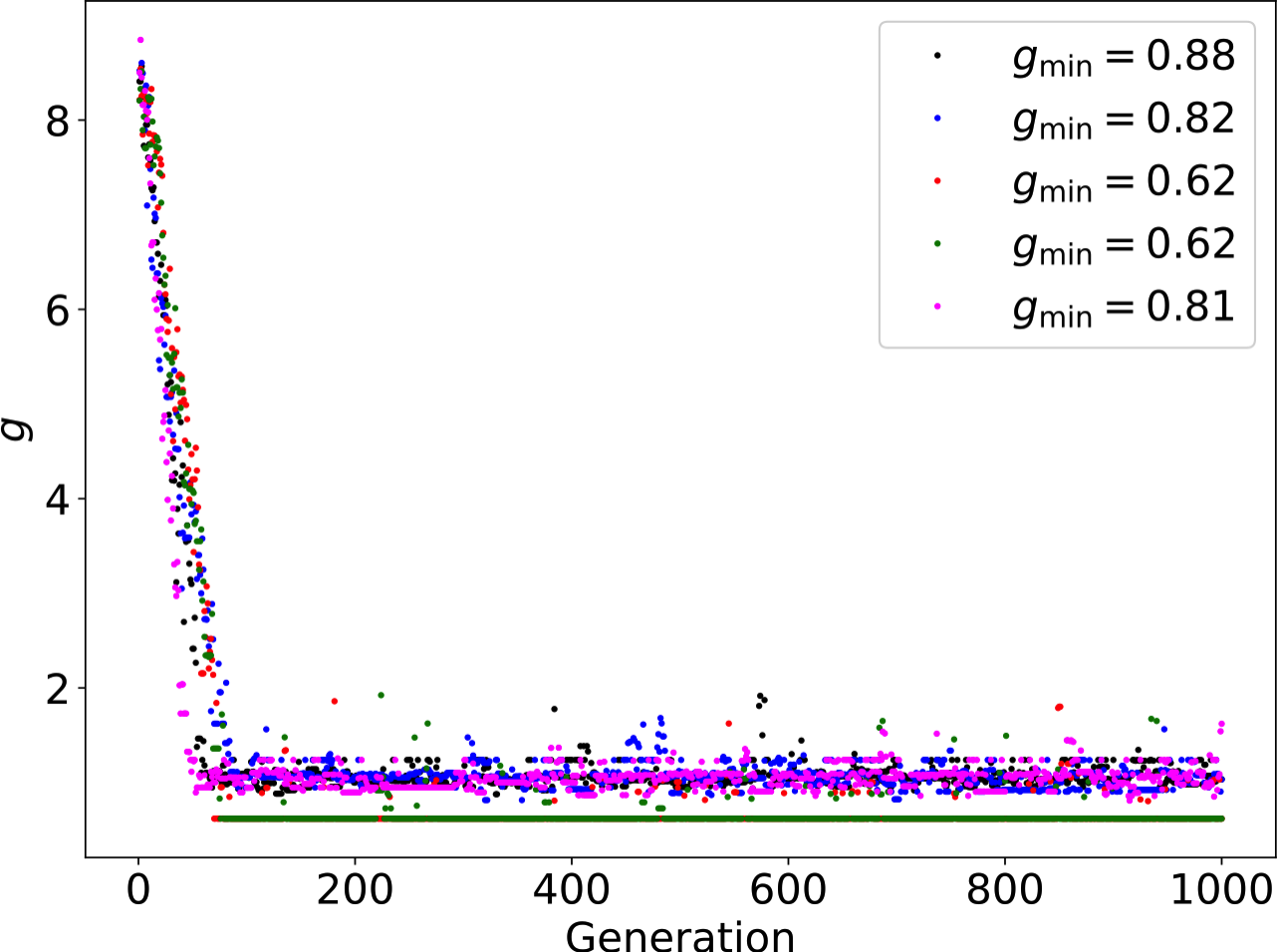}};
	\node[left=of img1,node distance=0cm,yshift=0.5cm,xshift=1.0cm,rotate=0]{(a)};
	\node[left=of img2,node distance=0cm,yshift=0.4cm,xshift=1.0cm,rotate=0]{(b)};
	\end{tikzpicture}
	\caption{Different runs of the genetic algorithm. Each color depicts a run with a randomly chosen initial population for $\gamma=0.31$ and all $J\ns_\delta=1$ for different system sizes: (a) $\Nx=\Ny=4$, population size 100. (b) $\Nx=\Ny=6$, population size 100. In these plots, $g_{\text{min}}$ refers to the minimum value of the gap encountered in the run.}
	\label{2d_genetic_algo_runs_sample}
\end{figure}

\begin{figure}[H]
	\centering
	\begin{tikzpicture}
	\node (img1){\includegraphics[width=8cm,height=6cm]{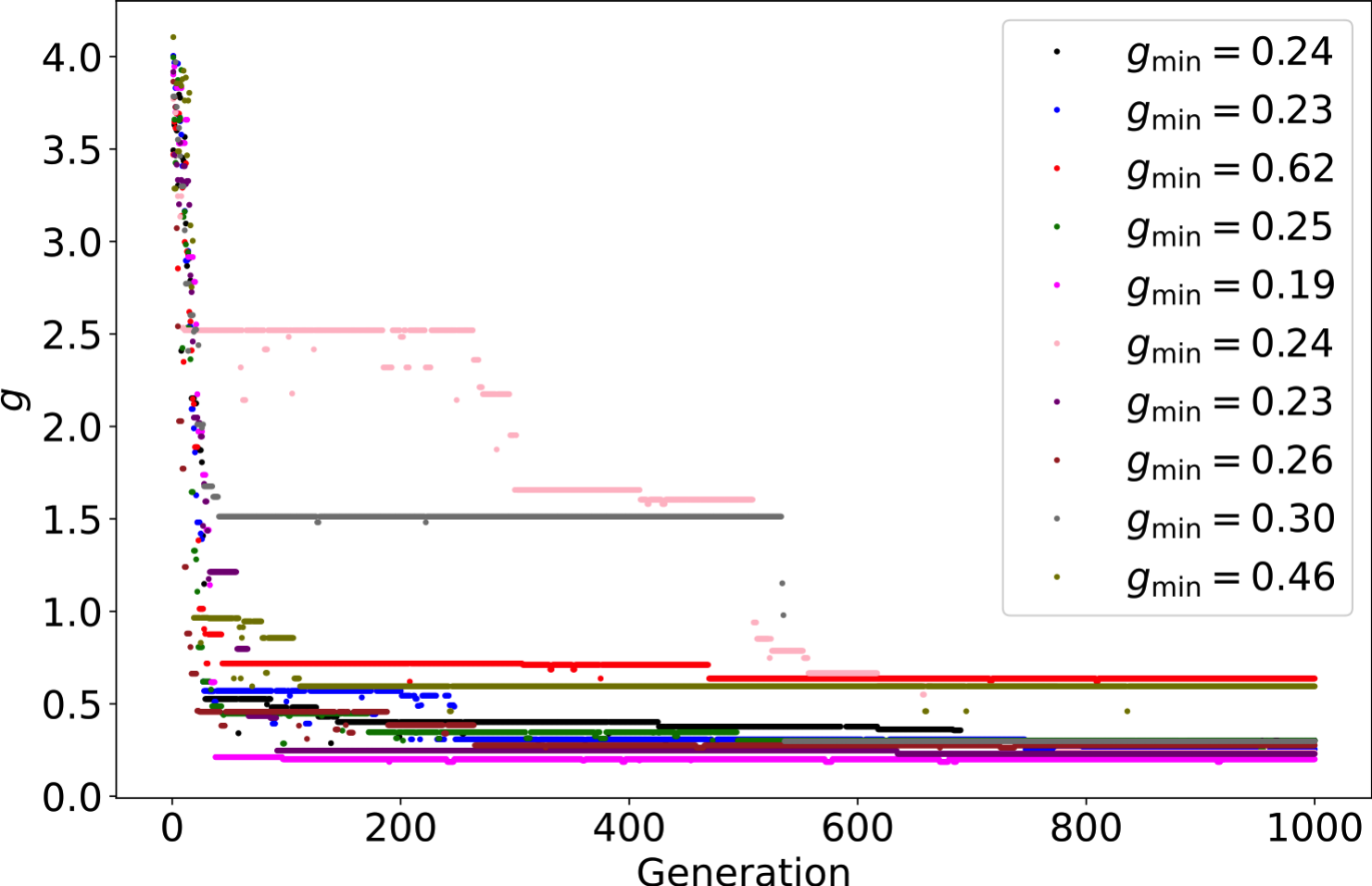}};
	\node (img2) [right= of img1,node distance=0cm,xshift=0cm]{\includegraphics[width=8cm,height=6cm]{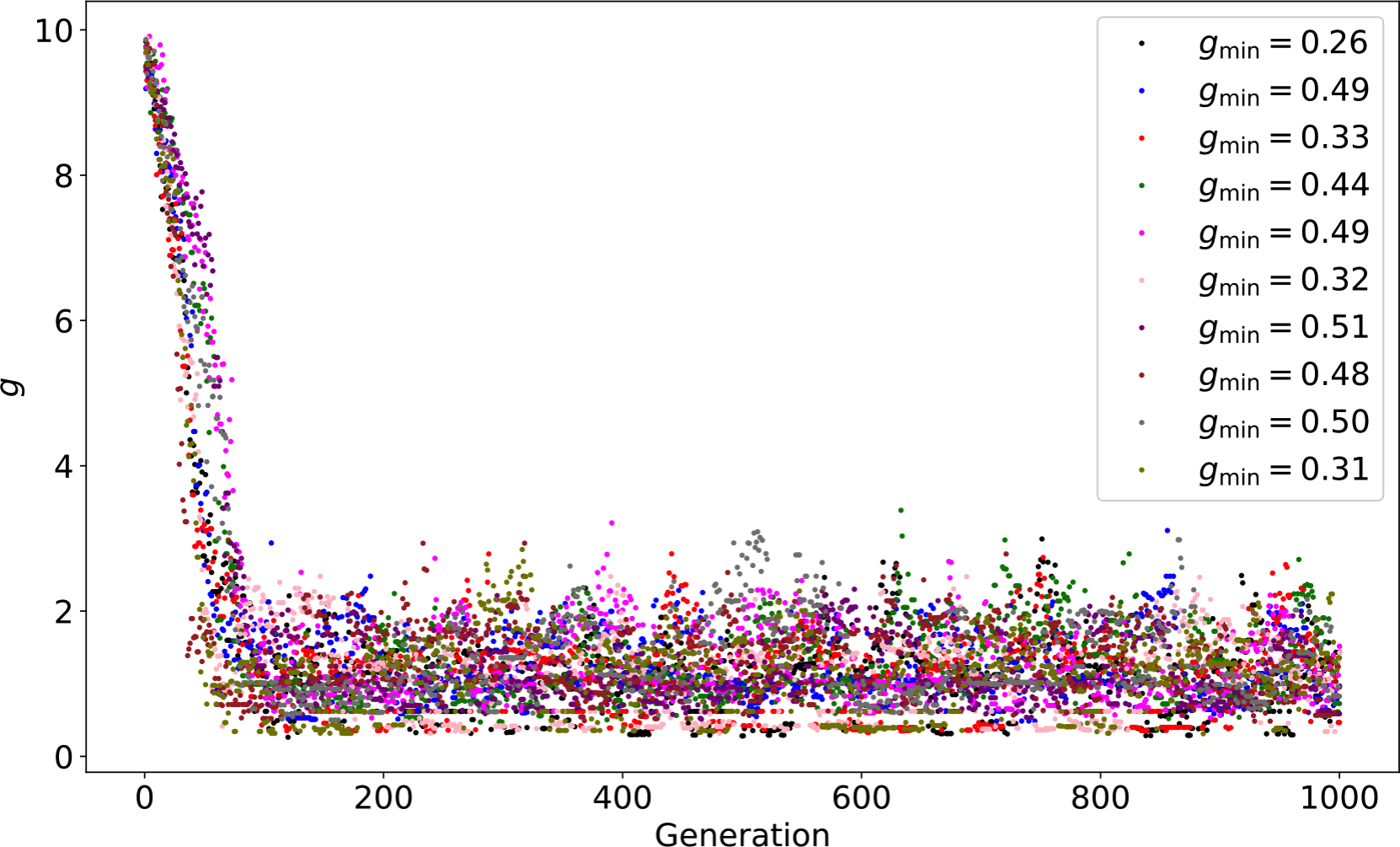}};
	\node[left=of img1,node distance=0cm,yshift=0.5cm,xshift=1.0cm,rotate=0]{(a)};
	\node[left=of img2,node distance=0cm,yshift=0.4cm,xshift=1.0cm,rotate=0]{(b)};
	\end{tikzpicture}
	\caption{Different runs of the genetic algorithm. Each color depicts a run with a randomly chosen initial population for $\gamma=0.31$ and $\left(J_1,J_2,J_3,J_4\right)=\left(3,4,1,2\right)$ for different system sizes: (a) $\Nx=\Ny=4$, population size 100. (b) $\Nx=\Ny=6$, population size 100. In these plots, $g_{\text{min}}$ refers to the minimum value of the gap encountered in the run.}
	\label{2d_genetic_algo_runs_sample_2}
\end{figure}

We do this over different runs, \ie\ by starting with different randomly chosen 
initial populations. We pick the minimum $g$ value obtained over different runs to
be the estimate of the gap. The values over different runs for a particular value 
of $\gamma$ are shown in figures \ref{2d_genetic_algo_runs_sample} and \ref{2d_genetic_algo_runs_sample_2} for different values of $J_1$, $J_2$, 
$J_3$ and $J_4$.

Increasing the population size, letting the simulation run for a larger number of generations, \etc\ can be used to reduce the spread of data over different runs.

\section{Some Flux Configurations from the Genetic Algorithm}\label{configs}
Here we list some flux configurations for first decay modes obtained from the 
genetic algorithm.  Relative to the NESS where all the gauge-invariant $\MZ\ns_2$
data are set to $-1$, we list all the $+1$ defects (\ie\ plaquette fluxes and 
Wilson phases).  As noted in the text, it may be that each of the states listed
here may be describable as having fewer defects with respect to others among the
exponentially many ($2^{3N+1}$) states in the NESS block.

\vskip 0.2in

For $N_x=N_y=4$, $J_1=J_2=J_3=J_4=1$:
\begin{itemize}
    \item \bm{$\gamma<0.41$}: $\Phi^+_{ 1 , 1 }$,
$\Phi^-_{ 4 , 2 }$,
$\Phi^-_{ 1 , 3 }$,
$\Phi^+_{ 4 , 2 }$,
$\Phi^-_{ 2 , 4 }$,
$\Phi^+_{ 3 , 3 }$,
$\Phi^+_{ 2 , 4 }$,
$\Phi^-_{ 3 , 1 }$,
$\Omega^+_{ 1 , 1 }$,
$\Omega^-_{ 1 , 3 }$,
$\Psi^-_{ 2 , 2 }$,
$\Psi^+_{ 4 , 2 }$,
$W_x$,
$W_y$
    \item \bm{$\gamma=0.41$}: $\Phi^+_{ 1 , 1 }$,
$\Phi^-_{ 2 , 2 }$,
$\Phi^+_{ 3 , 1 }$,
$\Phi^-_{ 4 , 2 }$,
$\Phi^-_{ 1 , 3 }$,
$\Phi^+_{ 2 , 2 }$,
$\Phi^-_{ 3 , 3 }$,
$\Phi^+_{ 1 , 3 }$,
$\Phi^-_{ 2 , 4 }$,
$\Phi^+_{ 3 , 3 }$,
$\Phi^-_{ 1 , 1 }$,
$\Phi^+_{ 2 , 4 }$,
$\Phi^-_{ 3 , 1 }$,
$\Phi^+_{ 4 , 4 }$,
$\Omega^+_{ 4 , 2 }$
    \item \bm{$\gamma>0.41$}: $\Phi^+_{ 1 , 3 }$,
$\Phi^+_{ 3 , 3 }$,
$\Phi^-_{ 4 , 4 }$,
$\Phi^+_{ 2 , 4 }$,
$\Phi^-_{ 3 , 1 }$,
$\Phi^+_{ 4 , 4 }$,
$\Omega^-_{ 4 , 4 }$
\end{itemize}

\vskip 0.2in

For $N_x=N_y=6$, $J_1=J_2=J_3=J_4=1$:
\begin{itemize}
    \item \bm{$\gamma<0.36$}: $\Phi^+_{ 3 , 1 }$,
$\Phi^-_{ 1 , 3 }$,
$\Phi^+_{ 2 , 2 }$,
$\Phi^-_{ 3 , 3 }$,
$\Phi^-_{ 5 , 3 }$,
$\Phi^+_{ 6 , 2 }$,
$\Phi^-_{ 2 , 4 }$,
$\Phi^+_{ 3 , 3 }$,
$\Phi^-_{ 6 , 4 }$,
$\Phi^-_{ 1 , 5 }$,
$\Phi^-_{ 3 , 5 }$,
$\Phi^+_{ 4 , 4 }$,
$\Phi^+_{ 6 , 4 }$,
$\Phi^-_{ 2 , 6 }$,
$\Phi^+_{ 5 , 5 }$,
$\Phi^-_{ 1 , 1 }$,
$\Phi^-_{ 5 , 1 }$,
$\Phi^+_{ 6 , 6 }$,
$W_y$
    \item \bm{$0.36 \leq \gamma \leq 0.41$}: $\Phi^+_{ 3 , 1 }$,
$\Phi^-_{ 1 , 3 }$,
$\Phi^+_{ 2 , 2 }$,
$\Phi^+_{ 1 , 3 }$,
$\Phi^+_{ 3 , 3 }$,
$\Phi^-_{ 6 , 4 }$,
$\Phi^+_{ 4 , 4 }$,
$\Phi^+_{ 1 , 5 }$,
$\Phi^-_{ 2 , 6 }$,
$\Phi^-_{ 4 , 6 }$,
$\Phi^+_{ 5 , 5 }$,
$\Phi^+_{ 2 , 6 }$,
$\Psi^-_{ 2 , 6 }$,
$W_x$
    \item \bm{$\gamma>0.41$}: $\Phi^+_{ 1 , 1 }$,
$\Phi^+_{ 2 , 2 }$,
$\Phi^+_{ 4 , 2 }$,
$\Phi^+_{ 6 , 2 }$,
$\Phi^+_{ 1 , 3 }$,
$\Phi^+_{ 3 , 3 }$,
$\Phi^-_{ 4 , 4 }$,
$\Phi^-_{ 6 , 4 }$,
$\Phi^+_{ 2 , 4 }$,
$\Phi^+_{ 6 , 4 }$,
$\Phi^-_{ 2 , 6 }$,
$\Phi^-_{ 1 , 1 }$,
$\Phi^-_{ 5 , 1 }$,
$\Phi^+_{ 6 , 6 }$,
$\Omega^+_{ 4 , 2 }$,
$W_x$,
$W_y$
\end{itemize}

\vskip 0.2in

For $N_x=N_y=4$, $(J_1,J_2,J_3,J_4)=(3,4,1,2)$:
\begin{itemize}
    \item \bm{$\gamma<0.86$}: $\Phi^+_{ 1 , 1 }$,
$\Phi^-_{ 2 , 2 }$,
$\Phi^+_{ 3 , 1 }$,
$\Phi^-_{ 4 , 2 }$,
$\Phi^-_{ 1 , 3 }$,
$\Phi^+_{ 2 , 2 }$,
$\Phi^-_{ 3 , 3 }$,
$\Phi^+_{ 4 , 2 }$,
$\Phi^+_{ 1 , 3 }$,
$\Phi^-_{ 2 , 4 }$,
$\Phi^-_{ 4 , 4 }$,
$\Phi^-_{ 1 , 1 }$,
$\Phi^+_{ 2 , 4 }$,
$\Phi^+_{ 4 , 4 }$,
$\Psi^-_{ 2 , 4 }$,
$\Psi^-_{ 4 , 4 }$,
$W_x$
    \item \bm{$0.86 \leq \gamma \leq 1.06$}: $\Phi^+_{ 1 , 1 }$,
$\Phi^-_{ 2 , 2 }$,
$\Phi^-_{ 4 , 2 }$,
$\Phi^-_{ 1 , 3 }$,
$\Phi^+_{ 2 , 2 }$,
$\Phi^+_{ 4 , 2 }$,
$\Phi^+_{ 1 , 3 }$,
$\Phi^+_{ 3 , 3 }$,
$\Phi^-_{ 4 , 4 }$,
$\Phi^-_{ 1 , 1 }$,
$\Phi^+_{ 2 , 4 }$,
$\Phi^+_{ 4 , 4 }$,
$\Psi^-_{ 4 , 4 }$,
$W_y$
    \item \bm{$\gamma>1.06$}: $\Phi^-_{ 4 , 2 }$,
$\Phi^-_{ 1 , 3 }$,
$\Phi^-_{ 3 , 3 }$,
$\Phi^+_{ 4 , 2 }$,
$\Phi^-_{ 2 , 4 }$,
$\Phi^+_{ 3 , 3 }$,
$\Phi^-_{ 4 , 4 }$,
$\Phi^-_{ 1 , 1 }$,
$\Phi^+_{ 2 , 4 }$,
$\Phi^+_{ 4 , 4 }$,
$\Psi^-_{ 4 , 2 }$,
$W_x$,
$W_y$
\end{itemize}

\vskip 0.2in

For $N_x=N_y=6$, $(J_1,J_2,J_3,J_4)=(3,4,1,2)$:
\begin{itemize}
    \item \bm{$\gamma<0.16$}: $\Phi^+_{ 1 , 1 }$,
$\Phi^-_{ 4 , 2 }$,
$\Phi^+_{ 5 , 1 }$,
$\Phi^+_{ 2 , 2 }$,
$\Phi^+_{ 4 , 2 }$,
$\Phi^+_{ 6 , 2 }$,
$\Phi^+_{ 1 , 3 }$,
$\Phi^-_{ 2 , 4 }$,
$\Phi^+_{ 3 , 3 }$,
$\Phi^-_{ 4 , 4 }$,
$\Phi^-_{ 6 , 4 }$,
$\Phi^-_{ 1 , 5 }$,
$\Phi^-_{ 3 , 5 }$,
$\Phi^+_{ 4 , 4 }$,
$\Phi^-_{ 5 , 5 }$,
$\Phi^-_{ 4 , 6 }$,
$\Phi^+_{ 5 , 5 }$,
$\Phi^-_{ 6 , 6 }$,
$\Phi^+_{ 2 , 6 }$,
$\Phi^-_{ 3 , 1 }$,
$\Omega^+_{ 3 , 1 }$,
$\Omega^-_{ 3 , 1 }$,
$\Psi^-_{ 3 , 1 }$,
$\Psi^+_{ 3 , 1 }$
    \item \bm{$0.16 \leq \gamma \leq 0.51$}: $\Phi^+_{ 1 , 1 }$,
$\Phi^-_{ 2 , 2 }$,
$\Phi^+_{ 3 , 1 }$,
$\Phi^+_{ 4 , 2 }$,
$\Phi^+_{ 6 , 2 }$,
$\Phi^+_{ 1 , 3 }$,
$\Phi^-_{ 2 , 4 }$,
$\Phi^+_{ 3 , 3 }$,
$\Phi^-_{ 4 , 4 }$,
$\Phi^+_{ 5 , 3 }$,
$\Phi^+_{ 2 , 4 }$,
$\Phi^-_{ 3 , 5 }$,
$\Phi^-_{ 5 , 5 }$,
$\Phi^+_{ 6 , 4 }$,
$\Phi^-_{ 2 , 6 }$,
$\Phi^+_{ 2 , 6 }$,
$\Phi^-_{ 3 , 1 }$,
$\Phi^+_{ 4 , 6 }$,
$\Psi^-_{ 3 , 3 }$,
$W_y$
    \item \bm{$\gamma>0.51$}: $\Phi^+_{ 1 , 1 }$,
$\Phi^+_{ 3 , 1 }$,
$\Phi^-_{ 4 , 2 }$,
$\Phi^+_{ 5 , 1 }$,
$\Phi^+_{ 2 , 2 }$,
$\Phi^+_{ 4 , 2 }$,
$\Phi^-_{ 5 , 3 }$,
$\Phi^-_{ 2 , 4 }$,
$\Phi^+_{ 3 , 3 }$,
$\Phi^+_{ 5 , 3 }$,
$\Phi^-_{ 6 , 4 }$,
$\Phi^+_{ 2 , 4 }$,
$\Phi^+_{ 1 , 5 }$,
$\Phi^+_{ 5 , 5 }$,
$\Phi^-_{ 6 , 6 }$,
$\Phi^-_{ 3 , 1 }$,
$\Phi^-_{ 5 , 1 }$,
$\Phi^+_{ 6 , 6 }$,
$\Psi^-_{ 1 , 1 }$,
$W_x$,
$W_y$
\end{itemize}

\end{document}